\documentclass[a4paper,review]{elsarticle}

\usepackage{lineno,hyperref}

\usepackage{fullpage}

\journal{Nuclear Physics B}

\usepackage[numbers]{natbib}

\usepackage[T1]{fontenc}
\usepackage{amssymb}
\usepackage{mathtools}
\usepackage{amsmath}
\usepackage{amsfonts}
\usepackage{upgreek}

\usepackage{booktabs}

\makeatletter
\newcommand*{\rom}[1]{\expandafter\@slowromancap\romannumeral #1@}
\makeatother

\begin{document}

\begin{frontmatter}

\title{Quasinormal Modes and Phase Transitions of Regular Black Holes}

\author{Chen Lan}
\ead{lanchen@nankai.edu.cn}

\author{Yan-Gang Miao\corref{cor1}}
\cortext[cor1]{Corresponding author}
\ead{miaoyg@nankai.edu.cn}

\author{Hao Yang}
\ead{yanghao4654@mail.nankai.edu.cn}

\address{School of Physics, Nankai University, Tianjin 300071, China}

\begin{abstract}
By applying the dimensionless scheme,
we investigate the quasinormal modes
and phase transitions analytically for three types of regular black holes.
The universal deviations to the first law of mechanics in regular black holes are proved.
Meanwhile, we verify that second order phase transitions
and Davies points still exist in these three models.
In addition, we calculate their quasinormal modes
in the eikonal limit by applying the light ring/quasinormal mode correspondence,
and discuss the spiral-like shapes
and the relations between the quasinormal modes and phase transitions.
As the main result, we show that spiral-like shapes in the complex frequency plane are closely related to the parameterization, namely in some particular units the spiral-like shapes will emerge in the models, which may not be of the spiral behaviors reported by other authors. We also discover a universal property of regular black holes, i.e., the imaginary parts of their QNMs do not vanish for the extreme cases, which
does not appear in singular black holes, such as the Reissner-Nordstr\"om and Kerr black holes, etc.
\end{abstract}

\begin{keyword}
Regular black hole\sep
quasinormal mode\sep
light ring/QNMs correspondence
\end{keyword}

\end{frontmatter}

\section{Introduction}
\label{sec:intr}

Black holes (BHs) as the prediction of general relativity (GR)
are of special properties, and expected to be the bridge between
a gravitational theory and a quantum theory.
Furthermore, the observation of gravitational waves from a binary black hole merger
reported by the LIGO Scientific and Virgo collaborations \cite{ligo2016} in 2016
provides a new window to study the BHs.
Thus, the
BH physics is now regarded as the core of modern physics. The quasinormal modes
(QNMs) \cite{konoplya2011} as a complex frequency of damped oscillations
from BH perturbations play an important role in the analysis of BH stability.
In particular, the gravitational waves are just the fundamental mode and thus carry the information of BHs.

The BH solutions of Einstein's equations are singular, which implies the
breakdown of completeness of spacetimes and embodies the shortage of Einstein's GR.
This singularity problem is unavoidable in Einstein's GR,
which was proved by Penrose \cite{penrose1965} and Hawking \cite{hawking1965} in 1960s.
As a result, it has been challenging to find such BH solutions that have no singularities or that are regular in other words. The first attempt was made by Bardeen \cite{bardeen1968}, where a kind of BH solutions without singularity in BH centers was obtained with the help of the relaxing of energy conditions. Later, the similar solutions were shown \cite{ayon-beato2000,ayon-beato1998,balart2014}
to be an outcome of the gravitational field coupled to a nonlinear electromagnetic field.
Another interesting attempt originated \cite{nicolini2005a,nicolini2005b,ansoldi2006,diGrezia2009,smailagic2010,modesto2010} in principle from the noncommutativity of spacetimes, but actually from a minimum length which is a natural inference of noncommutative (NC) spacetimes.
Such a class of BH solutions was thus named as the noncommutative spacetime inspired BHs or noncommutative BHs in short.
The most recent attempt was reported \cite{glavan2019,fernandes2020,konoplya2020}
that a novel $4$D Einstein-Gauss-Bonnet (EGB) BH provides a special class of ``regular'' solutions.
Here the phrase ``regular'' has a different meaning from that of the two classes of regular BHs mentioned above.
For the Bardeen BHs and noncommutative BHs,
the Ricci scalar, the contraction of two Ricci tensors and two Riemann tensors,
are finite in the physical domain.
However, for the $4$D EGB BH, its Ricci scalar is divergent though the metric is nonsingular at the center ($r=0$).
This feature gives rise to the physical process: When an infalling particle approaches to the center of the $4$D EGB BH,
the gravitational force becomes repulsive and tends to infinity,
such that the infalling particle can never reach the singular point.

As is known,
there exists \cite{jing2008,he2008} a close relation between QNMs and phase transitions of BHs,
i.e., the QNMs of a singular (traditional) BH present a spiral-like behavior on the complex frequency plane when the BH evolves to its Davies point. The Davies point was shown \cite{wei2019} to be located
at the maximum temperature $T$ on the $\Omega$-$T$ and $\lambda$-$T$ planes, where $\Omega$ and $\lambda$ denote the angular velocity and Lyapunov exponent, and determine the real and imaginary parts of QNM frequencies in the eikonal limit, respectively.
Since the QNMs are determined completely by the intrinsic properties of BHs that are associated with dynamics while
the phase transition is a phenomenon that belongs to thermodynamics, such a connection between QNMs and phase transitions
 enriches the relationship between the BH dynamics and BH thermodynamics.
Moreover, the regular BHs have different mechanical laws from that of singular ones,
such that they will have different thermodynamic rules,  e.g. there is a correction to entropy \cite{cai2009}.
Therefore, it is nontrivial to reexamine the relationship between dynamics and thermodynamics for the regular BHs.
In this paper,
we shall adopt the light ring/QNM correspondence \cite{cardoso2008,breton2016} to analytically calculate QNMs of regular BHs,
which provides an efficient estimation of QNMs in the eikonal limit.
With the help of such a correspondence, we can reveal new and deep relations between the BH dynamics and BH thermodynamics.
Although this correspondence may be violated \cite{konoplya2017} for gravitational perturbations,
we only focus on the perturbation of scalar fields in particular in our models.

This paper is organized as follows.
In Sec.\ \ref{sec:MP-BH},
based on the well-studied $5$D Myers-Perry BH,
we briefly elucidate the programs and technics we are going to use for investigating regular BHs in the following.
In addition, we give an explanation that why the Davies point is located at the
maximum temperature on the $\Omega$-$T$ and $\lambda$-$T$ planes.
We also show the existence of spiral-like shapes in the complex quasinormal mode plane,
where the real and imaginary components of QNMs are rescaled by angular momenta.
The treatment shown in this section is not only regarded as a typical example for the
BHs with singularity, but also an opposite case to Ref.\ \cite{wei2019},
where the QNMs of high dimensional BHs have no spiral structures.
In Sec.\ \ref{sec:motive} we verify that the regularity and traditional first law of BH mechanics cannot exist simultaneously in a BH model.
In other words, the first law of mechanics in any regular BH systems is broken,
even though the regularity is involved only in the metric, such as the 4D EGB BH model.
Then we consider in Sec.\ \ref{sec:BV-BH} our first regular BH associated with nonlinear electrodynamics
in terms of the dimensionless scheme,
and compute its QNMs in the eikonal limit.
We show that the spiral structure does not exist in the unit of mass $M$,
but it will emerge again if the charge $Q$ is used as unit.
Besides, we discover a novel phenomenon that the imaginary part of QNMs does not vanish as the horizon approaches to its extreme value. This does not appear in singular BHs and seems to be a universal property for all regular BHs considered in our current work.
In Sec.\ \ref{sec:noncom-SBH} we turn to our second regular BH, i.e., the noncommutative Schwarzschild BH, and make a parallel discussion to that of the BH with a nonlinear electrodynamic source. We study in Sec.\ \ref{sec:EGB-BH} the $4$D EGB BH by following the same procedure as that for the above two regular BHs.
We give our conclusions in Sec.\ \ref{eq:conclusion} where some comments and further extensions are included.

\section{5D Myers-Perry black holes}
\label{sec:MP-BH}

\subsection{Phase transition of  5D Myers-Perry black holes}

For the $5$D \emph{Myers-Perry} BH (MP BH) with only one nonzero angular momentum $a$
in the Boyer-Lindquist coordinates, the metric reads \cite{myers1986,emparan2008,cardoso2008}
\begin{eqnarray}
d s^2 &=& \frac{\Delta-a^2 \sin^2\vartheta}{\Sigma} d t^2
+\frac{2a(r^2+a^2-\Delta)\sin^2\vartheta}{\Sigma}d t d \varphi \nonumber \\
& & -\frac{(r^2+a^2)^2-\Delta a^2 \sin^2\vartheta}{\Sigma}d \varphi^2
-\frac{\Sigma}{\Delta}d r^2
-\Sigma d \vartheta^2
-r^2\cos^2\vartheta d \zeta^2,\label{eq:metric1}
\end{eqnarray}
where
$\Sigma=r^2+a \cos^2\vartheta$
and $\Delta= r^2 +a^2 - \mu$.
The physical mass and angular momentum are given in terms of mass parameter $\mu$ and angular momentum parameter $a$ as follows,
\begin{equation}
M=\frac{3 \mu }{8},\qquad
J=\frac{2 a M}{3}.\label{eq:MandJ}
\end{equation}
The area of the outer horizon, $r_+=\sqrt{\mu -a^2}$, can be computed by
the integral,
\begin{equation}
A=\int d\vartheta d\phi d\zeta\sqrt{-\sigma}=2\pi ^2 r_+ \left(a^2+r_+^2\right),
\end{equation}
where $\sigma$ is the induced metric obtained by the setting of $t={\rm const.}$ and $r=r_+$,
and the surface gravity can be computed \cite{litim2014} via two Killing vectors that are
associated with the time translation and the axisymmetry,
\begin{equation}
\kappa
=\frac{\sqrt{96 M^3-81 J^2}}{16 M^2}.
\end{equation}
Thus one can verify the first law of BH mechanics,
\begin{equation}
d M = \frac{\kappa}{8\pi} d A + \Omega d J,\qquad
\Omega = \frac{9 J}{16 M^2}.
\end{equation}
Moreover, it can be proved by the semiclassical method \cite{parikh1999,akhmedov2006,banerjee2008}
that the temperature without backreaction obeys the formula, $T=\kappa/2\pi$, namely,
\begin{equation}
T
=\frac{\sqrt{96 M^3-81 J^2}}{32 \pi ^2 M^2}.\label{eq:mptem}
\end{equation}
Therefore, the linear correspondence between the mechanic and thermodynamic variables is saved,
i.e.\ $T\propto \kappa$ and $A\propto S$.
The entropy is then obtained,
\begin{equation}
S=
\int \frac{d M}{T}
=\frac{A}{4}=\frac{1}{2} \pi ^2 r_+ \left(a^2+r_+^2\right),
\end{equation}
or it takes the following form in term of $M$ and $J$,
\begin{equation}
S=\frac{2}{3} \pi ^2 \sqrt{\frac{32 M^3}{3}-9 J^2}.
\end{equation}
The Smarr formula is $J \Omega +S T=2 M/3$.
By the definition of heat capacity, one has
\begin{equation}
C_J = T \left(\frac{\partial S}{\partial T}\right)_J
=\frac{16 \pi ^2 M^3 \sqrt{\frac{32 M^3}{3}-9 J^2}}{27 J^2-8 M^3}.
\end{equation}
The Davies point can be found from the solution of the algebraic equation, $1/C_J =0$,
\begin{equation}
J^*= \frac{\sqrt{\mu }}{2}
=\frac{2}{3} \sqrt{\frac{2}{3}} M^{3/2}.
\end{equation}

For the further discussions,
it is convenient to work with the usage of dimensionless variables.
First of all, one needs to introduce a factor $l^{\alpha_i}$ for each physical quantity,
where the exponent $\alpha_i$ is regarded as an index of scale factors, while
$l$ is an arbitrary positive real number.
We are going to make a transformation for every variable, e.g. $M\to l^{\alpha_1} M$ and $J\to l^{\alpha_2} J$,
such that all relevant equations and physical laws are invariant.
We observe that if the index of $M$ is set to be unit, i.e. ${\rm ind}(M)=1$,
the indices of the other variables can be fixed, see Tab.\ \ref{tab:scale1}.
\begin{table}[htp]
\caption{Indices of scale factors of the 5D MP BH.}
\begin{center}
\begin{tabular}{cc|cccccccc}
\toprule
    {$M$} & {$J$} & {$T$} & {$S$} & {$C_J$} & {$\kappa$} & {$r_+$} &{$A$} &{$\Omega$} & {$\lambda$} \\ \midrule
    1  & {$\frac{3}{2}$} & {$-\frac{1}{2}$} & {$\frac{3}{2}$} & {$\frac{3}{2}$} & {$-\frac{1}{2}$} & {$\frac{1}{2}$} & {$\frac{3}{2}$}& {$-\frac{1}{2}$}& {$-\frac{1}{2}$} \\ \bottomrule
\end{tabular}
\end{center}
\label{tab:scale1}
\end{table}

\begin{figure}[h!]
    \centering
    \includegraphics[width=.45\textwidth]{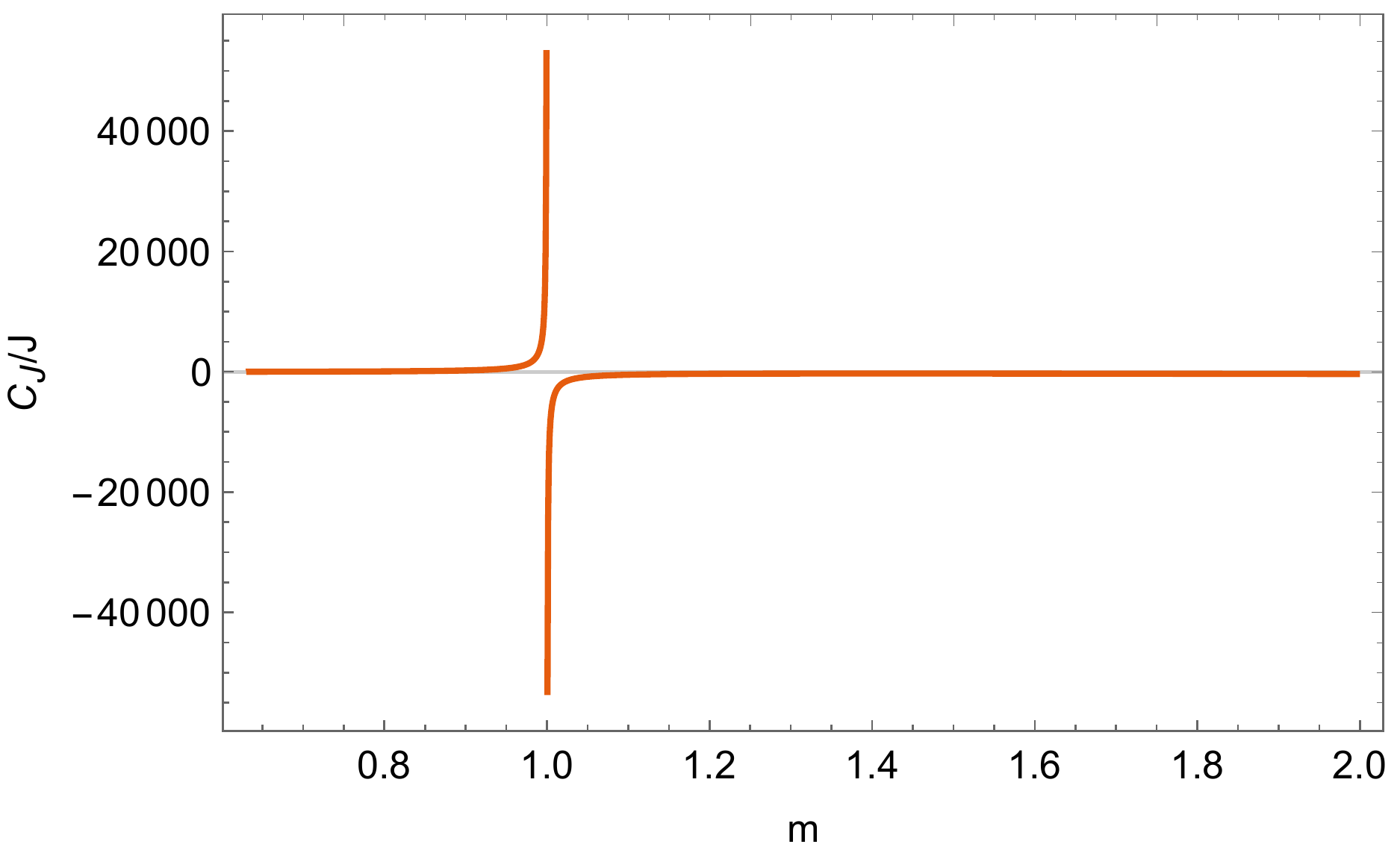}
    \caption{The heat capacity of the $5$D MP BH.}
    \label{fig:capacity-5d-MP}
\end{figure}

Secondly,
there are three ways to construct the dimensionless quantities since we have two independent parameters, $M$ and $J$, in the components of the metric, see Eqs.\ \eqref{eq:metric1}
and \eqref{eq:MandJ}.
The first way is the normalization by $M$,
the second by $J$,
and the third by the mixture of $M$ and $J$.
Based on the indices in Tab.\ \ref{tab:scale1},
we introduce a dimensionless quantity $m$ by
$M = \frac{3}{2} m J^{2/3}$,
where the fraction $3/2$ is introduced to normalize the Davies point to be unit,
but generally it is not necessary. Therefore,
we reconstruct the dimensionless formulations of the other variables that are of course associated with the initial variable $M$ or $J$.  For instance, the dimensionless heat capacity can be rewritten as $C_J/J$ or $C_J/M^{3/2}$,
\begin{equation}
C_J/J=-6 \pi ^2\,\frac{ m^3 \sqrt{4 m^3-1}}{m^3-1},\qquad
C_J/M^{3/2}= -4\sqrt{\frac{2}{3}} \,\pi ^2 \,\frac{ m^{3/2} \sqrt{ 4 m^3-1}}{m^3-1}.
\end{equation}
As a result, the Davies point can be represented as $m^*=1$, see Fig.\ \ref{fig:capacity-5d-MP},
where the heat capacity $C_J$ rescaled by the physical angular momentum is shown.
The advantage of this dimensionless scheme will be remarkable for most of regular BHs whose horizons and the other variables cannot be obtained analytically,
which will be shown obviously in the following models of regular BHs.

\subsection{Quasinormal modes in the eikonal limit}

\begin{figure}[h!]
    \centering
    \includegraphics[width=0.7\textwidth]{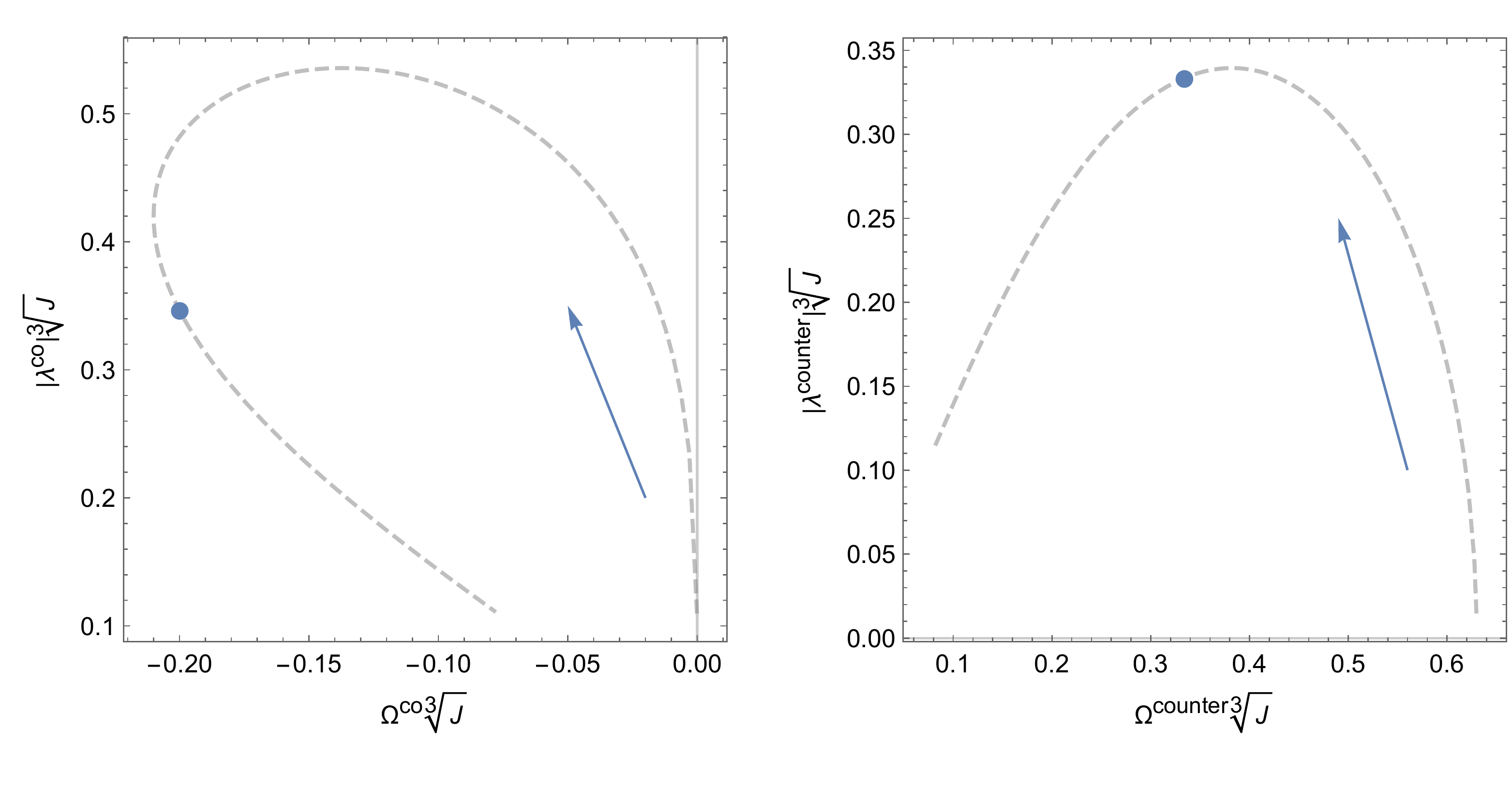}
    \caption{The QNMs of the $5$D MP BH
    on the $\Omega-\lambda$ plane,
    where the real and imaginary components of QNMs are rescaled by $J$.
    The arrows point to the increasing of parameter $m$,
    and the blue points are Davies points.
    }
    \label{fig:modes-MP-BH1}
\end{figure}

In order to give the quasinormal modes in the eikonal limit,
one can apply the light ring/QNM correspondence \cite{cardoso2008}. For the $5$D MP BH,
we have the equation for the photon sphere radius $r_c$,
\begin{equation}
r_c^2\pm 2 a \sqrt{\mu }-2 \mu =0,
\end{equation}
where the sign $\pm$ corresponds to the corotating and counterrotating orbits, respectively,
and solve the radius of circular null geodesics,
\begin{equation}
r_c = \sqrt{2\left(\mu \pm a \sqrt{\mu } \right)}.
\end{equation}
Since two $r_c$'s are larger than $r_+$, we have to consider two cases
for calculating QNMs, where
one is the corotating and the other the counterrotating. The results are as follows,
\begin{equation}
\Omega^{\rm co} \sqrt[3]{J} = -\frac{m}{4 m^{3/2}+1},\qquad
\Omega^{\rm counter} \sqrt[3]{J} = \frac{m}{4 m^{3/2}-1};
\end{equation}
\begin{equation}
\lambda^{\rm co} \sqrt[3]{J} = \frac{m\sqrt{m^{-3/2}+2} }{4 m^{3/2}+1},\qquad
\lambda^{\rm counter} \sqrt[3]{J} = \frac{m\sqrt{2-m^{-3/2}} }{1-4 m^{3/2}},
\end{equation}
which can be plotted on the complex frequency plane, where the horizontal and vertical axes
depict the real and imaginary parts of QNMs, respectively, see Fig.\ \ref{fig:modes-MP-BH1}.
Moreover, we note that $\{\Omega^{\rm co} \sqrt[3]{J},
\lambda^{\rm co} \sqrt[3]{J}\}$ converges to zero when $m\to 0$,
while $\Omega^{\rm counter} \sqrt[3]{J}=1/2^{2/3}$ and $\lambda^{\rm counter} \sqrt[3]{J}$ vanishes when $m$ goes to the critical point, $m_{\rm cp}=1/2^{2/3}$.
On the other hand, both $\{\Omega^{\rm co} \sqrt[3]{J},
\lambda^{\rm co} \sqrt[3]{J}\}$ and $\{\Omega^{\rm counter} \sqrt[3]{J},
\lambda^{\rm counter} \sqrt[3]{J}\}$ vanish when $m$ approaches to infinity.
This result implies that for the corotating case both the real and imaginary parts are definable
when $m$ takes values from zero to infinity,
while the situation for the counterrotating case is quite different, i.e., the imaginary part is undefinable if $m\in (1/2^{4/3}, 1/2^{2/3}]$, where the real part is well-defined.
In other words, when $m$ takes the values in this range, the $5$D MP
BH maintains oscillating without damping for the counterrotating case.

Similarly, since ${\rm ind}(\Omega)= {\rm ind}(\lambda)= -{\rm ind}(M)/2$,
$\Omega$ and $\lambda$ can also be rescaled by mass $M$, i.e. both
$\Omega \sqrt{M}$ and $\lambda  \sqrt{M}$ are dimensionless.
The discussions based on this choice of scale factors can be made in the same way as the above, so we do not repeat.

\begin{figure}[h!]
    \centering
    \includegraphics[width=0.7\textwidth]{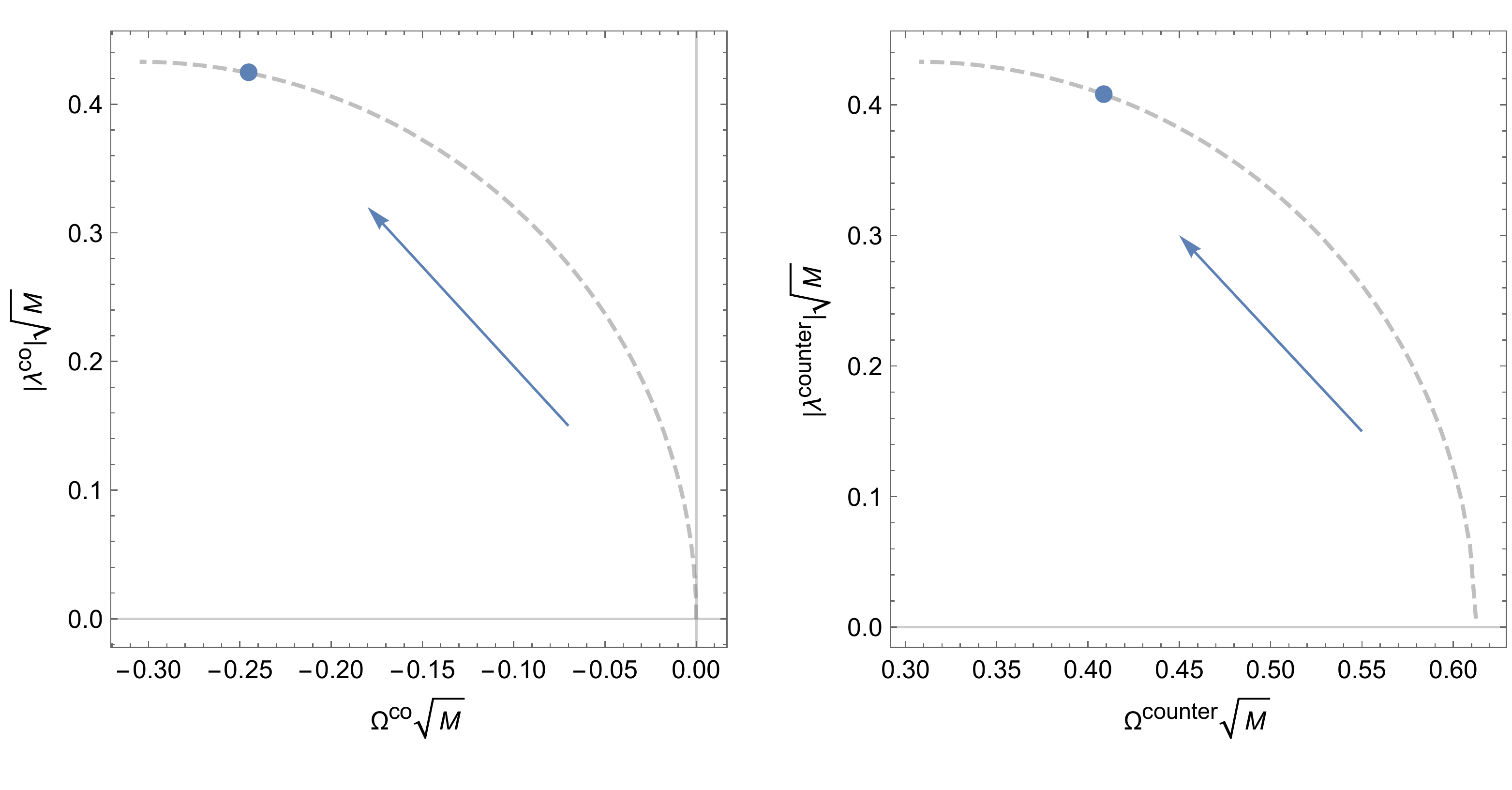}
    \caption{The QNMs of the $5$D MP BH
    in the complex frequency plane,
    where the components of QNMs are rescaled by $M$.
    The arrows point to the increasing of parameter $m$,
    and the blue points are Davies points.
    }
    \label{fig:modes-MP-BH2}
\end{figure}

However, it is worthy to mention that
the spiral-like shapes of QNMs in the unit $M$ or $J$ are much different.
As we showed in Fig.\ \ref{fig:modes-MP-BH1},
in the unit of $J$ the spiral-like shape obviously appears as $m$ increases,
it starts at the maximum point of rescaled $\lambda$ before the Davies point.
But in the unit of $M$,
the spiral behavior of complex QNMs does not exist, see Fig.\ \ref{fig:modes-MP-BH2}.

\subsection{Davies point as a saddle point of rescaled temperature}

\begin{figure}[h!]
    \centering
    \includegraphics[width=.7\textwidth]{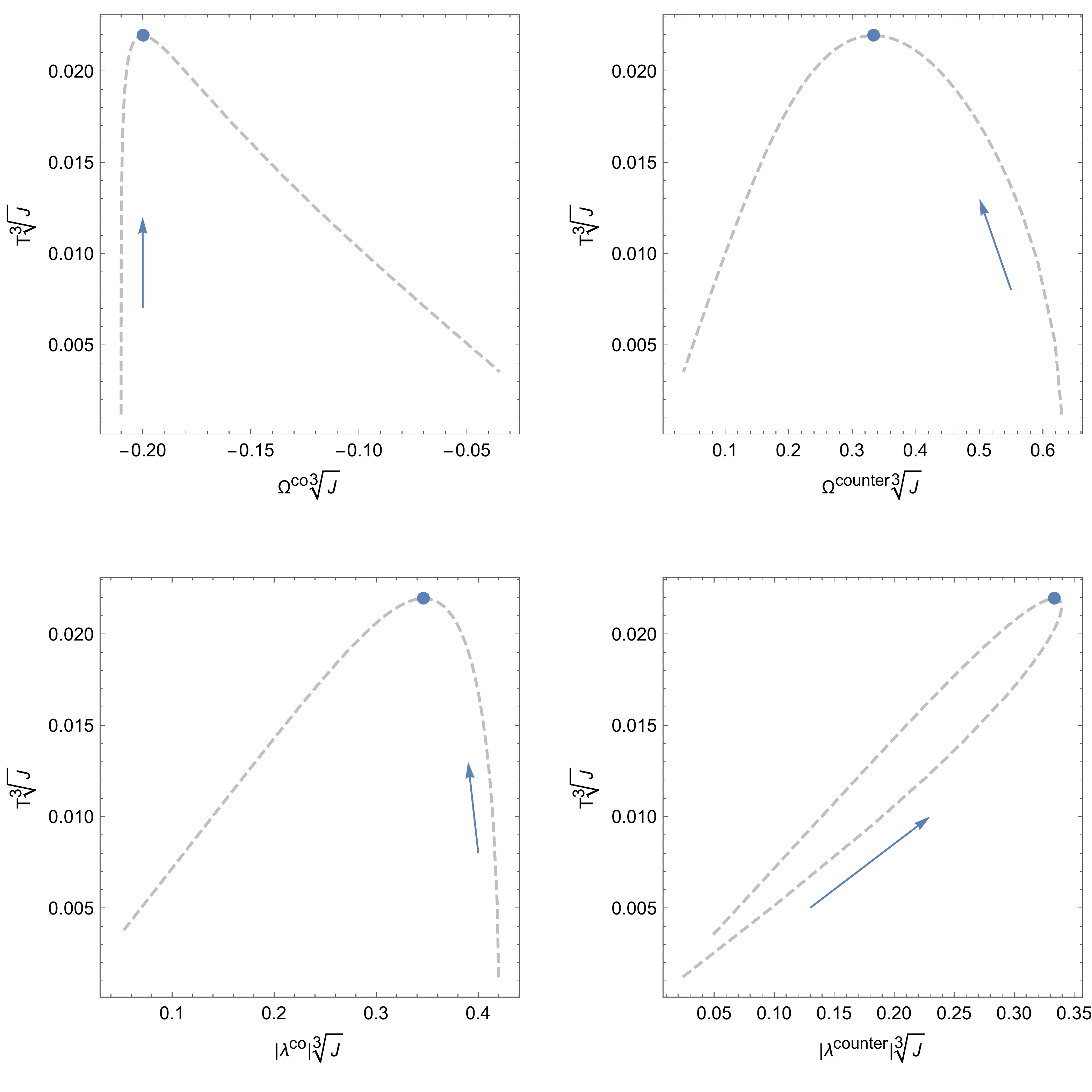}
    \caption{The QNMs of the 5D MP BH.
    The arrows point to the increasing of parameter $m$,
    and the blue points are Davies points.}
    \label{fig:modes-MP-BH}
\end{figure}

When the temperature, Eq.\ \eqref{eq:mptem}, is rescaled in terms of $m$,
\begin{equation}
\label{eq:MP-temperature}
T \sqrt[3]{J} =\frac{\sqrt{4 m^3-1}}{8 \pi^2  m^2},
\end{equation}
we can find that $T \sqrt[3]{J} $ reaches its maximum value $\sqrt{3}/(8 \pi ^2)$ at the Davies point $m^*=1$.
Moreover, the Davies point is also located at the maximum on the planes $\{\Omega\sqrt[3]{J}, T \sqrt[3]{J}\}$
and $\{\lambda\sqrt[3]{J}, T \sqrt[3]{J}\}$, which can be verified by the derivative test, see Fig.\ \ref{fig:modes-MP-BH}.
For a small rotation, the relations between QNMs and temperatures are
\begin{equation}
\Omega^{\rm co}=- \pi^2 T +\pi^4 T^2 a+ O(a^2)
,\qquad
\lambda^{\rm co}= \sqrt{2}\, \pi ^2 T+ O(a),
\end{equation}
and
\begin{equation}
\Omega^{\rm counter}=\pi^2 T +\pi^4 T^2 a+ O(a^2)
,\qquad
\lambda^{\rm counter}=\sqrt{2}\,  \pi ^2 T+ O(a),
\end{equation}
where the components of zero orders are consistent with that of the Schwarzschild BH.

In fact, for singular BHs with $n$ parameters
$\alpha_i$ except mass $M$, $i\in[1, n]$,
the Davies point must correspond to the saddle point of the temperature
with respect to the BH mass (or rescaled mass variable).
From the first law of black hole thermodynamics,
\begin{equation}
d M = T d S + \beta_i d\alpha_i,
\end{equation}
where $\beta_i$'s are the physical quantities of BHs rather than the temperature,
one can derive
\begin{equation}
\frac{1}{T}= \left(\frac{\partial S}{\partial M}\right)_{\alpha_i},
\end{equation}
which implies
\begin{equation}
C_{\alpha_i}=T\left(\frac{\partial S}{\partial T}\right)_{\alpha_i}
=\left(\frac{\partial M}{\partial T}\right)_{\alpha_i}.
\end{equation}
In other words, the Davies points as the roots of $1/C_{\alpha_i}=0$ must
be the saddle points of $T$ with respect to $M$, i.e., the Davies points satisfy the equation, $\left(\partial T/\partial M\right)_{\alpha_i}=0$.
On the other hand,
such a property of Davies points would be embodied in QNMs because QNMs are closely related to BH masses.

To identify whether the Davies points are a maximum or a minimum,
one needs to observe the second derivative of temperature with respect to mass, i.e.,
\begin{equation}
\frac{\partial^2 T}{\partial M^2} = -\frac{1}{C_{\alpha_i}^{2}} \frac{\partial C_{\alpha_i}}{\partial M}.
\end{equation}
The Davies points take the
maximum value of temperature if $\partial C_{\alpha_i}/\partial M >0$, while they correspond to the minimum value if $\partial C_{\alpha_i}/\partial M<0$.
Meanwhile, these two cases correspond to two different processes.
The former implies that the temperature increases at first if some amount of heat is given to a BH.
After the BH crosses the Davies point, as the amount of heat increases, the temperature decreases, where
the lost energy transforms to the Hawking radiation.
The latter denotes a completely inverse process, that is, the BH radiates before the Davie point,
after it crosses the critical point, the Hawking radiation stops, and then all the amount of heat given to the BH will show the increasing of temperature.

At the end of this section, we note that we have rescaled $T$ by multiplying $\sqrt[3]{J}$ in order to obtain
Eq.\ \eqref{eq:MP-temperature}. Alternatively, we can also rescale the temperature by multiplying the factor $\sqrt{M}$,
\begin{equation}
    T\sqrt{M} =\frac{1}{8 \pi ^2}
        \sqrt{6-\frac{3}{2 m^3}}.
\end{equation}
However, the Davies point ($m^*=1$) is no longer the maximum of $T\sqrt{M}$ under this type of rescaling.
The reason is obvious, i.e., $M$ is regarded as a variable but $J$ a constant in the definition of heat capacity, and such a rescaling changes the function of $T$ with respect to $M$. In other words, we have to avoid rescaling the temperature by using $M$ in order to make the Davies point be located in the saddle point of normalized temperatures.

\section{Regularity versus the first law of black hole mechanics}
\label{sec:motive}

For a spherically symmetric BH,
one can assume \cite{ansoldi2008} its shape function as follows,
\begin{equation}
f(r)=1-\frac{2M}{r}\sigma(M, r, \alpha),
\end{equation}
where $\alpha$ is the abbreviation of parameters rather than mass,
and then compute the surface gravity,
\begin{equation}
\kappa= \frac{f'(r_{\rm H})}{2}=
\frac{M }{ r_{\rm H}^2}\sigma(M, r_{\rm H},\alpha )
    -\frac{M }{r_{\rm H}}\frac{\partial}{\partial r_{\rm H}}\sigma(M,r_{\rm H},\alpha ).
\end{equation}
If the traditional first law of black hole mechanics is valid, which means $dM=\tilde{\kappa}dA/(8\pi)+\cdots$, where $A$ represents the area of BHs, one gets the surface gravity by an alternative way,
\begin{equation}
\tilde \kappa^{-1}=  r_{\rm H}\frac{\partial r_{\rm H}}{\partial M}
=
\frac{r_{\rm H}^2 }{ M }
\frac{M \frac{\partial}{\partial M}\sigma(M, r_{\rm H},\alpha )+\sigma(M, r_{\rm H},\alpha )}
{\sigma(M, r_{\rm H},\alpha )-r_{\rm H} \frac{\partial}{\partial r_{\rm H}}\sigma(M,r_{\rm H},\alpha )}.
\end{equation}
The ratio of $\kappa$ and $\tilde \kappa$ reads
\begin{equation}
\kappa/\tilde \kappa=M \frac{\partial}{\partial M}\sigma(M, r_{\rm H},\alpha )+\sigma(M, r_{\rm H},\alpha ).
\end{equation}
If a BH satisfies the traditional first law, one has $\kappa/\widetilde \kappa=1$, which leads to the following solution,
\begin{equation}
\sigma(M,r_{\rm H},\alpha )=1+ \frac{\zeta(r_{\rm H},\alpha )}{M},
\end{equation}
namely,
\begin{equation}
\sigma(M,r,\alpha )=1+ \frac{\zeta(r,\alpha )}{M},
\end{equation}
where $\zeta(r, \alpha)$ is an arbitrary function of $r$ and $\alpha$.
The shape function then takes the form,
\begin{equation}
f(r)=1-\frac{2M}{r}-\frac{2\zeta(r,\alpha )}{r}.
\end{equation}
Since the term $2M/r$ cannot be subtracted anyway,
$r=0$ remains to be the singular point of $f(r)$.
On the other hand, this implies that the regularity and the traditional first law of \emph{BH mechanics} cannot
exist simultaneously in a BH model.
In addition, if one rewrites the entropy from the first law as follows,
\begin{equation}
S = \int  \frac{d A}{4} (1+\rho),
\qquad
\rho = -1+\frac{8\pi}{\kappa(A)} \frac{d M(A)}{d A}
=-1+\frac{\tilde\kappa(A)}{\kappa(A)},
\end{equation}
it is easy to see that the linear relation $S=A/4$ no longer holds for the system whose first law breaks, i.e., $\rho\neq 0$.
There will be an additional term $\delta S$ in the entropy, $S=A/4+\delta S$.
In other words, the traditional first law of BH mechanics must be broken and the correction to
the entropy, $\delta S$, will be nontrivial for regular BHs.

Therefore, a natural question is whether there are second order phase transitions, i.e., the Davies points, in regular BHs where the traditional first law of  mechanics
 has been broken. If the Davies points appear, will the relationship between QNMs and phase transitions still exist? We shall study these issues in the three well-known regular BHs below.

\section{Regular black holes generated by nonlinear electrodynamics}
\label{sec:BV-BH}

Let us now take our first example of regular BHs from Ref.\ \cite{balart2014},
which is generated by nonlinear electrodynamics.
The shape function reads
\begin{equation}
f(r) = 1- \frac{2M}{r} e^{-\frac{q^2}{2Mr}},\label{eq:1stmf}
\end{equation}
where $q$ stands for electric charge. Thereinafter, we call this model the \emph{Balart-Vagenas} BH (BV BH).
This charged BH does not contain singularity, and has two event horizons,
\begin{equation}
\label{eq:horizon-BV-BH1}
r_+=-\frac{q^2}{2 M W_0\left(-\frac{q^2}{4 M^2}\right)},\qquad
r_-=-\frac{q^2}{2 M W_{-1}\left(-\frac{q^2}{4 M^2}\right)},
\end{equation}
where $W_0(z)$ and $W_{-1}(z)$ are Lambert's $W$ functions.
Because Lambert's $W$ functions are not homogenous, we only need to rescale $z$ to be dimensionless
in order to have a dimensionless horizon.
The scale factors of related variables are listed in Tab.\ \ref{tab:scale2}.
\begin{table}[htp]
\caption{Indices of scale factors of the BV BH.}
\begin{center}
\begin{tabular}{cc|cccccccc}
\toprule
    {$M$} & {$q$} & {$T$} & {$S$} & {$C_q$} & {$\kappa$} & {$r_{\pm}$}  & {$A$} & {$\Omega$} & {$\lambda$} \\ \midrule
    1  & 1 & {$-1$} & {$2$} & {$2$} & {$-1$} & 1 & 2 & {$-1$}& {$-1$} \\ \bottomrule
\end{tabular}
\end{center}
\label{tab:scale2}
\end{table}

Considering the characters of the model mentioned above,
we introduce the rescaled parameters $x$ and $Q$ as follows,
\begin{equation}
r\longrightarrow \frac{2 M x}{Q^2},\qquad
q\longrightarrow \frac{2 M}{Q},\label{eq:dlessrq}
\end{equation}
such that $f(r_{\rm H})=0$ becomes
\begin{equation}
\label{eq:horizon-BV-BH}
1-\frac{Q^2 e^{-1/x_{\rm H}}}{x_{\rm H}}=0.
\end{equation}
We then solve $Q$ from Eq.\ \eqref{eq:horizon-BV-BH},
\begin{equation}
Q^\pm = \pm \sqrt{x_{\rm H}}\, e^{\frac{1}{2 x_{\rm H}}}.
\end{equation}
This equality will be frequently used as the formula satisfied by the dimensionless horizon in the following.
Moreover, we can find the horizon radius and charge of the extremal BV BH,
\begin{equation}
x_{\rm ext}=1,\qquad
Q^\pm_{\rm ext} = \pm\sqrt{e},\label{eq:hrebh}
\end{equation}
We note that if the mass-to-charge ratio $|M/q|$ is less than $\sqrt{e}/2$,
no horizons exist, but there are two distinguished horizons if $|M/q|>|M_{\rm ext}/q_{\rm ext}|=\sqrt{e}/2$.

\subsection{Geometric quantity and regularity}

To verify the regularity of spacetime, we compute the Ricci scalar of the BV BH,
\begin{equation}
\mathcal{R}(x)
	 =\frac{Q^6}{4 M^2} \frac{e^{-1/x}}{ x^5},
\end{equation}
which is positive and finite when the radial coordinate is from zero to infinity, i.e., the  BV BH has a de Sitter core inside.
On the two boundaries, $x\to 0$ and $x\to \infty$, the Ricci scalar vanishes, $\mathcal{R}(0)\to 0$ and $\mathcal{R}(\infty)\to 0$,
and it reaches its maximum value at $x=1/5$,
\begin{equation}
\mathcal{R}\left(\frac{1}{5}\right) =\frac{3125 Q^6}{4 e^5 M^2}.
\end{equation}
The contraction of two Ricci tensors takes the form,
\begin{equation}
\mathcal{R}^{\mu\nu}(x)\mathcal{R}_{\mu\nu}(x)
=\frac{Q^{12}}{32 M^4}
\frac{ e^{-2/x}}{ x^{10}}\left(8 x^2-4 x+1\right),
\end{equation}
which is nonsingular on the two boundaries,
$\mathcal{R}^{\mu\nu}(0)\mathcal{R}_{\mu\nu}(0)$ and $\mathcal{R}^{\mu\nu}(\infty)\mathcal{R}_{\mu\nu}(\infty)$ vanish.
In addition, the Kretschmann scalar reads
\begin{equation}
\mathcal{R}^\rho_{\mu\nu\beta}(x)\mathcal{R}_\rho^{\mu\nu\beta}(x)=
\frac{Q^{12}}{16 M^4}
\frac{e^{-2/x}}{ x^{10}}
\big(12 x^4-24 x^3+24 x^2-8 x+1\big) ,
\end{equation}
which also maintains nonsingular at the center and infinity,
$\mathcal{R}^\rho_{\mu\nu\beta}(0)\mathcal{R}_\rho^{\mu\nu\beta}(0)\to 0$
and
$\mathcal{R}^\rho_{\mu\nu\beta}(\infty)\mathcal{R}_\rho^{\mu\nu\beta}(\infty)\to 0$. In summary, we have verified the regularity of the BV BH spacetime.

\subsection{Deformation of the first law of black hole mechanics}

The first law of singular (traditional) black holes breaks down in regular black holes. To search differences between the first laws of  singular and regular black holes, we investigate the first law of the BV BH.
At first, we compute the area,
\begin{equation}
A=4\pi r_+^2 =16 \pi M^2 e^{2 W_{0}\left(-\frac{q^2}{4 M^2}\right)}.
\end{equation}
Next, we give the surface gravity without backreaction,
\begin{equation}
\kappa
=-\frac{M }{q^2}
W_{0}\left(-\frac{q^2}{4 M^2}\right) \left[W_{0}\left(-\frac{q^2}{4 M^2}\right)+1\right],
\end{equation}
and derive the potential by integrating the electric field with respect to the radial coordinate,
\begin{equation}
\phi=\int^\infty_{r_+} d r\; E,\qquad
E= \frac{q }{r^2}\left(1-\frac{q^2}{8 M r}\right) \exp \left(-\frac{q^2}{2 M r}\right),
\end{equation}
namely,
\begin{equation}
\phi = \frac{1}{8 M q}\left[
\frac{3 q^2}{W_{0}\left(-\frac{q^2}{4 M^2}\right)}+12 M^2+q^2
\right].
\end{equation}
As we mentioned in Sec.\ \ref{sec:motive}, it turns out that these mechanic variables do not satisfy the first law, i.e., $d M \neq \kappa d A/(8\pi) +\phi d q$.
This character of the BHs generated by nonlinear electrodynamics has already been noticed in Refs.\ \cite{rasheed1997,zhang2016}.
By differentiating the area, we find
\begin{equation}
\label{eq:pre-first-law}
d M = \frac{\tilde\kappa}{8\pi} d A + \tilde\phi d q,
\end{equation}
where the coefficients of $d A$ and $d q$ can be computed explicitly,
\begin{equation}
\tilde\kappa=
\frac{4 M^3}{q^4}\left[W_{0}\left(-\frac{q^2}{4 M^2}\right)\right]^2\,
\frac{1+W_{0}\left(-\frac{q^2}{4 M^2}\right)}{1- W_{0}\left(-\frac{q^2}{4 M^2}\right)},
\qquad
\tilde\phi =\frac{2 M }{q}
\left[1-\frac{1}{1-W_{0}\left(-\frac{q^2}{4 M^2}\right)}\right].
\end{equation}
This shows that we cannot use the traditional first law of black holes to describe regular black holes and should rebuild a new first law of BH mechanics.

By rearranging the coefficients in Eq.\ \eqref{eq:pre-first-law}, we can give the modified first law of BH mechanics,
\begin{equation}
d M = \frac{\kappa}{8\pi} d A +\phi d q +\eta d M
-\frac{3}{4}\frac{\eta}{Q}d q,
\end{equation}
where the last two terms of the right hand side are corrected terms to the traditional first law, and the corresponding Smarr formula has the form,
\begin{equation}
 M = \frac{\kappa}{4\pi} A +\phi q +\eta  M
-\frac{3}{4}\frac{\eta}{Q} q,
\end{equation}
where
\begin{equation}
\eta \equiv 1-\frac{\kappa}{\tilde \kappa} =1-Q^2
-e^{W_0(-Q^2)}.
\end{equation}

Now we try to build the relationship among entropy, temperature and mass. At first, the equality,
\begin{equation}
\left(\frac{\partial S}{\partial \mathcal{E}}\right)_q=
	 \frac{1}{T},
\end{equation}
should hold for any thermodynamic systems,
where $\mathcal{E}$ is total energy.
Then, we can verify by the semiclassical method \cite{parikh1999,akhmedov2006,banerjee2008}
that the temperature without backreaction still obeys the formula,
\begin{equation}
T=\frac{\kappa}{2\pi}=\frac{f'(r_+)}{4\pi}.\label{eq:tembv}
\end{equation}
Therefore, we conclude that the linear correspondence,
$S\propto  A $,
no longer holds.

\begin{figure}[h!]
    \centering
    \includegraphics[width=0.4\textwidth]{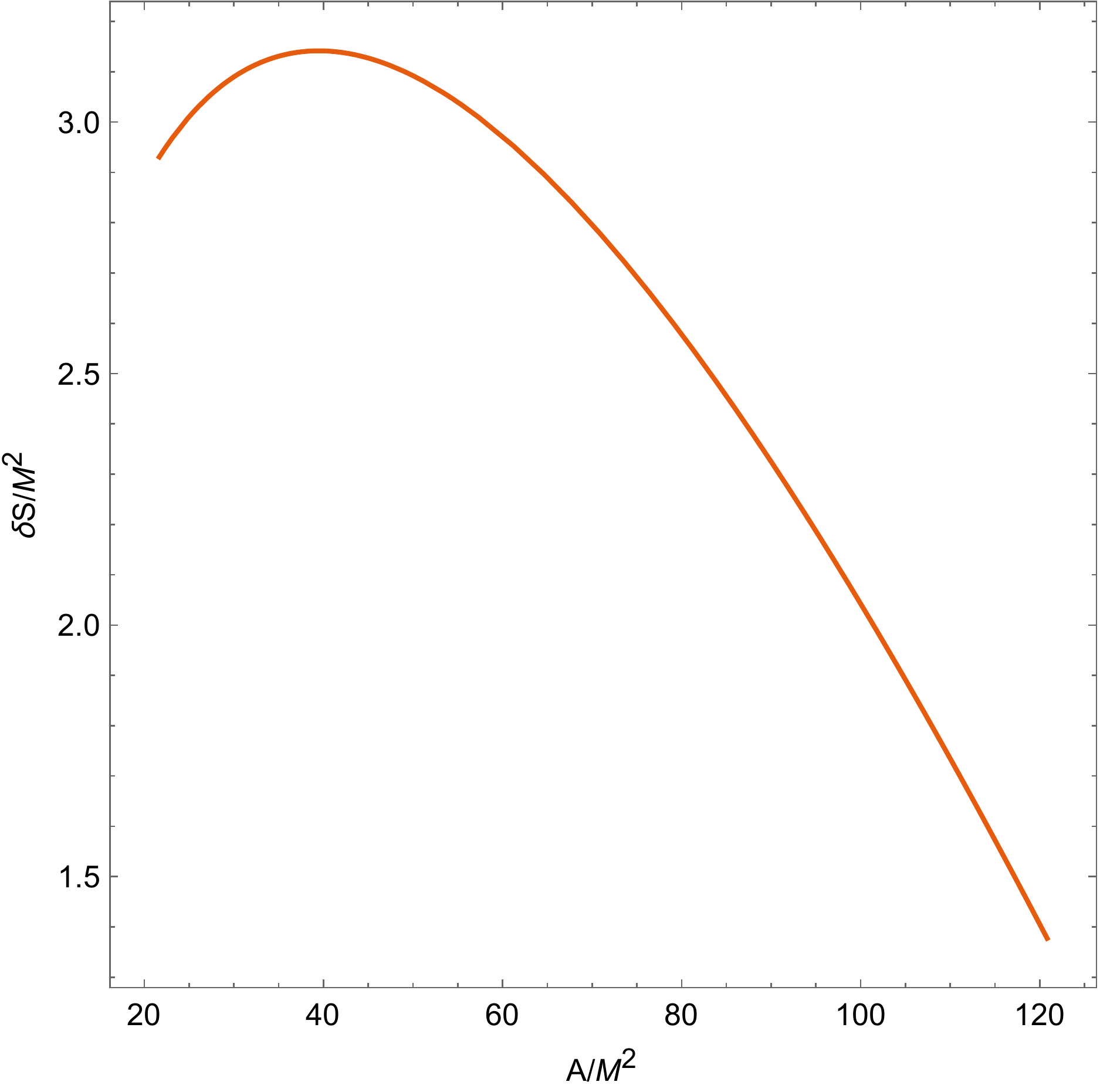}
    \caption{The dependence of entropy deviation on area for the BV BH.}
    \label{fig:derivation-entropy-BV-BH}
\end{figure}

Let us consider a deformation of entropy by following Ref.\ \cite{myung2007}.
If the linear correspondence, $\mathcal{E}\propto M$, still holds,
the first law becomes
\begin{equation}
d \mathcal{E}  = T d S +\Phi d q,
\end{equation}
where
\begin{equation}
d \mathcal{E} = d M,\qquad
d S= \frac{1}{4}d A+\frac{\eta}{T}d M,\qquad
\Phi=\phi -\frac{3}{4}\frac{\eta}{Q}=\frac{q}{2 M}.
\end{equation}
In the integration of $d S$, we calculate the second part at first,
\begin{equation}
\delta S \equiv \int \frac{\eta}{T}d M =
\frac{\pi q^2 }{W_{0}\left(-\frac{q^2}{4 M^2}\right)}
\left[e^{W_{0}\left(-\frac{q^2}{4 M^2}\right)}-1\right],\label{deltaS}
\end{equation}
and then have the total entropy,
\begin{equation}
\label{eq:entropy-BV-BH}
S=\frac{A}{4}+\delta S
=4 \pi  M^2 e^{W_0\left(-\frac{q^2}{4 M^2}\right)}.
\end{equation}
We plot the dependence of entropy deviation on area in Fig.\ \ref{fig:derivation-entropy-BV-BH}, which represents a nonlinear relation.

Since $\delta S= \frac{1}{4}A_{\rm Sch} \left(\sqrt{A/A_{\rm Sch}}-A/A_{\rm Sch}\right)$ is nonnegative, where $A_{\rm Sch}$ stands for  the area of the Schwarzschild BH,
we can conclude that the entropy bound of the BV BH must be altered, i.e., it is different from the Bekenstein bound, $S\ge A/4$,
for singular BHs.
Moreover, due to the property of Lambert's $W$ functions,
we have $Q\ge \sqrt{e}$. Thus, on the one hand, $A/M^2 \to 16 \pi ^2/e^2$
and $\delta S/M^2\to 4 (e-1) \pi /e^2$
as $Q\to \sqrt{e}$;
and on the other hand, $A/M^2\to 16\pi^2$ and $\delta S/M^2 \to 0$ as $Q\to \infty$.
In other words, as the BV BH turns back to the Schwarzschild BH, the correction to entropy vanishes.

\subsection{Heat capacity and Davies points}
\begin{figure}[h!]
    \centering
    \includegraphics[width=0.45\textwidth]{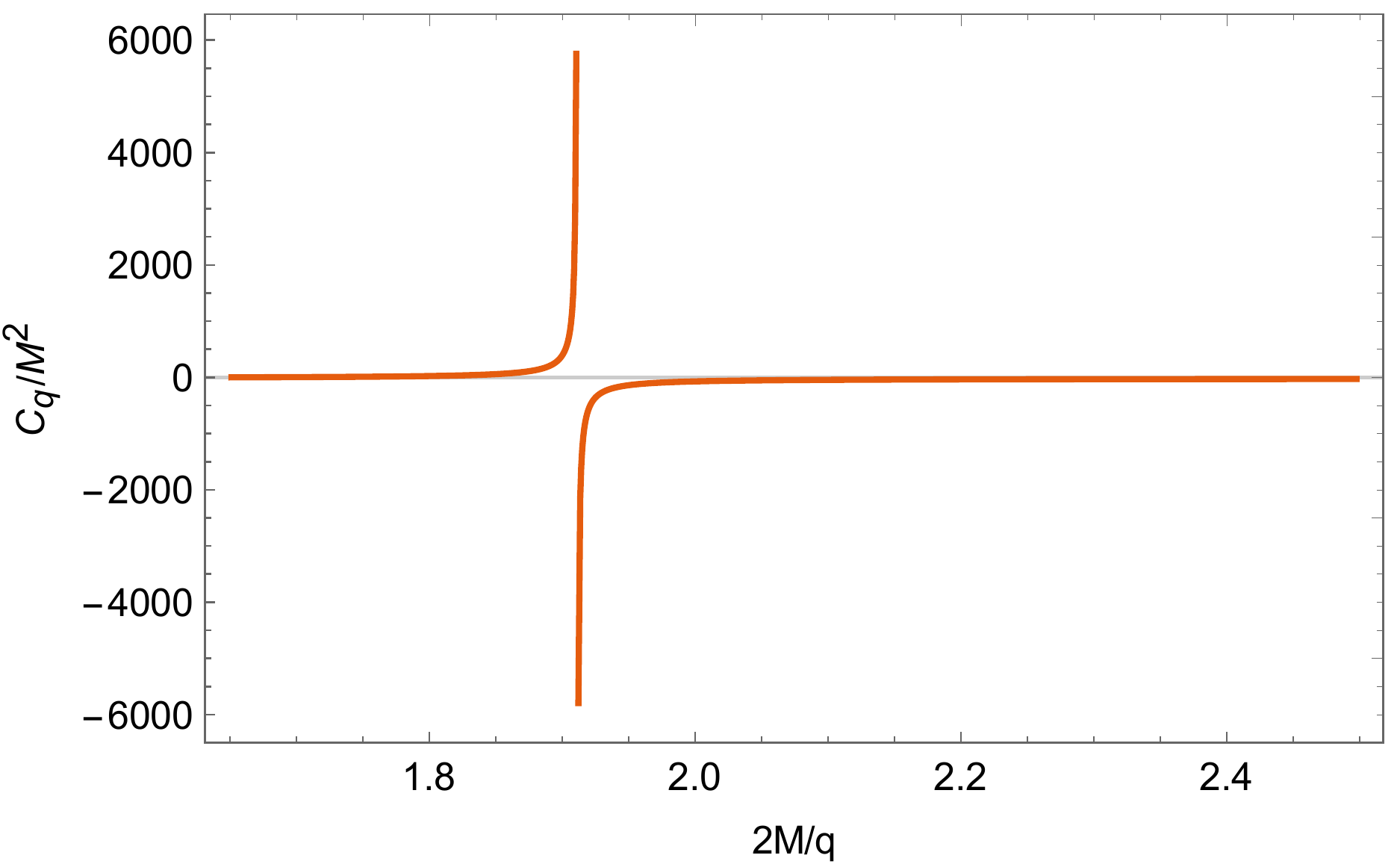}
    \caption{The heat capacity versus $2M/q$ (rescaled parameter $Q$) for the BV BH. The heat capacity diverges at $Q^*\approx \pm1.91$, where only the positive case is shown in the figure.}
    \label{fig:capacity-BV-BH-M}
\end{figure}

Let us search whether there exists a second order phase transition in the BV BH which
has a different first law from that of singular BHs.
According to the indices of scale factors,
we recast the temperature in the dimensionless form,
\begin{equation}
T M =-\frac{Q^2 }{8 \pi}
W_0\left(-\frac{1}{Q^2}\right)
\left[W_0\left(-\frac{1}{Q^2}\right)+1\right],\label{tempmass}
\end{equation}
where mass $M$ is regarded as unit,
then we obtain the dimensionless heat capacity in the unit of $M$,
\begin{equation}
C_q/M^2 
=\frac{8 \pi
	\left[e^{W_0\left(-\frac{1}{Q^2}\right)}-\frac{1}{Q^2}\right]}
{\left[W_0\left(-\frac{1}{Q^2}\right)-2\right]W_0\left(-\frac{1}{Q^2}\right)-1},
\end{equation}
and plot the relation between the heat capacity and the rescaled parameter $Q$ in Fig.~ \ref{fig:capacity-BV-BH-M}.

From Fig.\ \ref{fig:capacity-BV-BH-M}, we find that there is still a Davies point at which a second order phase transition happens in the BV BH although its entropy related to area is different from that of singular black holes. By solving the algebra equation, $1/C_q=0$,  we get the Davies point,
\begin{equation}
Q^*=\pm  \frac{e^{\frac{1}{\sqrt{2}}-\frac{1}{2}}}{\sqrt{\sqrt{2}-1}}\approx\pm1.91,
\end{equation}
at which the horizons of the BV BH can be calculated,
\begin{equation}
(x^{*}_{\rm H1})^{-1}=\sqrt{2}-1
,\qquad
(x^{*}_{\rm H2})^{-1}=-W_{-1}\left[ \left(1-\sqrt{2} \right)e^{1-\sqrt{2}}\right].
\end{equation}
Numerically, we have $x^{*}_{\rm H1}\approx 2.41>x_{\rm ext}$
and
$x^{*}_{\rm H2}\approx 0.51<x_{\rm ext}$,
which implies that only $x^{*}_{\rm H1}$ is physical.

Alternatively, the temperature and heat capacity can also be rewritten in the unit of charge $q$,
\begin{equation}
T q = -\frac{Q }{4 \pi }
W_0\left(-\frac{1}{Q^2}\right) \left[W_0\left(-\frac{1}{Q^2}\right)+1\right],\label{tempcharge}
\end{equation}
and
\begin{equation}
C_q/q^2=\frac{2 \pi  \left[W_0\left(-\frac{1}{Q^2}\right)+1\right]}
{-\left[W_0\left(-\frac{1}{Q^2}\right)\right]^3
+2 \left[W_0\left(-\frac{1}{Q^2}\right)\right]^2+W_0\left(-\frac{1}{Q^2}\right)}.
\end{equation}
For the further discussions in the next subsection,
it is convenient to represent the temperature in terms of horizon $x_{\rm H}$ with the help of
Eq.\ \eqref{eq:horizon-BV-BH}. That is, using $Q^2=x_{\rm H}\,e^{1/x_{\rm H}} $ to replace $Q^2$
in Eqs.~\eqref{tempmass} and \eqref{tempcharge}, we compute the temperature,
\begin{equation}
TM=-\frac{x_{\rm H}\,e^{\frac{1}{x_{\rm H}}}  }{8 \pi }
W_0\left(-\frac{e^{-\frac{1}{x_{\rm H}}}}{x_{\rm H}}\right)
\left[W_0\left(-\frac{e^{-\frac{1}{x_{\rm H}}}}{x_{\rm H}}\right)+1\right],
\end{equation}
and
\begin{equation}
Tq = -
\frac{\sqrt{x_{\rm H}}\,e^{\frac{1}{2 x_{\rm H}}}  }{4 \pi}
W_0\left(-\frac{e^{-\frac{1}{x_{\rm H}}}}{x_{\rm H}}\right)
\left[W_0\left(-\frac{e^{-\frac{1}{x_{\rm H}}}}{x_{\rm H}}\right)+1\right].
\end{equation}
When $x_{\rm H}= x_{\rm ext}$, the temperatures in the two units vanish, i.e. $TM=0$ and $Tq = 0$, which is similar to the case of the extreme RN black hole.
Moreover, since $x_{\rm H}\ge 1$,
considering the property of Lambert's $W$ functions,
we simplify the above temperatures to be
\begin{equation}
TM=\frac{e^{\frac{1}{x_{\rm H}}}  }{8 \pi }
\left(1-\frac{1}{x_{\rm H}}\right),\qquad
Tq =
\frac{e^{\frac{1}{2 x_{\rm H}}}  }{4 \pi \sqrt{x_{\rm H}}}
\left(1-\frac{1}{x_{\rm H}}\right).\label{tempmq}
\end{equation}

\subsection{Quasinormal modes in the eikonal limit}
\label{subsec:QNMs-BV-BH}
\begin{figure}[h!]
    \centering
    \includegraphics[width=0.4\textwidth]{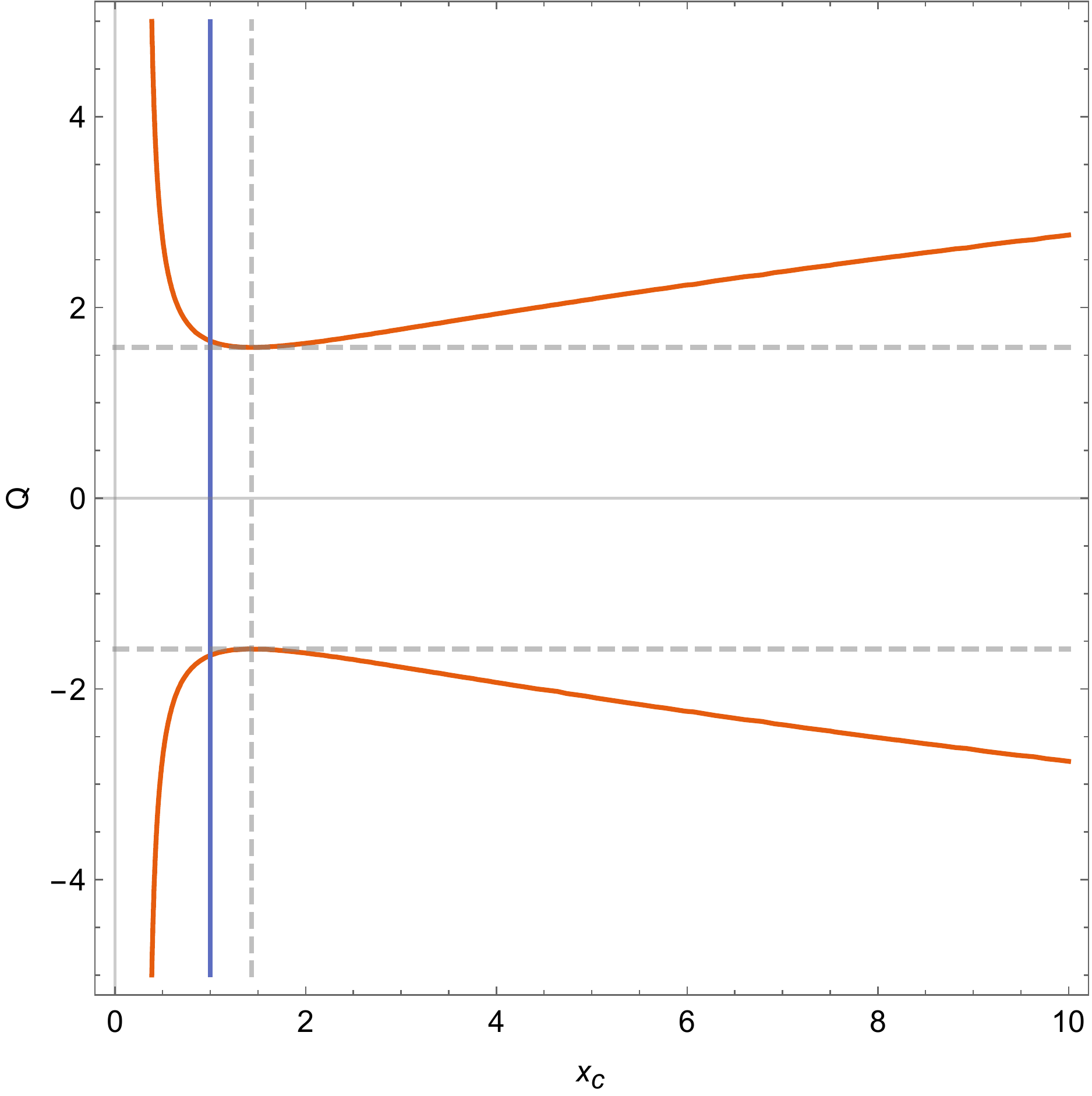}
    \caption{The dependence of the rescaled charge on the rescaled radius of photon spheres in the BV BH.
    The orange curve denotes Eq.\ \eqref{eq:null-geodesics-rescale},
    while the blue line $x_c=x_{\rm ext}=1$.
    }
    \label{fig:PS-BV-BH}
\end{figure}

To calculate the QNMs, we start
with the equation of photon spheres \cite{cardoso2008},
which takes the form for the BV BH,
\begin{equation}
\label{eq:null-geodesics}
 \left(\frac{q^2}{r_c}-6 M\right)e^{-\frac{q^2}{2 M r_c}}+2 r_c=0.
\end{equation}
The radius of photon spheres cannot be solved analytically from the above equation.
Thus we use the variables $x$ and $Q$ and
simplify Eq.\ \eqref{eq:null-geodesics} to be
\begin{equation}
\label{eq:null-geodesics-rescale}
Q^2 (3 x_c-1)=2 x_c^2\, e^{1/x_c}.
\end{equation}
It can be visualized in the $x_c-Q$ plane, see Fig.\ \ref{fig:PS-BV-BH}, where $Q$ reaches its extreme values at $x_0= \left(5+\sqrt{13}\right)/6\approx 1.43>x_{\rm ext}=1$. For a positive charge,
if ${Q^+}<Q^+_0\equiv  Q^+(x_0)$,
the radius of photon spheres does not exist;
if ${Q^+}=Q^+_0$, there is one single root, $x_0$;
if ${Q^+}>Q^+_0$, there are two photon sphere radii,
one is inner, $r^{-}_{\rm c}$, and the other is outer, $r^{+}_{\rm c}$.

\begin{figure}[h!]
    \centering
    \includegraphics[width=.7\textwidth]{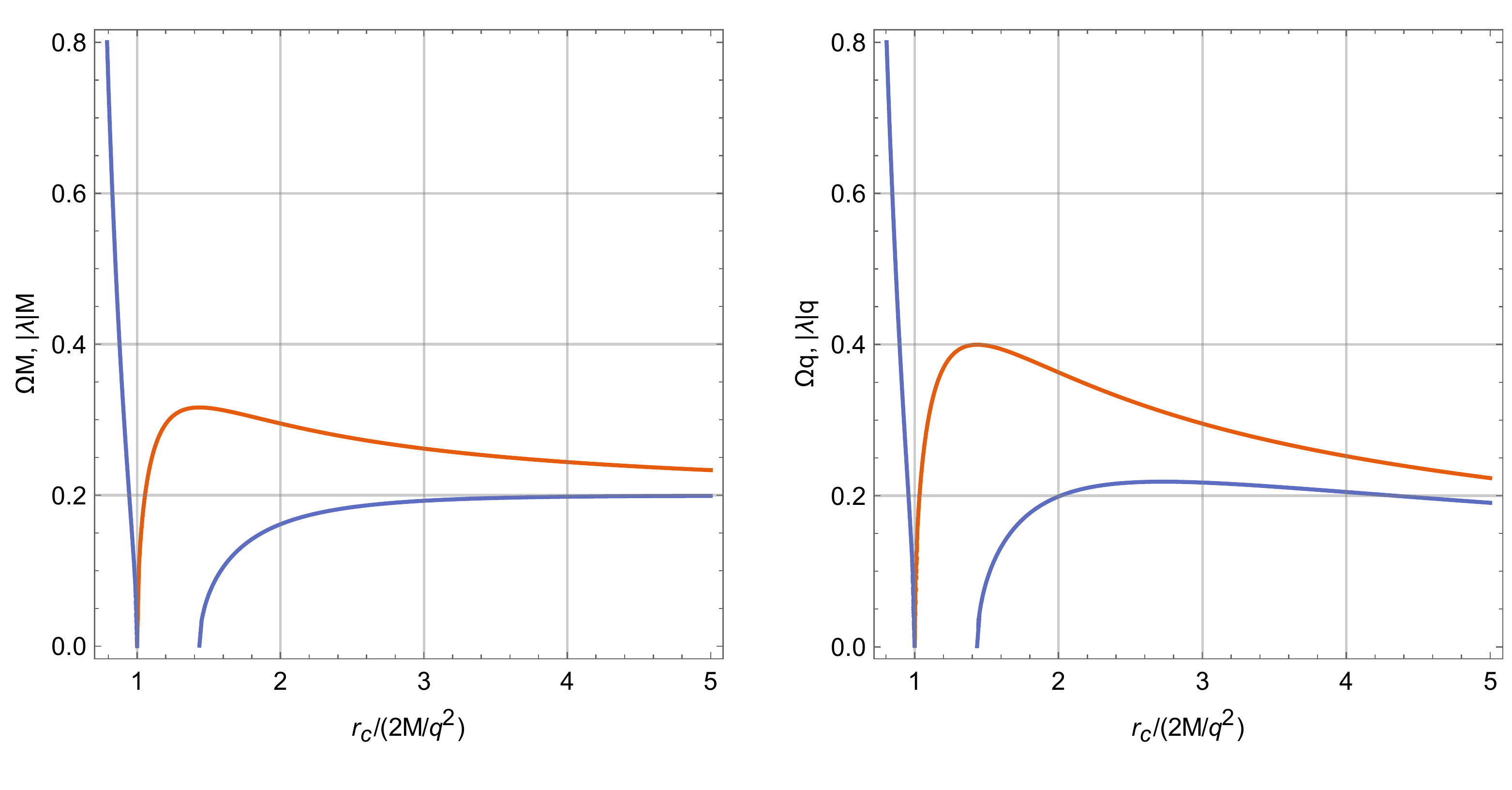}
    \caption{The QNMs of the BV BH in two different units.
    The orange curves correspond to real parts $\Omega M$ and $\Omega q$ of QNMs,
    while the blue curves to imaginary parts $|\lambda|M$ and $|\lambda|q$.}
    \label{fig:QNMs-BV-BH-MQ}
\end{figure}

The real and imaginary components of QNMs can be found in the two units, $M$ and $q$, respectively,
\begin{equation}
\Omega M =\frac{ x_c\,e^{{1}/{x_c}} }{3 x_c-1}\sqrt{\frac{x_c-1}{3 x_c-1}},\qquad
|\lambda| M=\frac{ x_c\,e^{{1}/{x_c}} }{3 x_c-1}
	\sqrt{\frac{(x_c-1) [x_c (3 x_c-5)+1]}{ x_c(3 x_c-1)}},
\end{equation}
and
\begin{equation}
\begin{split}
\Omega q=\frac{ \sqrt{2(x_c-1)}\,e^{{1}/{(2 x_c)}}}{3 x_c-1},\qquad
|\lambda|  q =\frac{e^{{1}/{(2 x_c)}}}{{3 x_c-1}}
	\sqrt{\frac{2(x_c-1) [x_c (3 x_c-5)+1]}{ x_c(3 x_c-1)}}.
\end{split}
\end{equation}
The QNMs in the two different units are shown in Fig.\ \ref{fig:QNMs-BV-BH-MQ}. Note that the imaginary parts of QNMs  disappear in the range of $1< x_{c}<x_0\approx 1.43$. Does it mean that there are only normal modes of perturbation for the BV BH? We shall answer this question by analyzing the relationship between the photon sphere radius $x_c$ and the horizon radius $x_{\rm H}$ below.

\begin{figure}[h!]
    \centering
    \includegraphics[width=0.4\textwidth]{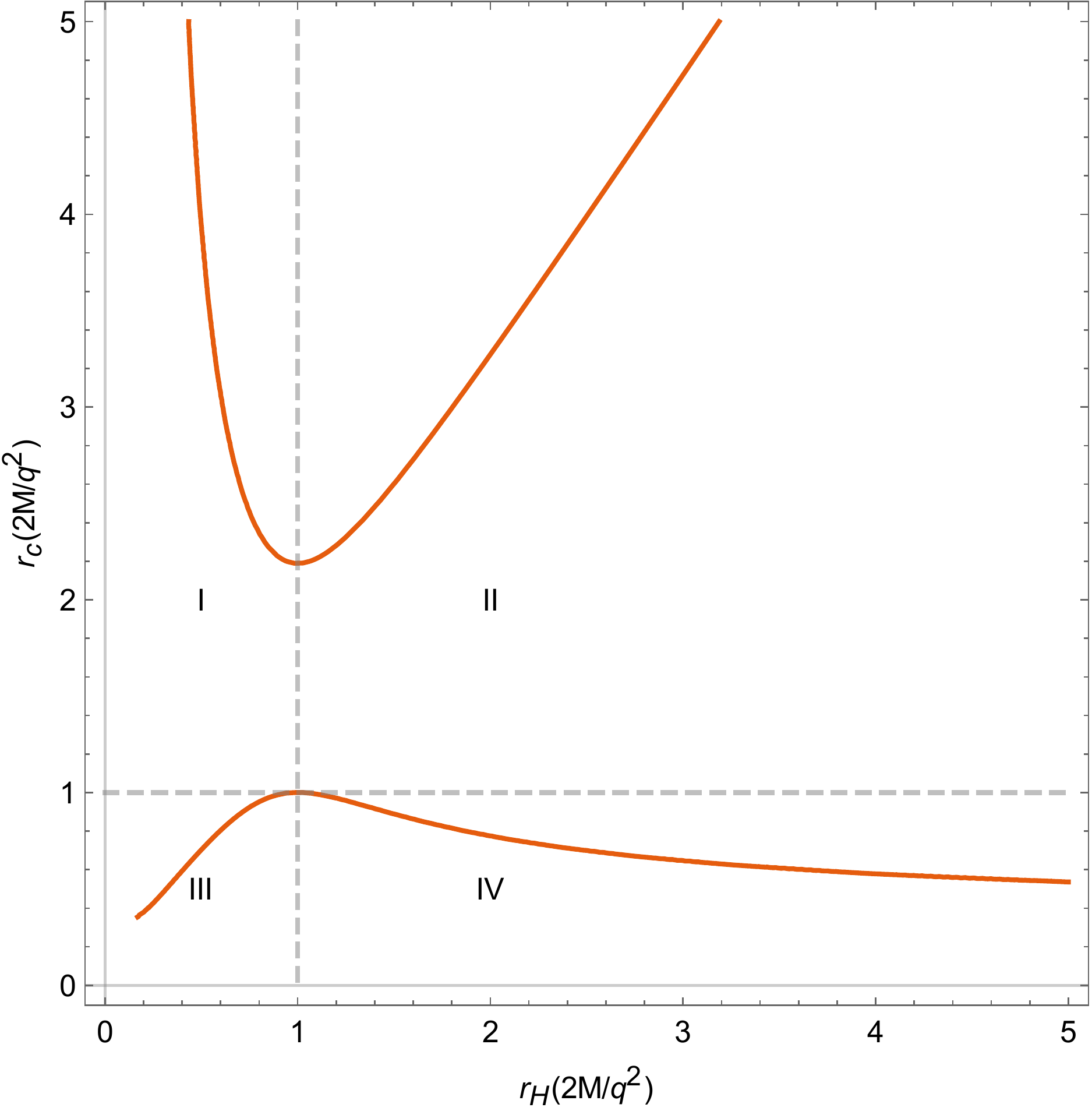}
    \caption{The PS-BH cone of the BV BH.
    The horizontal and vertical dashed gray lines, $x_c=1$ and $x_{\rm H}=1$,
    correspond to the photon sphere radius and the horizon radius in the extremal BV BH, respectively.
    The two lines separate the $x_{\rm H}-x_c$ plane into four regions,
    which are labeled by the Roman numerals.}
    \label{fig:BH-PS-cone-nonlinear}
\end{figure}

To investigate the dependence of temperature on QNMs,
we have to represent the temperature and QNMs in a consistent way,
namely, we replace $x_{\rm H}$ in the temperature by $x_c$ via the following relation,
\begin{equation}
x_{\rm H}\,e^{{1}/{x_{\rm H}}} +\frac{2x_c^2\, e^{{1}/{x_c}} }{1-3 x_c}=0,
\end{equation}
which is dubbed ``\emph{black hole-photon sphere cone}'' (BH-PS cone) and obtained by combining $f(x_{\rm H})=0$ with the photon sphere radius, Eq.\ \eqref{eq:null-geodesics-rescale}.
This name comes from the similarity to the Dirac cone in form.
The BH-PS cone is plotted in Fig.\ \ref{fig:BH-PS-cone-nonlinear}.
Since an extremal BH has the minimal horizon radius and is surrounded by a photon sphere,
only region \rom{2} in Fig.\ \ref{fig:BH-PS-cone-nonlinear} is physical.
The BH-PS cone equation can be solved exactly,
\begin{equation}
x_{\rm H1}^{-1}= -W_0\left[\frac{\left(1-3 x_c\right)e^{-{1}/{x_c}} }{2 x_c^2}\right],\qquad
x_{\rm H2}^{-1}= -W_{-1}\left[\frac{\left(1-3 x_c\right)e^{-{1}/{x_c}} }{2 x_c^2}\right].
\end{equation}
According to the property of Lambert's $W$ functions \cite{olver2010},
we find $x_{\rm H1}>1$ and $0<x_{\rm H2}<1$,
which means that only $x_{\rm H1}$ belongs to region \rom{2}.
Moreover, the imaginary part of QNMs is defined in the whole physical domain if the upper and lower parts of the cone connect without gap.

The BH-PS cone equation (the $x_{\rm H}-x_c$ relationship) can be used to answer
whether the QNMs can reduce to normal modes in the damping process of the BV BH, or whether the imaginary part of QNMs can disappear in the certain range of photon sphere radii, $x_{\rm ext}<x_{c}<x_0$, see the horizontal gap between the two blue curves of Fig.\ \ref{fig:QNMs-BV-BH-MQ}. To this end,
let us numerically compute the minimal photon sphere radius allowed by the physical region in Fig.\ \ref{fig:BH-PS-cone-nonlinear}, $x_c^{\rm min}\approx 2.19$. As a result, the BV BH can never cross to $x_c^{\rm min}$ in its damping process even though $x_c^{\rm min}>x_0 >x_{\rm ext}$. That is to say, the phenomenon of normal modes or of disappearance of imaginary part of QNMs
will never happen because the range of $x_{\rm ext}<x_{c}<x_0$ is outside the physical region.
It is worthy to mention here that this is much different from the situation of $5$D MP BH in Sec.\ \ref{sec:MP-BH}, where
the imaginary part of QNMs for counterrotating case can reach to the minimum of upper BH-PS cone, as the horizon approaches to the extreme value.

\begin{figure}[h!]
    \centering
    \includegraphics[width=.7\textwidth]{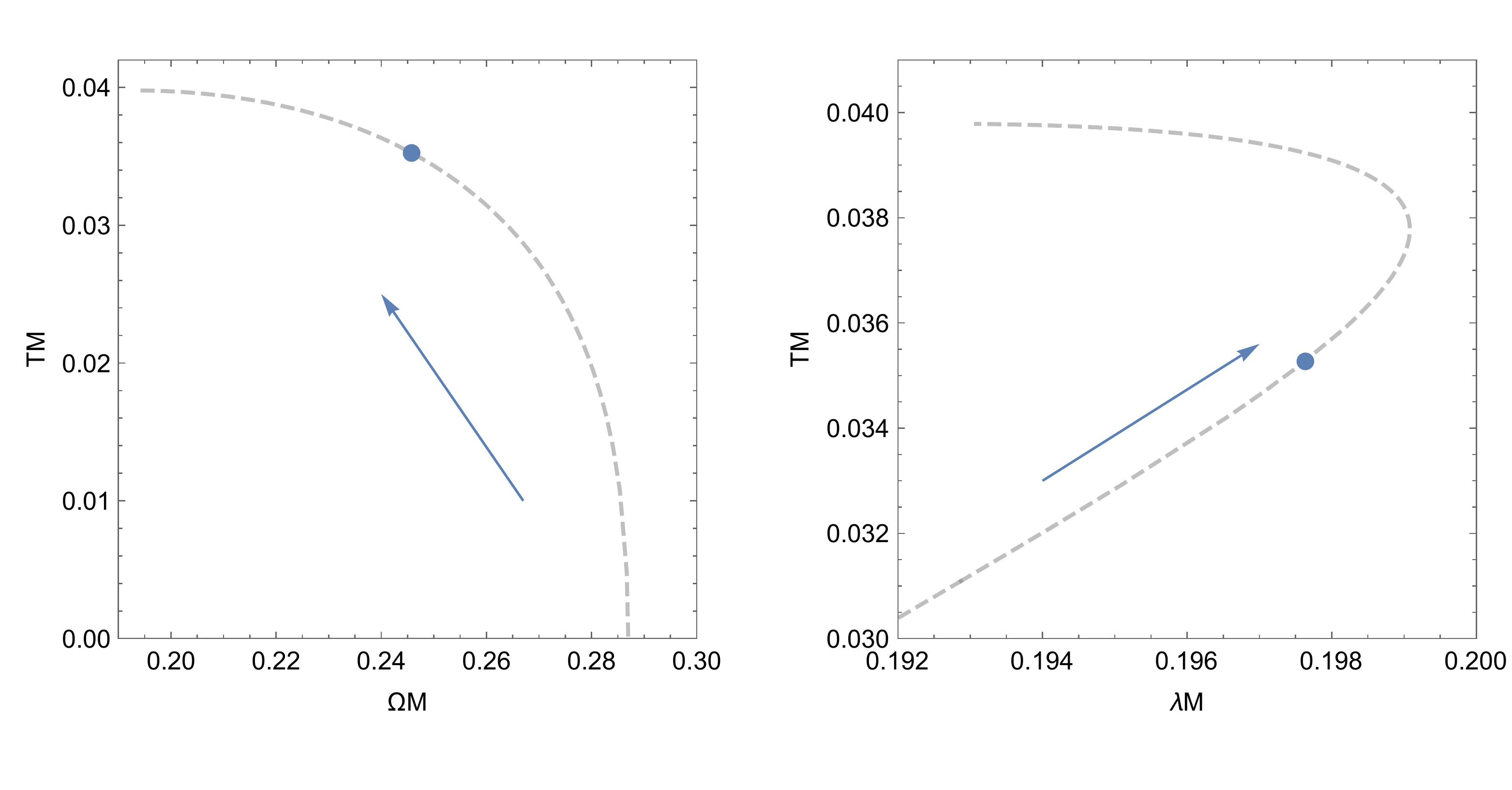}
    \caption{The dependence of temperature on QNMs rescaled by mass in the BV BH.
    The arrow points the direction of increasing $x_c$. The blue points are Davies points.}
    \label{fig:Davies-BV-BH}
\end{figure}

\begin{figure}[h!]
    \centering
    \includegraphics[width=.7\textwidth]{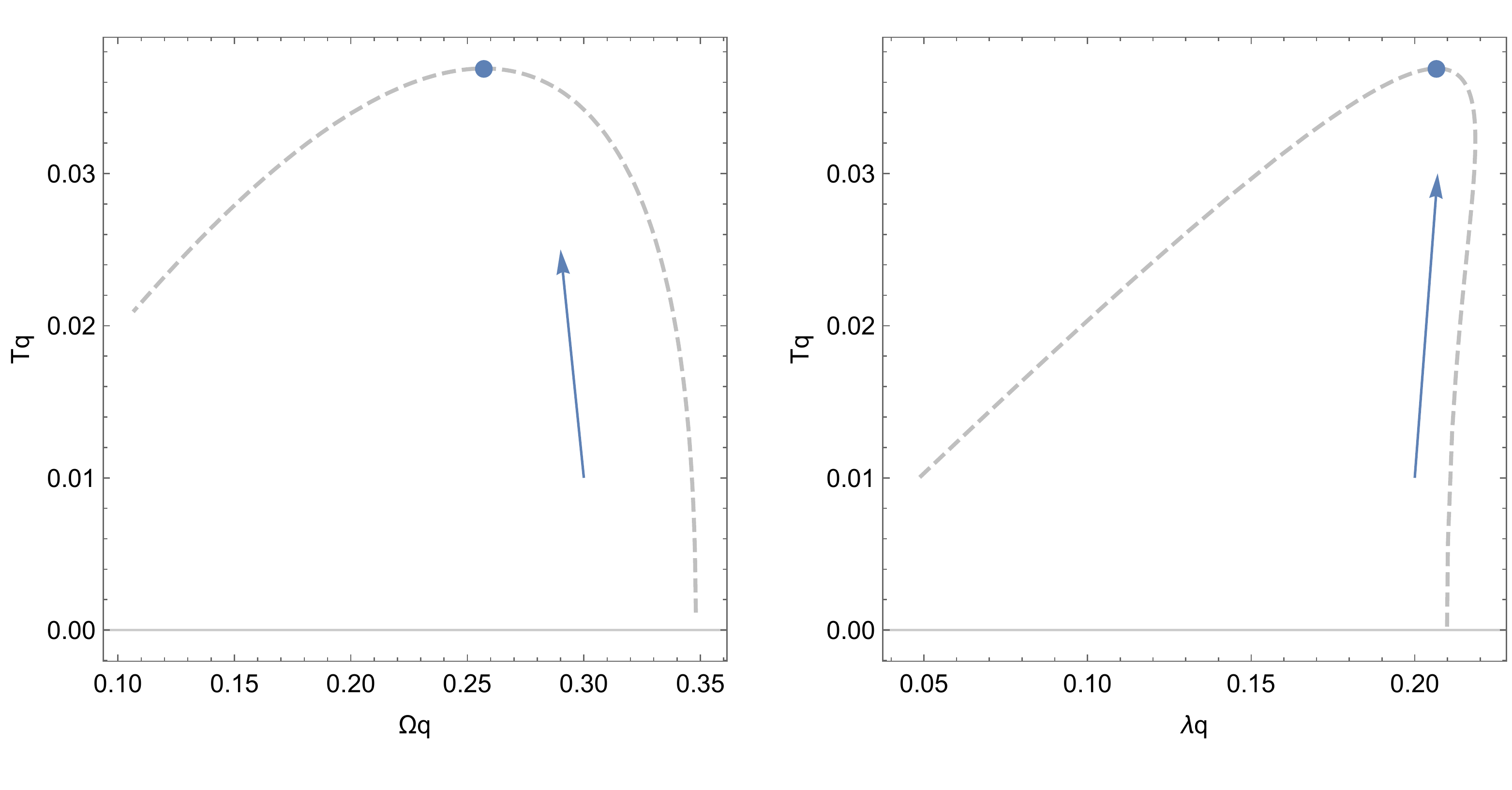}
    \caption{The dependence of temperature on QNMs rescaled by charge in the BV BH.
    The arrow points the direction of increasing $x_c$. The Davies points (blue dots) are located at the maxima.}
    \label{fig:Davies-BV-BH2}
\end{figure}

Omitting the subscript $1$ of $x_{\rm H1}$ and replacing $x_{\rm H}$ by $x_c$ with the help of the BH-PS cone equation,
we rewrite the temperatures Eq.~\eqref{tempmq} in the units of $M$ and $q$, respectively, as a function of $x_c$,
\begin{equation}
T M =
\frac{e^{-W_0\left[\frac{\left(1-3 x_c\right)e^{-{1}/{x_c}} }{2 x_c^2}\right]} }{8 \pi }
\left[W_0\left(\frac{\left(1-3 x_c\right)e^{-{1}/{x_c}} }{2 x_c^2}\right)+1\right],
\end{equation}
and
\begin{equation}
T q=\frac{e^{-\frac{1}{2} W_0\left[\frac{\left(1-3 x_c\right)e^{-{1}/{x_c}} }{2 x_c^2}\right]}}{4 \pi }
\sqrt{-W_0\left[\frac{\left(1-3 x_c\right)e^{-{1}/{x_c}} }{2 x_c^2}\right]}
\left[W_0\left(\frac{\left(1-3 x_c\right)e^{-{1}/{x_c}} }{2 x_c^2}\right)+1\right].
\end{equation}
Then, we depict the dependences of temperature on the real and imaginary parts of QNMs in Fig.\ \ref{fig:Davies-BV-BH}
and Fig.\ \ref{fig:Davies-BV-BH2} in the units of $M$ and $q$, respectively. As we demonstrated in the previous section,
the Davies points are still located at the maxima in the planes $\Omega q-T q$ and
$\lambda q-T q$, see Fig.\ \ref{fig:Davies-BV-BH2}, but such a phenomenon does not happen in the planes $\Omega M-T M$ and $\lambda M-T M$, see Fig.\ \ref{fig:Davies-BV-BH}.


\begin{figure}[h!]
    \centering
    \includegraphics[width=.7\textwidth]{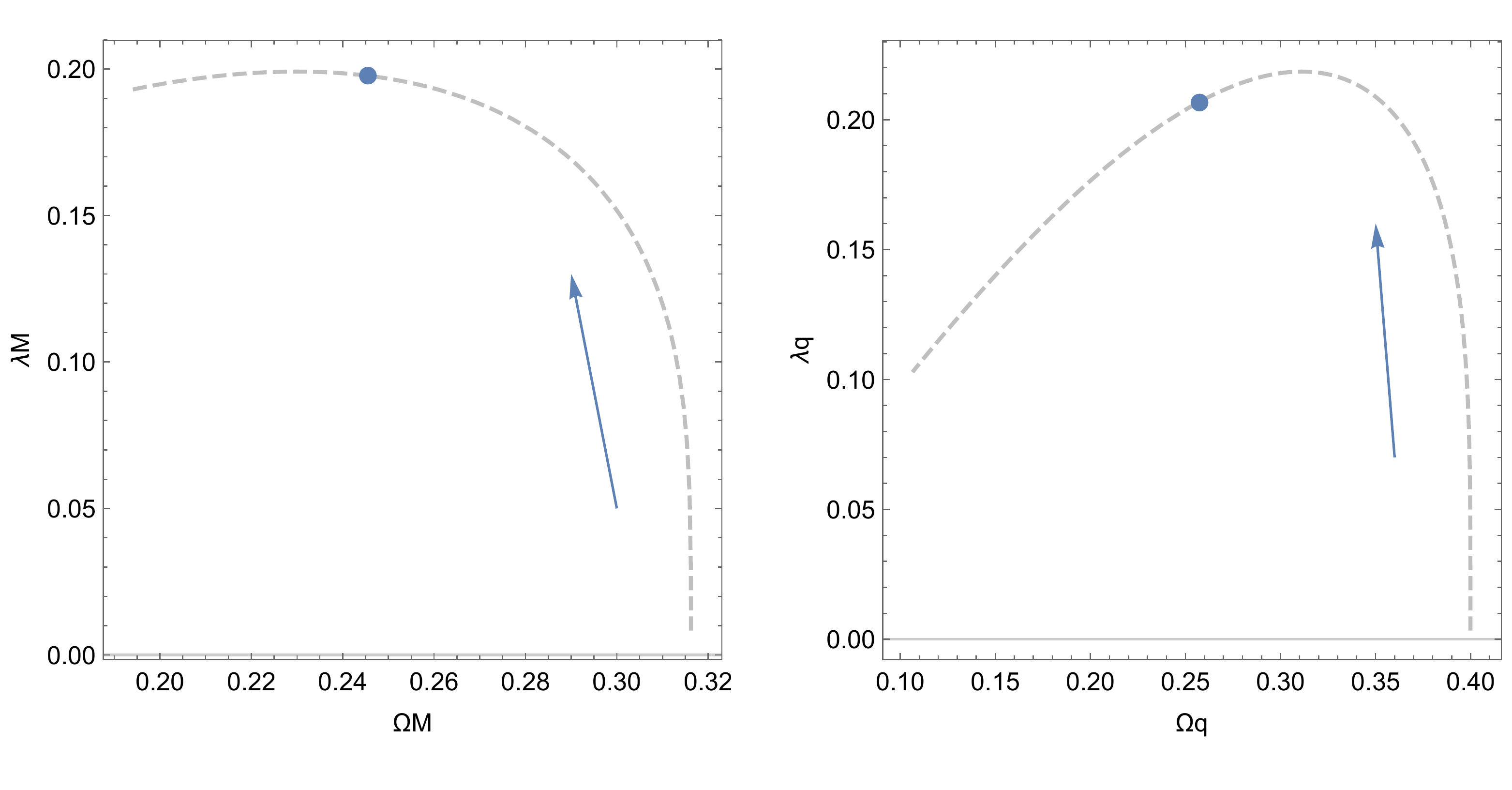}
    \caption{The QNMs of the BV BH in two different units in the complex frequency plane.
    The blue points are Davies points, and the arrows point the direction of increasing $x_c$.}
    \label{fig:mode-plane-BV-BH}
\end{figure}

For the complex frequency plane, the QNMs are exhibited in Fig.\ \ref{fig:mode-plane-BV-BH},
where the Davies points are displayed as well.
The spiral-like shape of QNMs is shown in the unit of $q$,
but it is not so apparent in the unit of $M$.

\section{Noncommutative Schwarzschild black holes}
\label{sec:noncom-SBH}

Now we turn to consider our second example,
the noncommutative Schwarzschild BH \cite{nicolini2005b} whose
action is\footnote{See \ref{app:lagrangian-anisotropic} in detail for the notations and derivations.}
\begin{equation}
S=\int d^{4}x\sqrt{-g}\left(
\frac{1}{16\pi} R -
 \mathcal{L}
\right),
\end{equation}
with
\begin{equation}
\mathcal{L}=-\rho_\theta+p_r-p_\perp-j^\mu a_{\mu}.
\end{equation}
Then from Einstein's equation, $R_{\mu\nu}-\frac{1}{2}g_{\mu\nu}R=8\pi  (T_\theta)_{\mu\nu}$,
one can obtain the shape function,
\begin{equation}
f(r)= 1-\frac{4M}{\sqrt{\pi}\,r}\gamma\left(\frac{3}{2},\frac{r^2}{4\theta}\right),\label{eq:ncbhmf}
\end{equation}
where $\theta$ denotes the noncommutative parameter and $\gamma\left(\frac{3}{2},\frac{r^2}{4\theta}\right)$ the lower incomplete Gamma function,
\begin{equation}
\gamma\left(\frac{3}{2},\frac{r^2}{4 \theta }\right)=\int^{\frac{r^2}{4\theta}}_0 d t \;t^{\frac{1}{2}} e^{-t}.
\end{equation}
The indices of scale factors for the relevant variables are shown in Tab.\ \ref{tab:scale3}.
\begin{table}[htp]
\caption{Indices of scale factors of the noncommutative Schwarzschild BH.}
\begin{center}
\begin{tabular}{cc|cccccccc}
\toprule
    {$M$} & {$\theta$} & {$T$} & {$S$} & {$C_\theta$} & {$\kappa$} & {$r$}  & {$A$} & {$\Omega$} & {$\lambda$} \\ \midrule
    1  & 2 & {$-1$} & {$2$} & {$2$} & {$-1$} & 1 & 2 & {$-1$}& {$-1$} \\ \bottomrule
\end{tabular}
\end{center}
\label{tab:scale3}
\end{table}

Therefore, we have two methods to rescale the variables.
Firstly, $\theta$ is regarded as unit, so $r$ and $M$ as initial variables can be rescaled as follows,
\begin{equation}
r /\sqrt{\theta}= u(x),\qquad
M/\sqrt{\theta} =\alpha(\Theta),
\end{equation}
where $u(x)$ and $\alpha(\Theta)$ are two arbitrary functions which will be fixed below.
Secondly, $M$ is regarded as unit, then $r$ and $\theta$ as initial variables can correspondingly be rescaled by
\begin{equation}
r/M= g(x),\qquad
\theta/M^2= \zeta(m),
\end{equation}
where $g(x)$ and $\zeta(m)$ are arbitrary, and $x$, $\Theta$ and $m$ are dimensionless.

\begin{figure}[h!]
    \centering
    \includegraphics[width=0.4\textwidth]{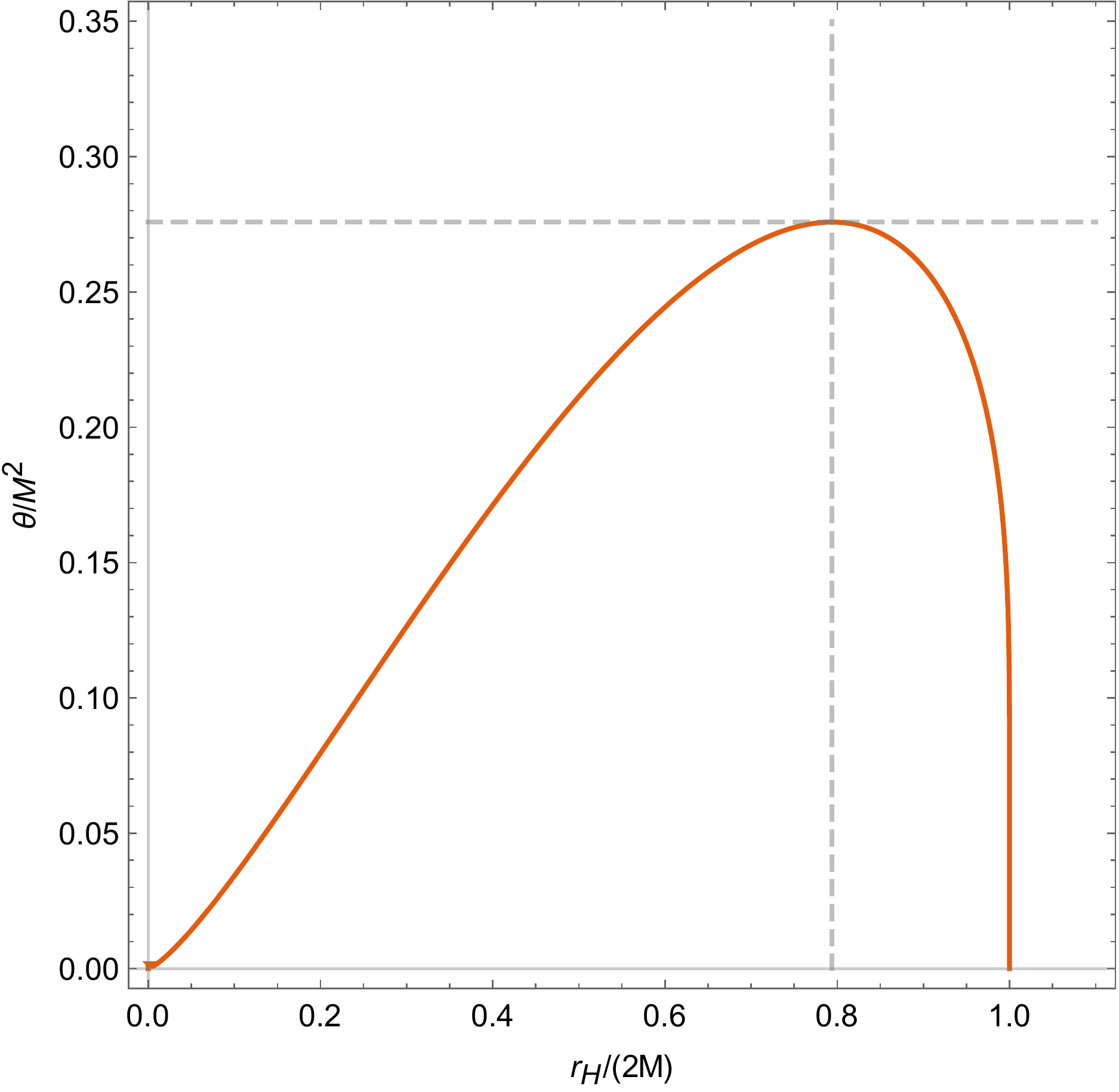}
    \caption{The orange curve depicts Eq.\ \eqref{eq:horizon-noncomm-SBH} or Eq.\ \eqref{eq:horizon-noncomm-SBH2}. The dashed lines are $r_{\rm H}/(2M)=0.79$ and $\theta/M^2=0.28$, respectively.}
    \label{fig:horizon-noncomm-SBH}
\end{figure}

Considering the specific property of Gamma functions
and requiring the linearity of rescaling,
we prefer to adopt the first rescaling method and obtain the concise relations: $u(x)\propto x$
and $\alpha(\Theta)\propto  \Theta^{-1}$.
Based on the relations, we find that it is convenient to make the following rescaling for $r$ and $\theta$,
\begin{equation}
\label{eq:rescale}
r=2x M \Theta ,\qquad
\theta=\Theta^2 M^2,
\end{equation}
where $x\ge0$ and $\Theta\ge0$. Note that the relation of $r$ and $x$ is linear and so is the relation of $\sqrt{\theta}$ and $\Theta$.
As a result, the horizon radius can be obtained from $f(r_{\rm H})=0$,
\begin{equation}
\label{eq:horizon-noncomm-SBH}
1-\frac{2 \gamma \left(\frac{3}{2},x_{\rm H}^2\right)}{\sqrt{\pi}\,x_{\rm H} \Theta} =0,
\end{equation}
or
\begin{equation}
\label{eq:horizon-noncomm-SBH2}
\Theta =\frac{2 }{\sqrt{\pi}\,x_{\rm H}} \gamma \left(\frac{3}{2},x_{\rm H}^2\right).
\end{equation}
Here $x_{\rm H}$ denotes the normalized horizon, where $x_{\rm H}=0$ corresponds to the center and $x_{\rm H}=\infty$ to $\frac{r_{\rm H}}{2M}=1$, which can be seen clearly from Eq.\ \eqref{eq:horizon-noncomm-SBH} or Eq.\ \eqref{eq:horizon-noncomm-SBH2}.
Moreover, $\Theta(x_{\rm H})$ as a function of $x_{\rm H}$ is of a global maximum, see Fig.\ \ref{fig:horizon-noncomm-SBH}, which implies no horizons when $\Theta>\Theta_{\rm max}$, but two event horizons when $0<\Theta<\Theta_{\rm max}$.
$\Theta_{\rm max}$ corresponds to the extremal noncommutative
Schwarzschild BH and can be numerically computed by the first-derivative test,
\begin{equation}
\label{eq:dif-Theta}
\Theta'(x_{\rm H})=
\frac{4  x_{\rm H}}{\sqrt{\pi }}e^{-x_{\rm H}^2} -\frac{2 }{\sqrt{\pi}\, x_{\rm H}^2 }\gamma \left(\frac{3}{2},x_{\rm H}^2\right)=0.
\end{equation}
The root is located at $x_{\rm ext}\approx 1.51$ which corresponds to
the maximum value $\Theta_{\rm max}\approx 0.53$.
Then we get the event horizon for the extremal noncommutative
Schwarzschild BH by using Eq.\ \eqref{eq:rescale},
\begin{equation}
r_{\rm ext}\approx 3.02  \sqrt{\theta},
\end{equation}
when $M= M_{\rm ext} \approx 1.90 \sqrt{\theta}$.
Moreover,
the area of the extremal case is $A_{\rm ext}\approx 114.61 \theta$, which coincides with the results in Refs.\ \cite{nicolini2005b, nicolini2008}.

\subsection{Geometric quantities and regularity }
\begin{figure}[h!]
    \centering
    \includegraphics[width=0.45\textwidth]{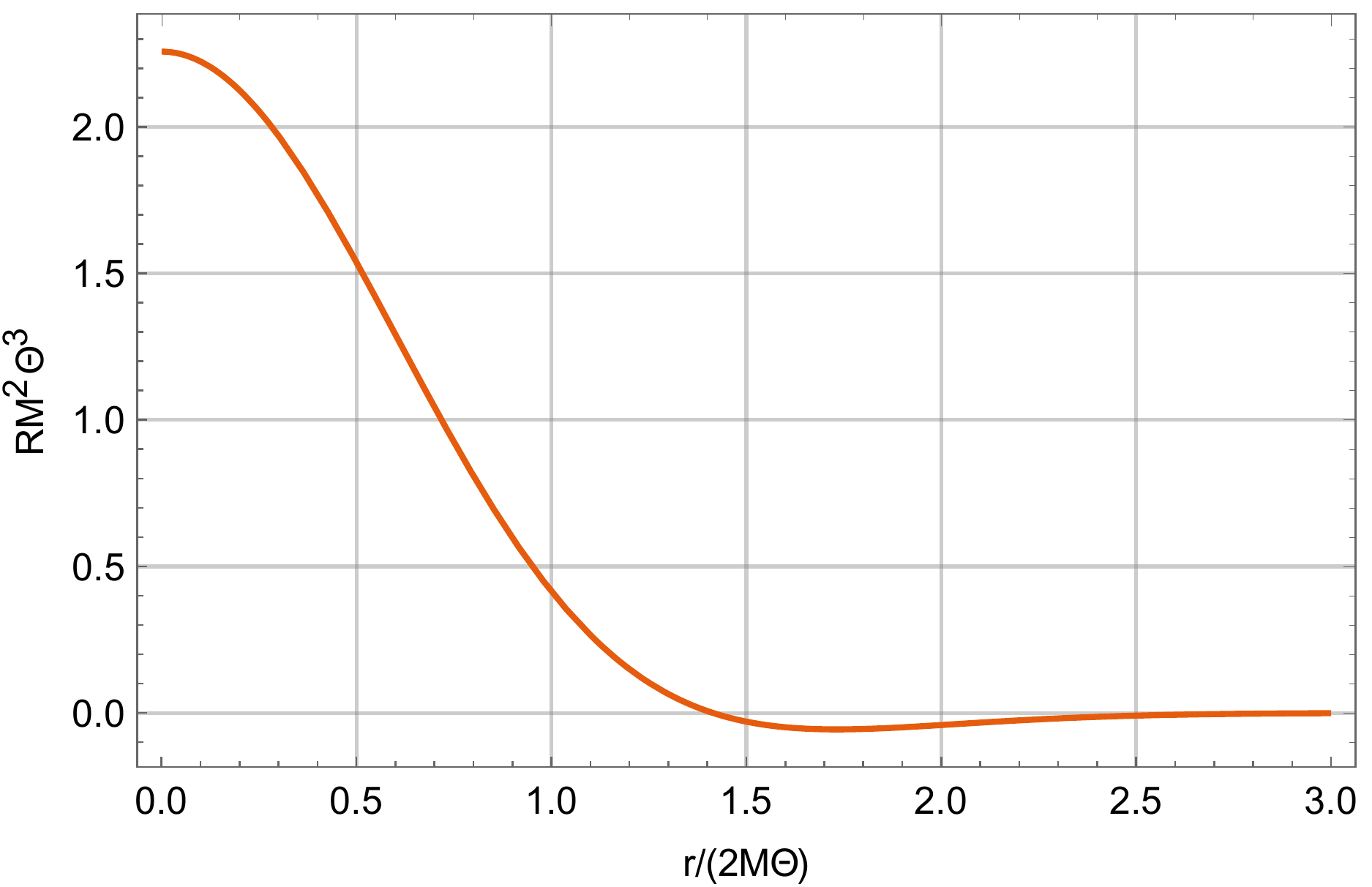}
    \caption{The dependence of the Ricci scalar with respect to the radial coordinate in the noncommutative Schwarzschild BH.}
    \label{fig:RicciS-noncom-SBH}
\end{figure}

The regularity of black holes is related to geometric quantities of spacetime. The most immediate quantity is the Ricci scalar,
\begin{equation}
\mathcal{R}(x)
=
\frac{2 e^{-x^2} \left(2-x^2\right)}{  \sqrt{\pi }\,M^2\Theta ^3 },
\end{equation}
which approaches to the following forms in the two limits, $x\to 0$ and $x\to\infty$,
\begin{equation}
\mathcal{R}(0) = \frac{4}{\sqrt{\pi}M^2 \Theta^3 }>0,\qquad
\mathcal{R}(\infty) = 0.
\end{equation}
In addition, $\mathcal{R}$ has a critical value at $x_0=\sqrt{2}$,  i.e. $\mathcal{R}(\sqrt{2})=0$,
and changes its sign as $r$ increasing from the inside
to the outside range of the noncommutative Schwarzschild BH,
\begin{equation}
\mathcal{R}(x)\begin{cases}
>0, & x<x_0\\
<0, & x>x_0\\
=0, & x=x_0,\; \infty
\end{cases}
\end{equation}
which implies that the noncommutative Schwarzschild BH has a dS core and an AdS outer horizon.
For other geometric quantities, such as the ``square'' of the Ricci tensor,
one has $\mathcal{R}_{\mu\nu}R^{\mu\nu}= 4/(\pi M^4\Theta^6)$ on the one side $x\to 0$, and $\mathcal{R}_{\mu\nu}R^{\mu\nu}\to 0$ on the other side $x\to \infty$; for the Kretschmann scalar, $\mathcal{R}^\rho_{\mu\nu\beta}\mathcal{R}_\rho^{\mu\nu\beta}= 8/(3 \pi \Theta ^6 M^4)$ as $x\to 0$,
and $\mathcal{R}^\rho_{\mu\nu\beta}\mathcal{R}_\rho^{\mu\nu\beta}\to 0$ as $x\to \infty$.
We notice that all the above geometric quantities are regular, so the noncommutative black hole is regular everywhere.
The above demonstrations can alternatively be seen clearly in Fig.\ \ref{fig:RicciS-noncom-SBH}.

\subsection{Deformation of the first law of black bole mechanics}
\begin{figure}[h!]
    \centering
    \includegraphics[width=0.45\textwidth]{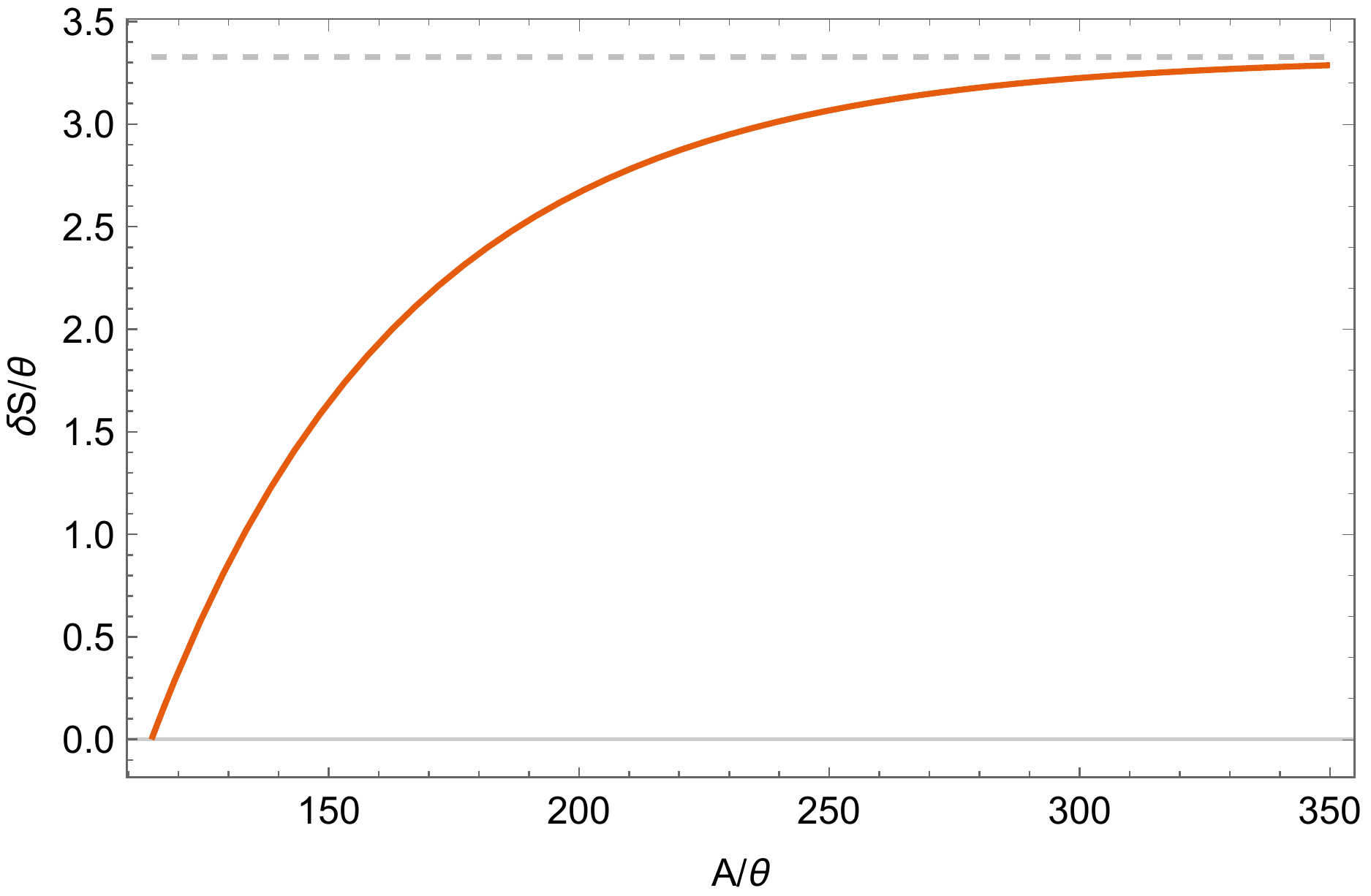}
    \caption{The dependence of entropy deviation with respect to the BH area for the noncommutative Schwarzschild BH.}
    \label{fig:entropy-deviation-SBH}
\end{figure}

Using the same method as in the BV BH, we try to rebuild the relationship among entropy, temperature and mass in the noncommutative Schwarzschild BH.
The surface gravity without the backreaction reads
\begin{equation}
\kappa = \frac{f'(r_{\rm H})}{2}=\frac{1}{2r_{\rm H}}\left(
1
-\frac{r_{\rm H}^3 e^{-\frac{r_{\rm H}^2}{4 \theta }}}{4 \theta ^{3/2} \gamma _{\rm H}}
\right),\label{ncbhsg}
\end{equation}
where $\gamma_{\rm H}$ can be represented via the horizon $r_{\rm H}$,
\begin{equation}
\gamma_{\rm H}\equiv\gamma\left(\frac{3}{2},\frac{r_{\rm H}^2}{4\theta}\right) =\frac{\sqrt{\pi }\,r_{\rm H}}{4M}.\label{bcbhga}
\end{equation}
Using Eqs.\ \eqref{ncbhsg} and \eqref{bcbhga},  we calculate the variation of mass,
\begin{equation}
d M =
\frac{\sqrt{\pi }}{4 \gamma _{\rm H}}
\left(
1-\frac{r_{\rm H}^3 e^{-\frac{r_{\rm H}^2}{4 \theta }}}{4 \theta ^{3/2} \gamma _{\rm H}}
\right)
d r_{\rm H}.
\end{equation}
Then considering the variation of area, $d A = 8\pi r_{\rm H} d r_{\rm H}$,
we obtain the deformation of the first law of BH mechanics,
\begin{equation}
\label{eq:fistlaw-noncomm-SBH}
\frac{r_{\rm H}}{2M}d M
= \frac{\kappa}{8\pi} d A.
\end{equation}
Note that $\frac{r_{\rm H}}{2M}\to 1$ as $\theta\to 0$, i.e., the deformation disappears.
To reconstruct the first law of BH thermodynamics,
let us apply $T=\kappa/2\pi$ and rearrange Eq.\ \eqref{eq:fistlaw-noncomm-SBH} to be
\begin{equation}
d M = T d S,
\qquad
d S = \frac{d A}{4}+\frac{2\pi}{\kappa}\left(1-\frac{r_{\rm H}}{2M}\right)d M.
\end{equation}
Since $0\le r_{\rm H} \le 2M$, the second part in $d S$ is positive.
This implies that $d S \ge d A/4$.
The entropy can be obtained by the integral,
\begin{equation}
S=\int \frac{d M}{T}=
\frac{A}{4}+
\delta S,\qquad
 \delta S=8 \pi \theta
\int_{z_{\rm ext}}^{z} d \tilde z
\frac{\sqrt{\pi }-2 \gamma \left(\frac{3}{2},\tilde z\right)}{4 \gamma \left(\frac{3}{2},\tilde z\right)},\label{nchbent}
\end{equation}
where $z=r_{\rm H}^2/(4 \theta)=A/(16\pi \theta)$ and $z_{\rm ext}=r_{\rm ext}^2/(4 \theta)\approx 2.28$, see Fig.\ \ref{fig:entropy-deviation-SBH}.

As $\theta$ approaches to zero, the upper limit in the integral of entropy deviation tends to infinity, i.e. $z\to\infty$, and the integral is finite,
\begin{equation}
\int_{z_{\rm ext}}^{\infty} d \tilde z
\frac{\sqrt{\pi }-2 \gamma \left(\frac{3}{2},\tilde z\right)}{4 \gamma \left(\frac{3}{2},\tilde z\right)}
\approx 3.33,
\end{equation}
which implies that the relation between $S$ and $A$ reduces to the standard form, $S=A/4$, due to the vanishing coefficient of $ \delta S$. That is to say, the noncommutative Schwarzschild BH turns back to the normal Schwarzschild BH.
Alternatively, the entropy can be obtained by the integral \cite{smailagic2012,Miao:2016ipk},
\begin{equation}
S=\int \frac{d M}{T}=8 \pi ^{3/2} \theta  \int_{z_{\rm ext}}^{z} \frac{d \tilde z}{ \gamma \left(\frac{3}{2},\tilde z\right)},
\end{equation}
which coincides with Eq.\ \eqref{nchbent}.

\subsection{Heat capacity and Davies points}

\begin{figure}[h!]
    \centering
    \includegraphics[width=0.45\textwidth]{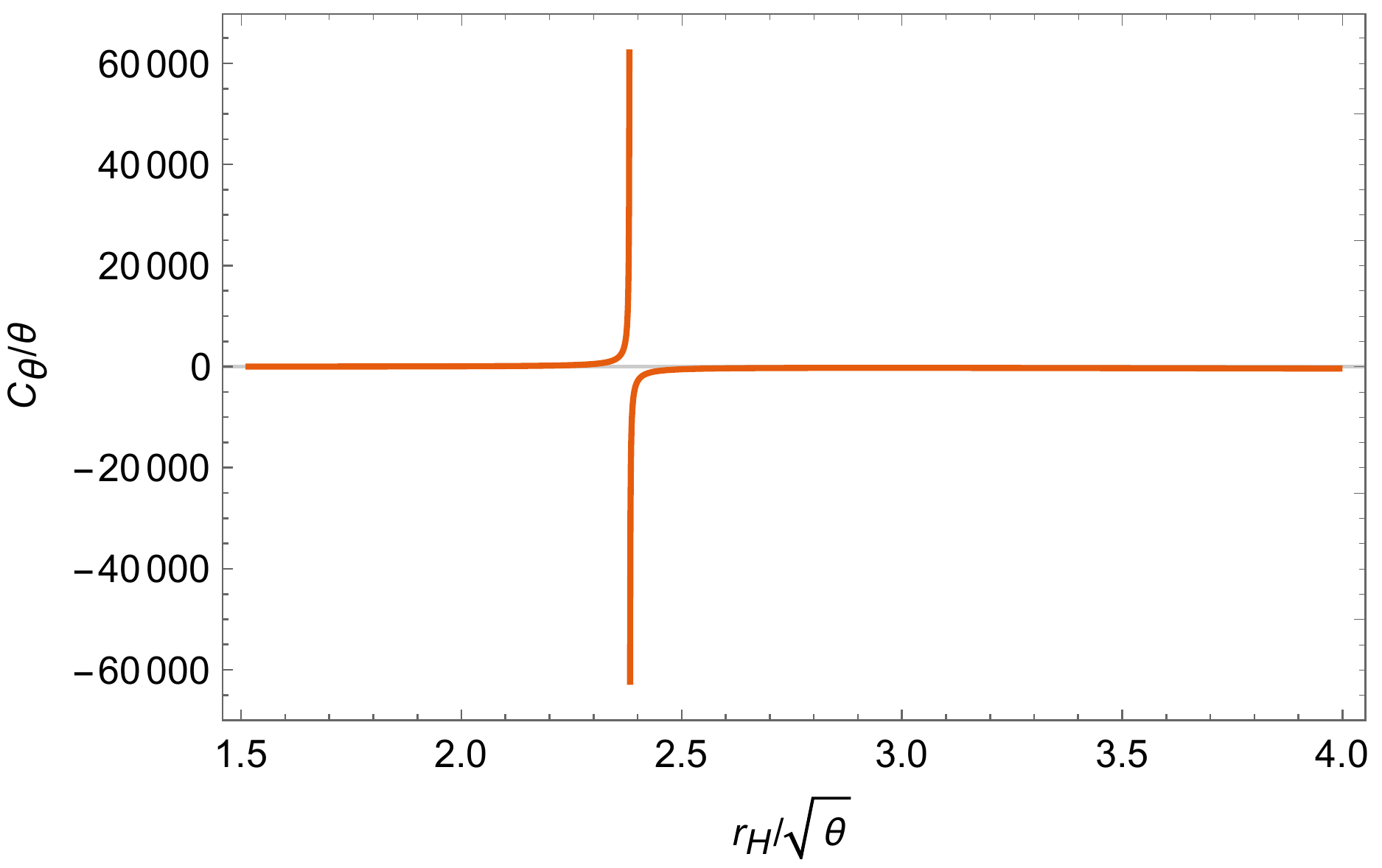}
    \caption{The heat capacity versus the horizon radius for the noncommutative Schwarzschild BH in the unit of $\theta$.}
    \label{fig:capacity-noncomm-SBH}
\end{figure}

The temperature of the noncommutative Schwarzschild BH in the unit of $\theta$ reads
\begin{equation}
T\sqrt{\theta} =
\frac{1}{8 \pi  x_{\rm H}}-\frac{x_{\rm H}^2e^{-x_{\rm H}^2} }{4 \pi  \gamma_{\rm H} },
\end{equation}
and the heat capacity takes the form in terms of the rescaled horizon $x_{\rm H}$,
\begin{equation}
C_\theta/\theta =
\frac{T}{\theta}\cdot\left(\frac{\partial S}{\partial T}\right)_\theta
=\frac{4 \pi ^{3/2}  x_{\rm H}^2 \left( \gamma_{\rm H} -2 x_{\rm H}^3e^{-x_{\rm H}^2}\right)}
	{- \gamma^2_{\rm H}
		+4  \left(x_{\rm H}^2-1\right) x_{\rm H}^3 \gamma_{\rm H} \,e^{-x_{\rm H}^2}
		+4 x_{\rm H}^6e^{-2x_{\rm H}^2}}.
\end{equation}
The Davies point can be found numerically, i.e. $x_{\rm H}^*\approx 2.38$ or $r_{\rm H}^*\approx 4.76\sqrt{\theta}>r_{\rm ext}\approx 3.02\sqrt{\theta}$.
See Fig.\ \ref{fig:capacity-noncomm-SBH} for the graph of the heat capacity versus the horizon radius rescaled by $\theta$.

In addition, we give the dimensionless temperature in the unit of $M$,
\begin{equation}
T M=\frac{\gamma_{\rm H} -2 x_{\rm H}^3 e^{-x_{\rm H}^2} }
{8 \sqrt{\pi } \, \gamma_{\rm H}  \left(3  \gamma_{\rm H} -2 x_{\rm H}^3e^{-x_{\rm H}^2}\right)},
\end{equation}
and the heat capacity in the same unit,
\begin{equation}
C_\theta/M^2=
\frac{16  \sqrt{\pi } \, \gamma^2_{\rm H}
\left( \gamma_{\rm H} -2 x_{\rm H}^3e^{-x_{\rm H}^2}\right)}
{-\gamma^2_{\rm H}
+4 \left(x_{\rm H}^2-1\right) x_{\rm H}^3 \gamma_{\rm H}\, e^{-x_{\rm H}^2}
+4 x_{\rm H}^6e^{-2x_{\rm H}^2}}.
\end{equation}

\subsection{Quasinormal modes in the eikonal limit}

\begin{figure}[h!]
    \centering
    \includegraphics[width=0.4\textwidth]{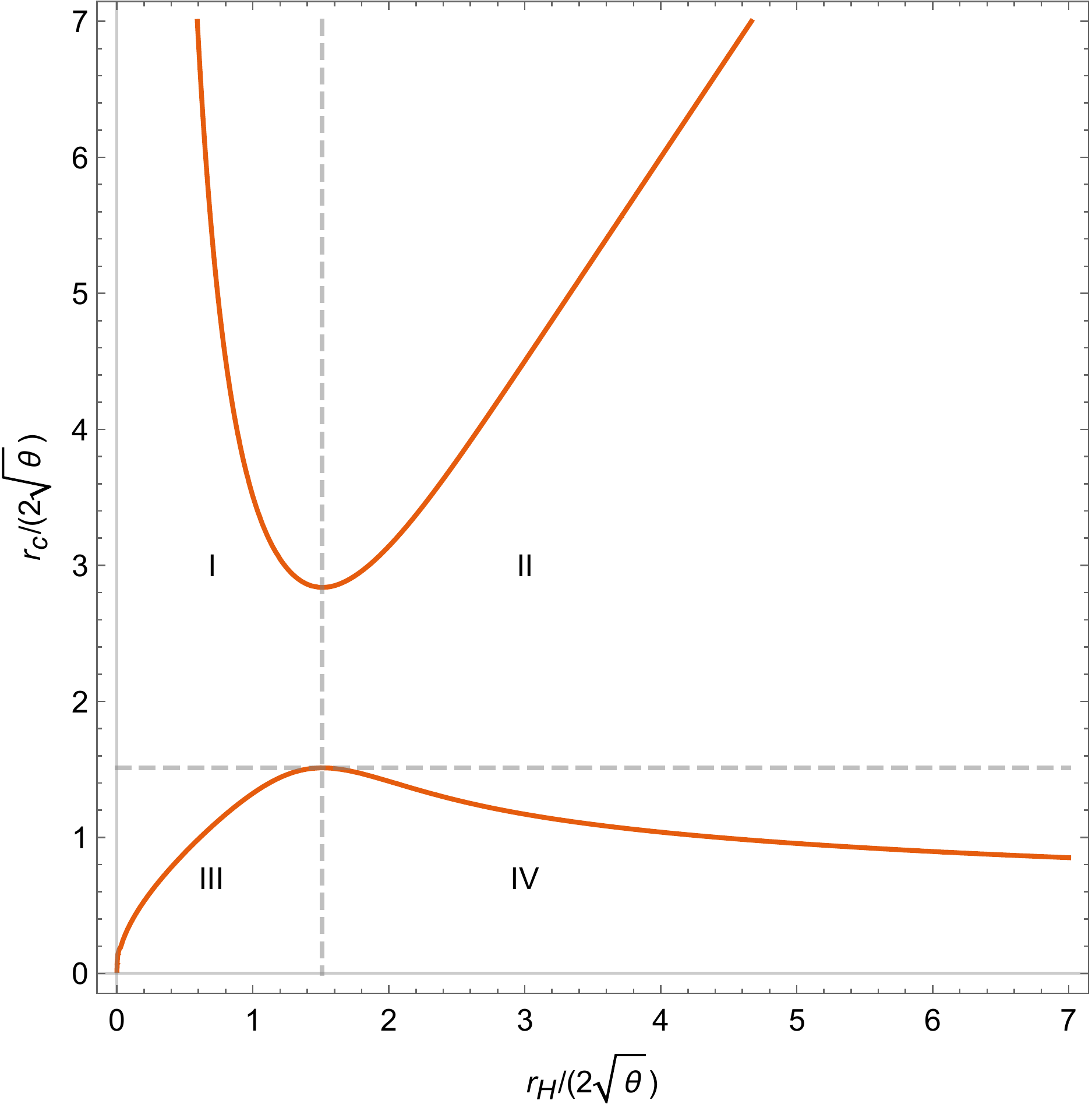}
    \caption{The BH-PS cone of the noncommutative Schwarzschild BH.
    The orange curves denote the BH-PS cone,
    while the dashed gray lines correspond to the extremal value which is approximately equal to $1.51$.}
    \label{fig:BH-PS-cone-noncomS}
\end{figure}

Although the radius of photon spheres, $r_{c}$, cannot be solved exactly from the following equation in the noncommutative Schwarzschild BH,
\begin{equation}
2  \left(\sqrt{\pi }\, r_c-6 M \gamma_c \right)
+M\theta ^{-3/2}  r_c^3  e^{-\frac{r_c^2}{4 \theta }}=0,
\end{equation}
where $\gamma_c\equiv\gamma\left(\frac{3}{2},\frac{r_c^2}{4 \theta }\right)$,
we can still represent the physical quantities in terms of $r_c$.
Using the transformation Eq.\ \eqref{eq:rescale}, we simplify the above equation to be
\begin{equation}
 \left(\sqrt{\pi}\, \Theta  x_c -3 \gamma_c \right)
 +2 x_c^3e^{-x_c^2}=0,\label{ncbhpsr}
\end{equation}
where the normalized radius $x_c$ connects to the horizon $x_{\rm H}$ through $\Theta$.
Comparing Eq.\ \eqref{ncbhpsr} with Eq.\ \eqref{eq:horizon-noncomm-SBH}, we derive the BH-PS cone equation,
\begin{equation}
\label{eq:relation-horizon-radius}
x_{\rm H}^{-1} \gamma_{\rm H}
=\frac{3 }{2 }x_c^{-1}\gamma_c
- x_c^2\,e^{-x_c^2}.
\end{equation}
See Fig.\ \ref{fig:BH-PS-cone-noncomS} for the graph of the horizon radius versus the photon sphere radius in the unit of $\theta$.

Using the BH-PS cone equation, we derive the derivative of $x_c$ with respect to $x_{\rm H}$,
\begin{equation}
x_c'(x_{\rm H})= -\frac{-3 \gamma_c +4  x_{\rm H}^2x_c\,e^{-x_{\rm H}^2} +2 x_c^3\,e^{-x_c^2} }{\left(4 x_c^2+2\right) \gamma_{\rm H} -6 x_{\rm H} x_c \gamma_c },
\end{equation}
where the stationary points correspond to the roots of the equation $x_c'(x_{\rm H})=0$.
With the help of Eq.\ \eqref{eq:relation-horizon-radius},
we obtain the reduced constraint for the stationary points,
\begin{equation}
2 x_{\rm H}^3e^{-x_{\rm H}^2} -\gamma_{\rm H}=0.
\end{equation}
Because the left hand side of the above equation is proportional to Eq.\ \eqref{eq:dif-Theta}, we conclude that the above equation has the same roots
as that of $\Theta'(x_{\rm H})=0$, i.e., $x_{\rm H}=x_{\rm ext}\approx 1.51$ is associated with two extremal values of $x_c$.
One is $x_c=x_{\rm ext}$, which can be verified by directly substituting it into Eq.\ \eqref{eq:relation-horizon-radius};
the other solution can be numerically computed, $x_c\approx 2.84$.
Since the outer horizon radius and the photon sphere radius should be greater than $r_{\rm ext}$, only region \rom{2} in Fig.\ \ref{fig:BH-PS-cone-noncomS} is physical.

As done in the previous section, we calculate the real and imaginary parts of QNMs of the noncommutative Schwarzschild BH in the unit of $M$, respectively,
\begin{equation}
\Omega M=\frac{\sqrt{\pi}}{2}
\frac{\sqrt{ \gamma_c -2 x_c^3e^{-x_c^2}}}
{\left(3  \gamma_c -2 x_c^3e^{-x_c^2}\right)^{3/2}},
\qquad
\lambda M=
\frac{\sqrt{\pi}}{2}
\frac{\sqrt{\left(\gamma_c -2 x_c^3e^{-x_c^2} \right)
\left[3 \gamma_c -2  \left(x_c^3+2 x_c^5\right)e^{-x_c^2}\right]}}
{ \left(3\gamma_c-2 x_c^3 e^{-x_c^2} \right)^2}.
\end{equation}
We note that the same factor, $\sqrt{\gamma_c -2 x_c^3e^{-x_c^2}}$, appears in the numerators of $\Omega M$ and $\lambda M$ and that it is formally proportional to Eq.\ \eqref{eq:dif-Theta}.
This implies that both $\Omega$ and $\lambda$ vanish as $x_c\to x_{\rm ext}\approx 1.51$.
\begin{figure}[h!]
    \centering
    \includegraphics[width=0.34\textwidth]{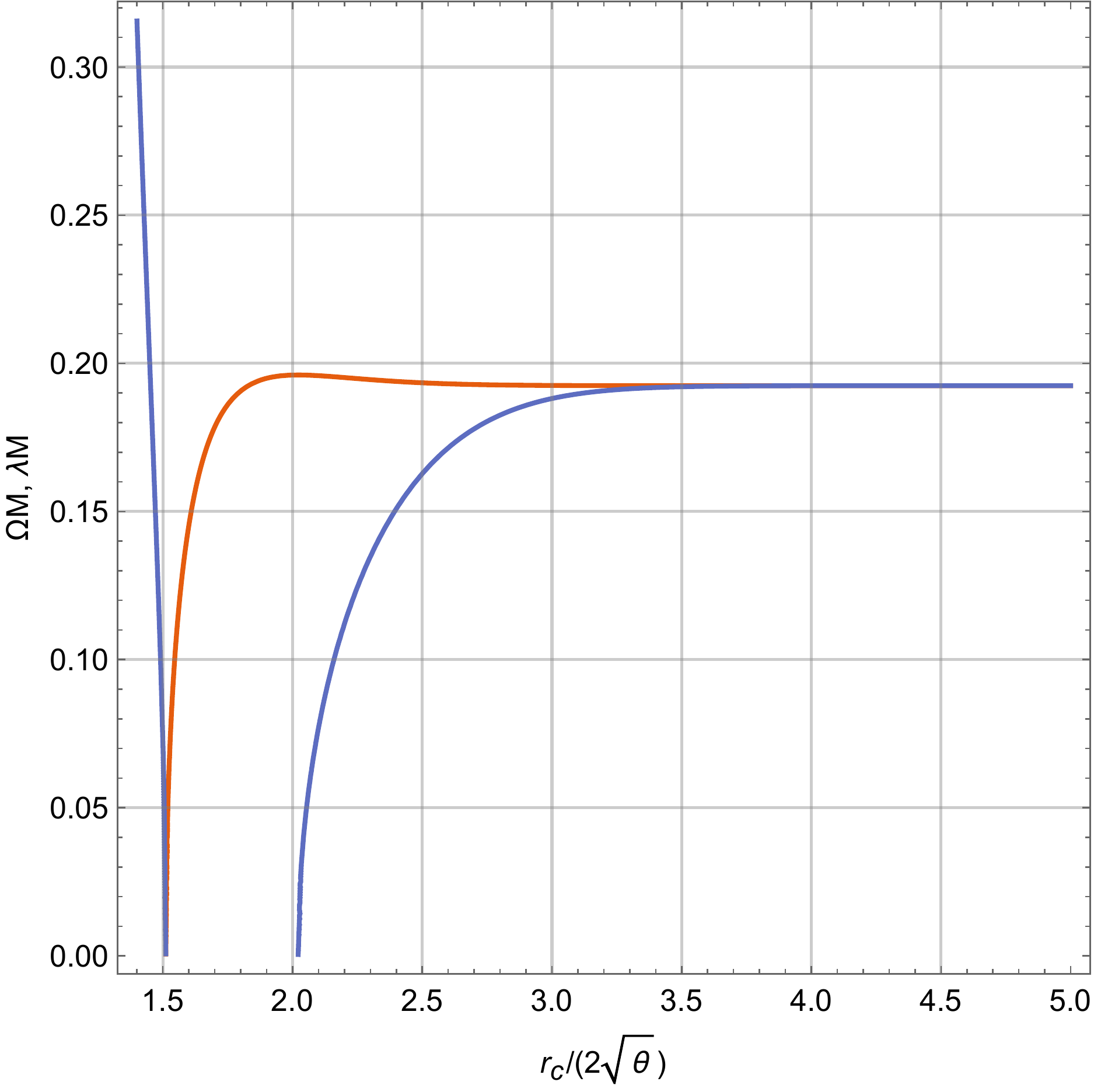}
    \includegraphics[width=0.35\textwidth]{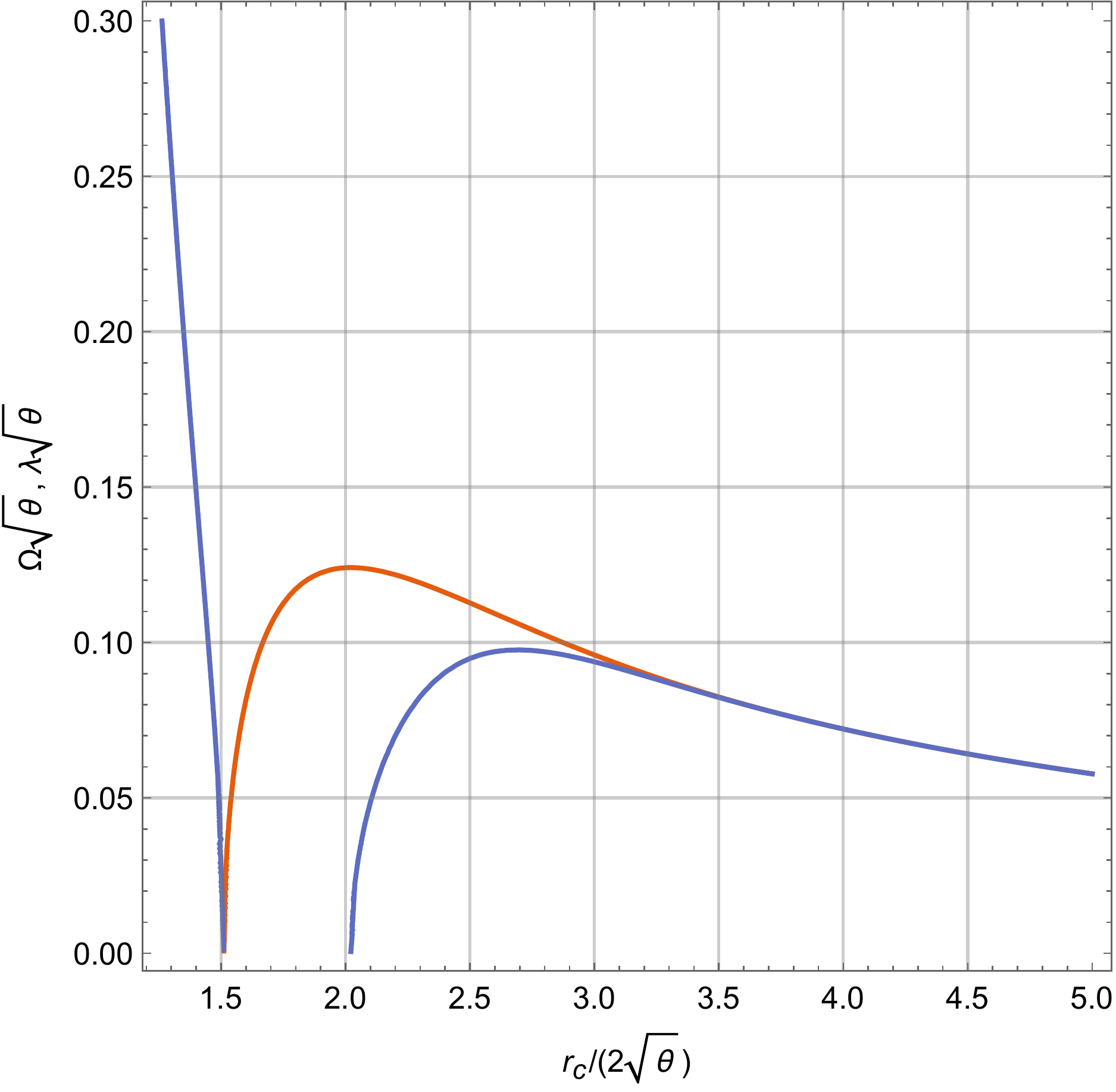}
    \caption{The QNMs with respect to the photon sphere radius in the unit of $M$ (left) and $\theta$ (right) for the noncommutative Schwarzschild BH.
    The orange curves denote the real part of QNMs,
    while the blue curves the imaginary part which
    is defined in the range of $x_c\equiv r_c/(2\sqrt{\theta})>2.02$.}
    \label{fig:QNMs-noncom-SBH-theta}
\end{figure}
In addition, $\lambda$ is of the other zero at the critical point $x_c= x_0\approx 2.02$. That is to say, $\lambda$ vanishes in the gap of $x_c$ from $x_{\rm ext}$ to  $x_0$. Considering $x_c^{\rm min}\approx 2.84$ (the minimum point in the physical region \rom{2},  see Fig.\ \ref{fig:BH-PS-cone-noncomS}). larger than $x_0$, we deduce that the physical QNMs in the eikonal limit will never be zero.
See the left graph of Fig.\ \ref{fig:QNMs-noncom-SBH-theta} for the details.

Moreover, the real and imaginary parts of QNMs in the unit of $\theta$ can be computed, respectively,
\begin{equation}
\Omega \sqrt{\theta}=
\frac{1}{2 x_c}
\sqrt{\frac{\gamma_c-2 x_c^3e^{-x_c^2} }{3 \gamma_c-2 x_c^3e^{-x_c^2} }},
\qquad
\lambda \sqrt{\theta}=
\frac{
	\sqrt{3  \gamma_c^2
	-4 \left(2+x_c^2\right) x_c^3 \gamma_c\, e^{-x_c^2}
	+4  \left(x_c^6+2 x_c^8\right)e^{-2x_c^2}}}
	{6 x_c \gamma_c -4 x_c^4 e^{-x_c^2}}.
\end{equation}
The shape of QNMs in this unit does not change too much when
compared with that in the unit of $M$, see the right graph of Fig.\ \ref{fig:QNMs-noncom-SBH-theta}.

\begin{figure}[h!]
    \centering
        \includegraphics[width=.7\textwidth]{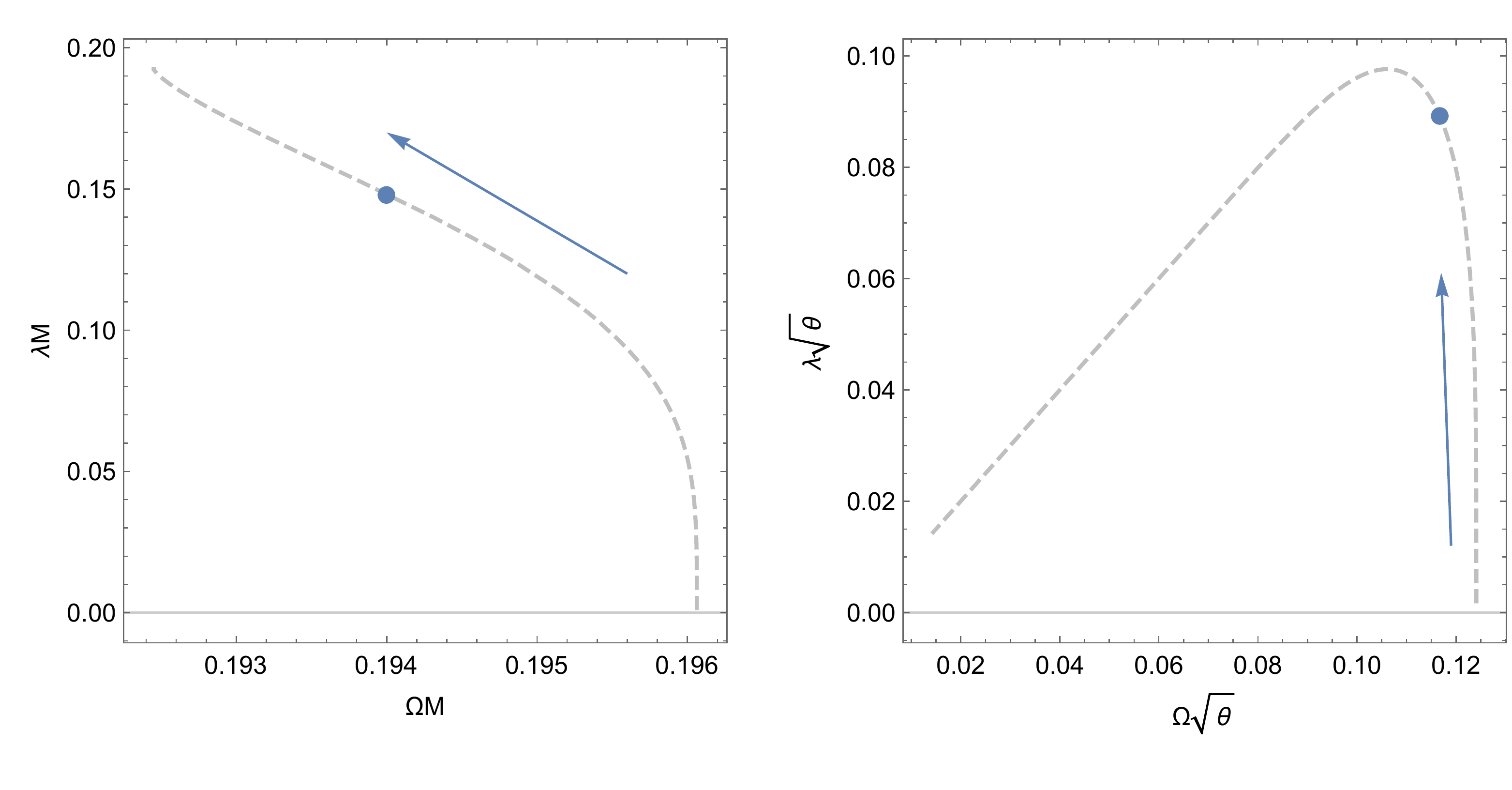}
    \caption{The real part versus the imaginary part of QNMs for the noncommutative Schwarzschild BH in the unit of $M$ (left) or $\theta$ (right).
    The arrows point to the direction of increasing $x_c$. The blue dots are Davies points.}
    \label{fig:Davies-SBH-M}
\end{figure}

As to the spiral-like shape in the noncommutative Schwarzschild BH, we have a quite interesting discovery. The spiral-like shape does not exist when the QNMs are rescaled in the unit of $M$, but the obvious spiral-like shape appears when the QNMs are rescaled in the unit of $\theta$. This implies that the spiral-like shape depends on the way of rescaling. See Fig.\ \ref{fig:Davies-SBH-M} for the details.

\begin{figure}[h!]
    \centering
    \includegraphics[width=.7\textwidth]{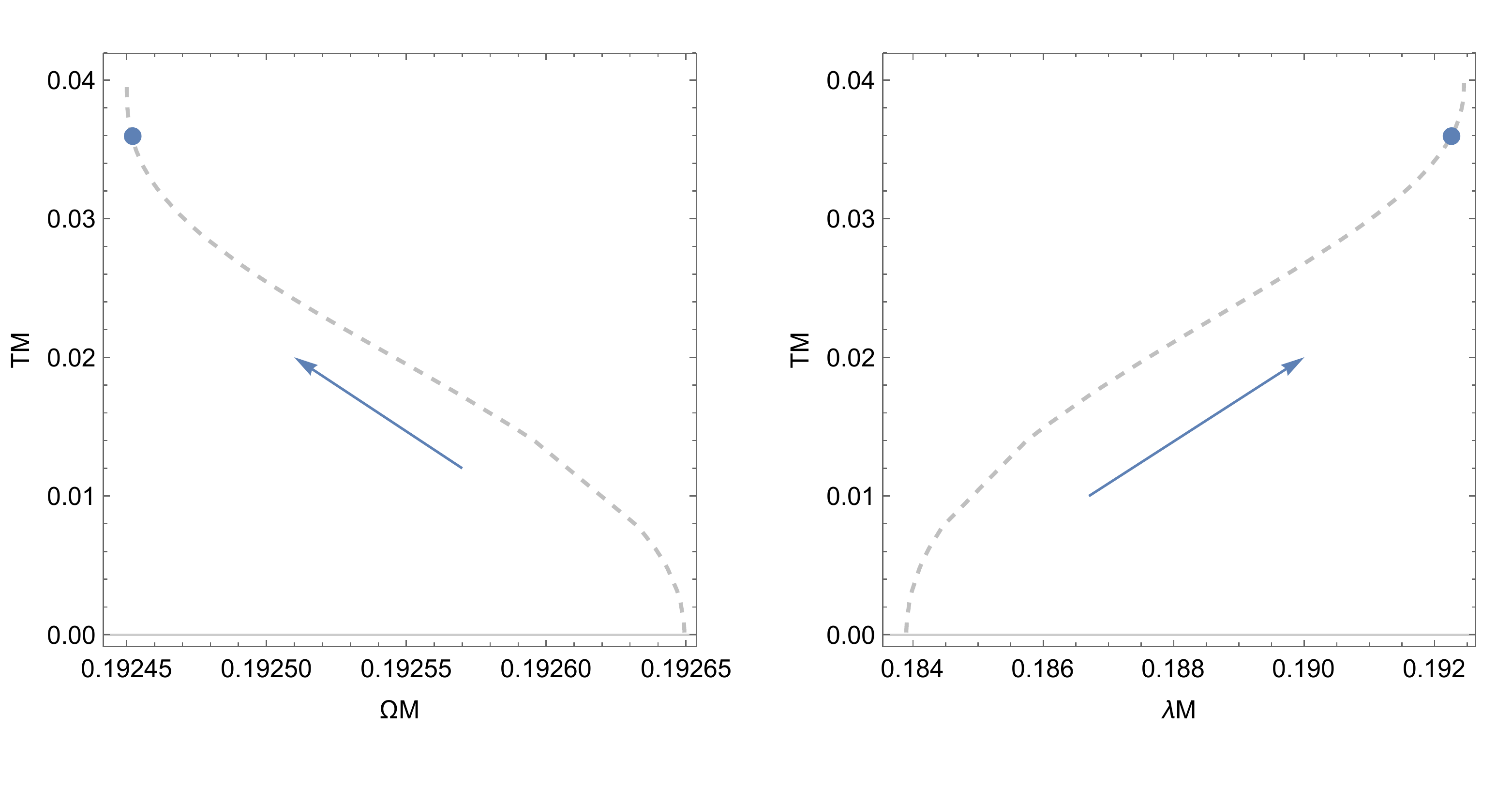}
    \caption{The dependence of temperature on  the real and imaginary parts of QNMs, respectively, in the unit of $M$ for the noncommutative Schwarzschild BH.
    The arrows point to the direction of increasing $x_c$. The blue points are Davies points.}
    \label{fig:QNMs-noncom-SBH-M}
\end{figure}

The dependence of temperature on the real and imaginary parts of QNMs in the unit of $M$ is shown in Fig.\ \ref{fig:QNMs-noncom-SBH-M}.
As we have demonstrated, the Davies point is not located at the maximum of the curves.

\begin{figure}[h!]
    \centering
        \includegraphics[width=0.35\textwidth]{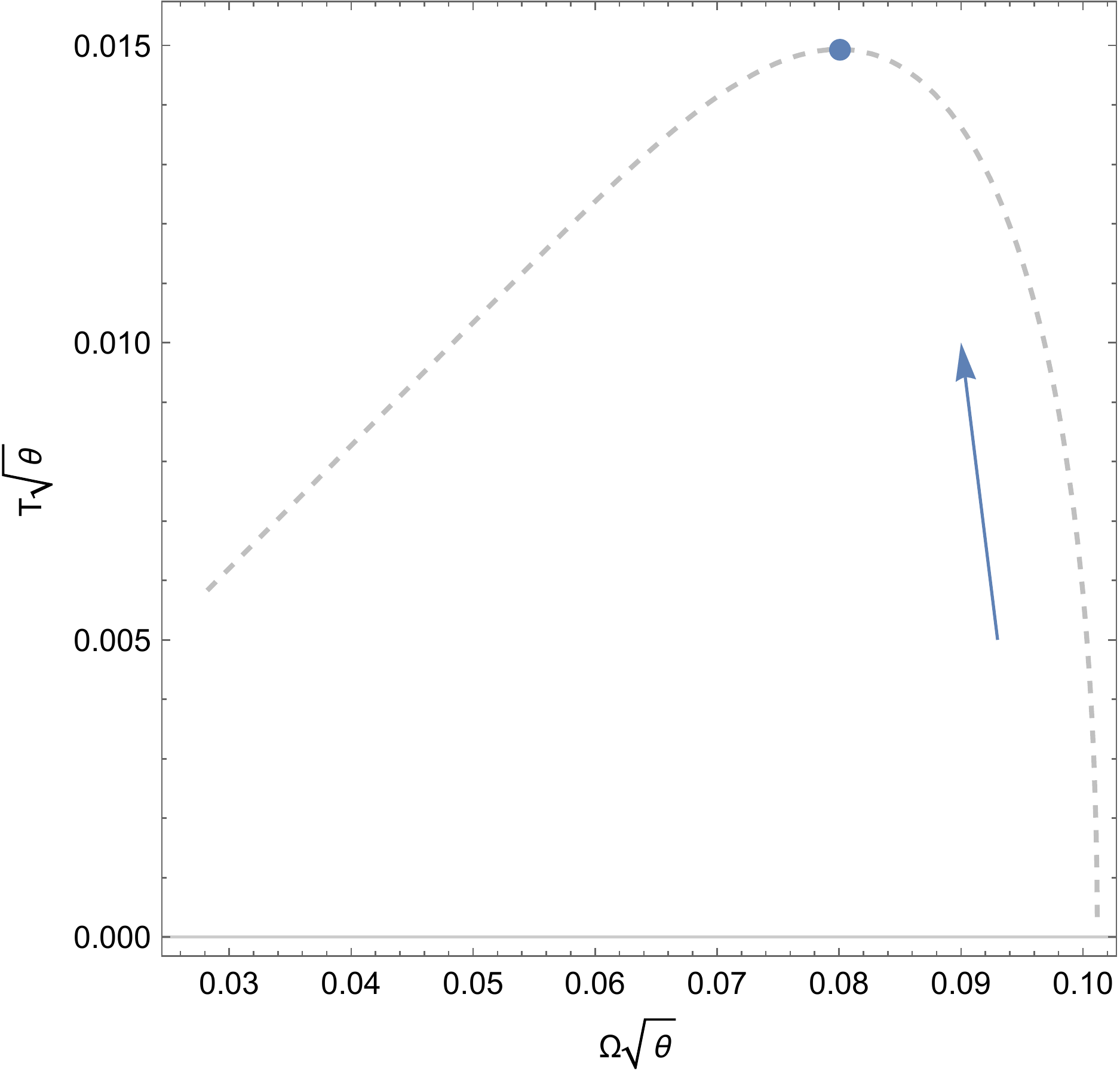}
    \includegraphics[width=0.35\textwidth]{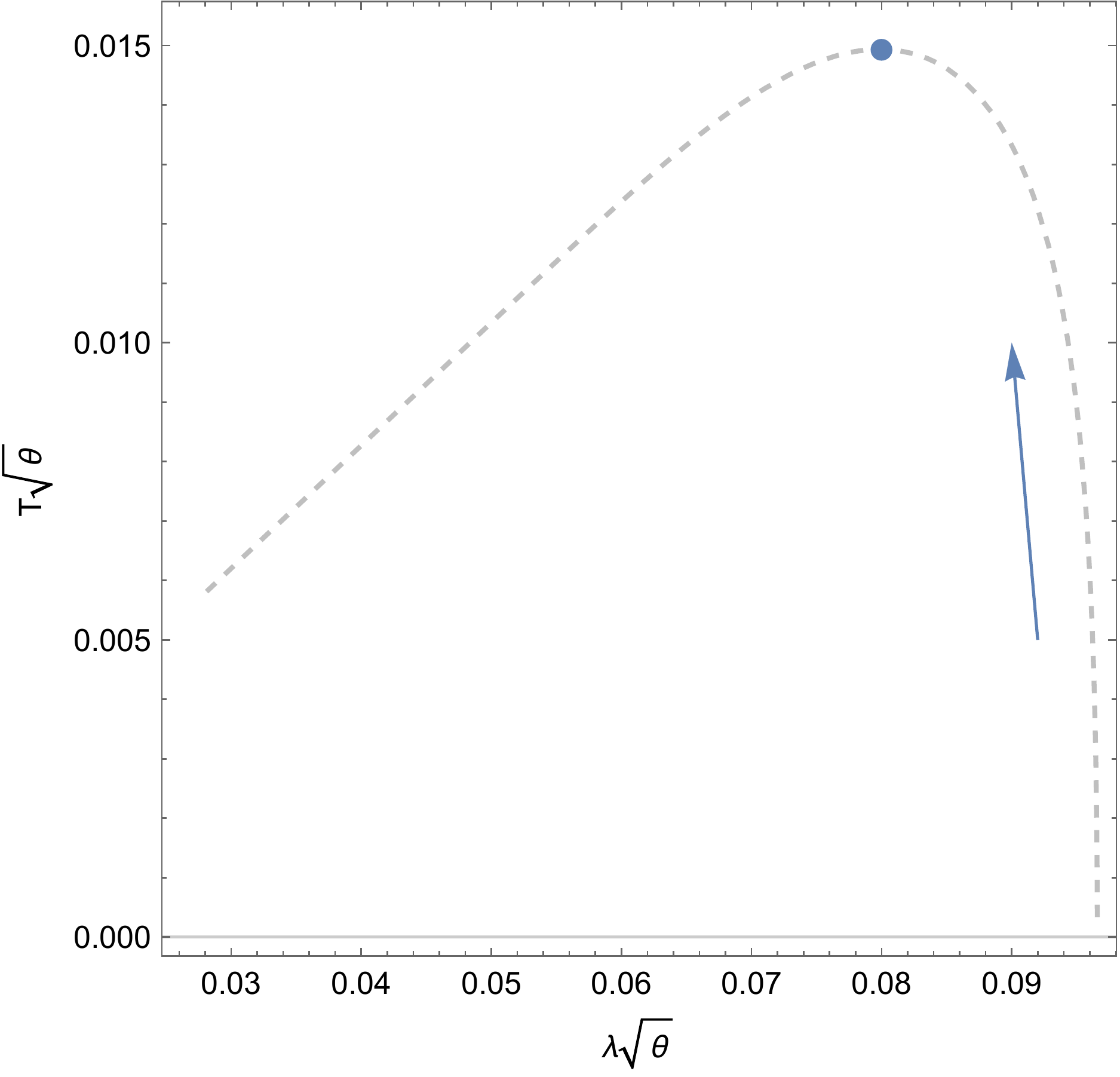}
    \caption{The dependence of temperature on the real and imaginary parts
    of QNMs, respectively, in the unit of $\theta$ for the noncommutative Schwarzschild BH.
    The arrows point to the direction of increasing $x_c$. The Davies points (blue points) are located at the maximum.}
    \label{fig:Davies-SBH2}
\end{figure}

The dependence of temperature on the real and imaginary parts of QNMs in the unit of $\theta$ is shown in
Fig.\ \ref{fig:Davies-SBH2}, where the Davies points are located at the maximum of the curves as expected.

\section{4D Einstein-Gauss-Bonnet black holes}
\label{sec:EGB-BH}

\subsection{Horizons and regularity}

The 4D EGB BH was given in Ref.\ \cite{cai2009} where a conformal gravity  with a trace anomaly was considered. After a reparametrization, we can write the shape function,\footnote{This BH solution was rediscovered~\cite{{glavan2019}} by means of a rescaling of a higher dimensional  EGB model to the four dimensional one, but such a rescaling was proved~\cite{metin2020there} to be inconsistent with the equation of motion of the four dimensional model.}
\begin{equation}
\label{eq:egb-solution-4}
f(r)=1+\frac{r^2}{2\alpha}\left(
1\pm
\sqrt{1+\frac{8M \alpha}{r^3}}
\right),
\end{equation}
where $\alpha$ is Gauss-Bonnet coupling constant, and we can verify that the corresponding metric is finite at the center, $r=0$.

The branch with ``$+$'' sign is unstable as discussed in Ref.\ \cite{cai2009}.
Here we just consider the branch with ``$-$'' sign, and write the outer and inner horizons as follows,
\begin{equation}
\label{eq:horizon_a}
r_{\pm}=M\pm\sqrt{M^2-\alpha }.
\end{equation}
Note that the 4D EGB BH turns back to the Schwarzschild BH when $\alpha \to 0$. Moreover, one has the horizon radius for the extreme case,
$r_{\rm ext}=\sqrt{\alpha}$, and the corresponding area, $A_{\rm ext} = 4\pi \alpha$.

Now we calculate the scale indices for the 4D EGB BH and  list them in Tab.\ \ref{tab:appscale4}.
\begin{table}[htp]
\caption{Indices of scale factors of the 4D EGB BH.}
\begin{center}
\begin{tabular}{cc|cccccc}
\toprule
    {$M$} & {$\alpha$} & {$r$} & {$A$} & {$S$} & {$C$} & {$\kappa$} & {$T$}   \\ \midrule
    1  & {$2$} & {$1$} & {$2$} & {$2$} & {$2$} & {$-1$} & {$-1$}  \\ \bottomrule
\end{tabular}
\end{center}
\label{tab:appscale4}
\end{table}
Based on this table, we can choose the following rescaling for $r$ and $\alpha$, respectively,
\begin{equation}
r\to a M x,\qquad
\alpha \to a^2 M^2,
\end{equation}
where $x$ and $a$ are dimensionless, and obtain the rescaled outer horizon,
\begin{equation}
x_{\rm H}= \sqrt{\frac{1}{a^2}-1}+\frac{1}{a},
\end{equation}
from which we can fix the range of $a$, $0<a\le1$.
The above equation can be rewritten as the horizon equation,
\begin{equation}
\label{eq:a_xH}
a= \frac{2 x_{\rm H}}{x_{\rm H}^2+1},
\end{equation}
from which we get the horizon for the extreme 4D EGB BH, $x_{\rm ext}=1$.

Around the center, $x=0$, the leading order of the Ricci scalar takes the form,
\begin{equation}
\mathcal{R}=\frac{15}{2 a^{5/2} M^2 x^{3/2}}-\frac{12}{a^2 M^2}
+O\left(\sqrt{x}\right),
\end{equation}
while in the limit of $x\to \infty$, it reads
\begin{equation}
\mathcal{R}=-\frac{12}{a^4 M^2 x^6}
+O\left(x^{-7}\right).
\end{equation}
Namely, the 4D EGB BH is of dS core near the center and has asymptotic AdS behavior far out of the horizon. Although the Ricci scalar is divergent when $x\rightarrow 0$,  the gravitational force is repulsive at a short distance and thus an
infalling particle never reaches the center as mentioned in Ref.\ \cite{glavan2019}. Therefore, the regularity of 4D EGB BHs can be saved.

\subsection{Deformation of the first law of black hole mechanics}

\begin{figure}[h!]
    \centering
    \includegraphics[width=0.45\textwidth]{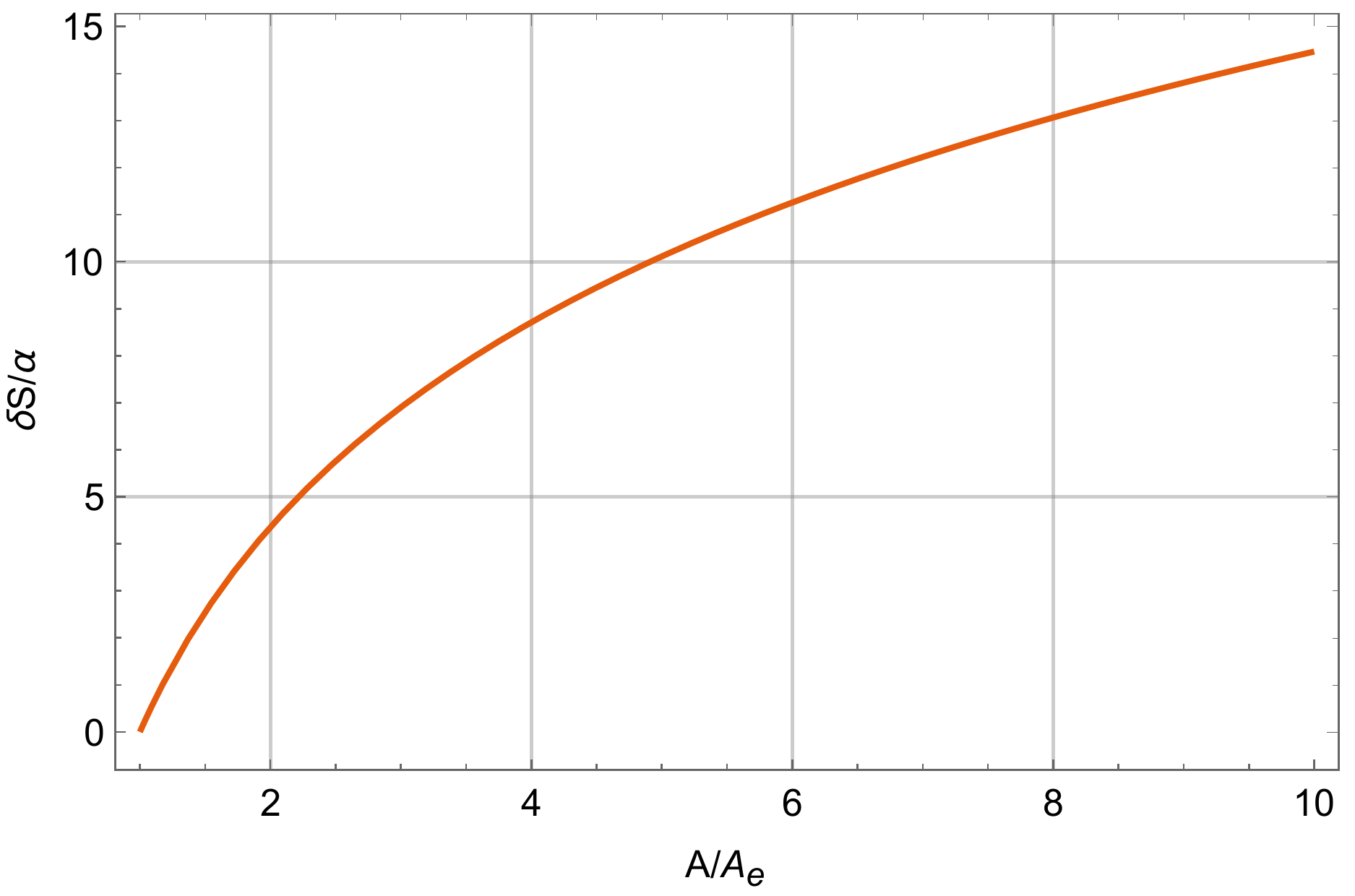}
    \caption{The dependence of entropy deviation on the BH area for the 4D EGB BH.}
    \label{fig:entropy-deviation-GB-BH}
\end{figure}

As we have known, the regularity of 4D EGB BHs will deform the first law of mechanics. The surface gravity of this BH equals
\begin{equation}
\kappa
=\frac{r^2_{\rm H}-\alpha }{2 r_{\rm H}\left(r^2_{\rm H}+2 \alpha  \right)}.
\end{equation}
Using Eq.~(\ref{eq:horizon_a}), we calculate the variation of mass,
\begin{equation}
\frac{r_{\rm H}}{\sqrt{M^2-\alpha}}d M=d r_{\rm H}.
\end{equation}
Then considering the variation of area, $d A = 8\pi r_{\rm H} d r_{\rm H}$,
we obtain the deformation of the first law of BH mechanics,
\begin{equation}
d M = \frac{\kappa}{8\pi} d A
+\frac{2\alpha}{r_{\rm H}^2+2\alpha} d M,
\end{equation}
 where the deformation disappears when $\alpha\rightarrow 0$. Correspondingly, the first law of BH thermodynamics reads

\begin{equation}
d M = T d S,\qquad
d S = \left(\frac{1}{4}+\frac{2\pi\alpha}{A}\right)d A.
\end{equation}
We thus obtain the entropy,
\begin{equation}
S=\frac{A}{4} +\delta S,
\qquad
\delta S =2 \pi \alpha \ln\left(\frac{A}{A_{\rm ext}}\right),
\end{equation}
where $A$ is bounded by $A_{\rm Sch}=16\pi M^2$.
$\delta S$ is regarded as the correction to the entropy, and it vanishes as $\alpha\to 0$,
see Fig.\ \ref{fig:entropy-deviation-GB-BH}.

\subsection{Heat capacity and Davies points}

\begin{figure}[h!]
    \centering
    \includegraphics[width=0.45\textwidth]{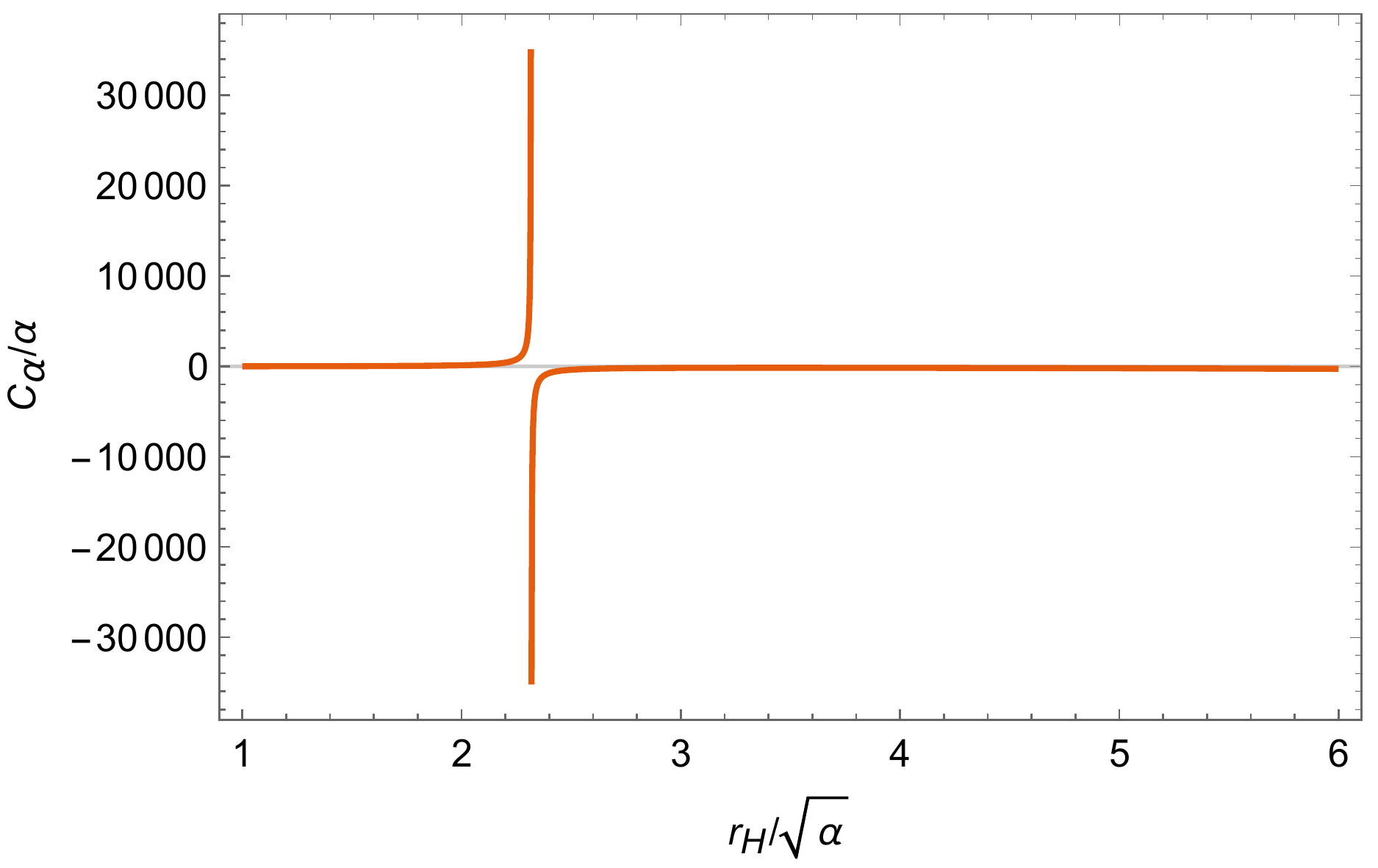}
    \caption{The heat capacity of the 4D EGB BH in the unit of $\alpha$.}
    \label{fig:capacity-GB-BH}
\end{figure}

The temperature can be represented in the unit of $M$ as
\begin{equation}
TM = \frac{x_{\rm H}^4-1}{8 \pi  x_{\rm H}^2 \left(x_{\rm H}^2+2\right)},
\end{equation}
and in the unit of $\alpha$ as
\begin{equation}
T\sqrt{\alpha} =
\frac{x_{\rm H}^2-1}{4 \pi x_{\rm H} \left(x_{\rm H}^2+2 \right)}.
\end{equation}
Moreover, the heat capacity can be rewritten in the unit of $M$ to be
\begin{equation}
C_\alpha/M^2 = -\frac{8 \pi  x_{\rm H}^2 \left(x_{\rm H}^6+3 x_{\rm H}^4-4\right)}{\left(x_{\rm H}^2+1\right)^2 \left(x_{\rm H}^4-5 x_{\rm H}^2-2\right)},
\end{equation}
and in the unit of  $\alpha$ to be
\begin{equation}
C_\alpha/\alpha =
-\frac{2 \pi  \left(x_{\rm H}^2-1\right) \left(x_{\rm H}^2+2\right)^2}{x_{\rm H}^4-5 x_{\rm H}^2-2}.
\end{equation}
Thus, the Davies point can be calculated analytically from $\alpha/C_\alpha=0$,
\begin{equation}
x^*_{\rm H}\equiv r^*_{\rm H}/\sqrt{\alpha }=  \sqrt{\frac{1}{2} \left(\sqrt{33}+5\right)}.
\end{equation}
We plot the heat capacity with respect to the horizon in Fig.\ \ref{fig:capacity-GB-BH}, which shows the existence of a second order phase transition in the 4D EGB BH.

\subsection{Quasinormal modes in the eikonal limit}

\begin{figure}[h!]
    \centering
    \includegraphics[width=0.4\textwidth]{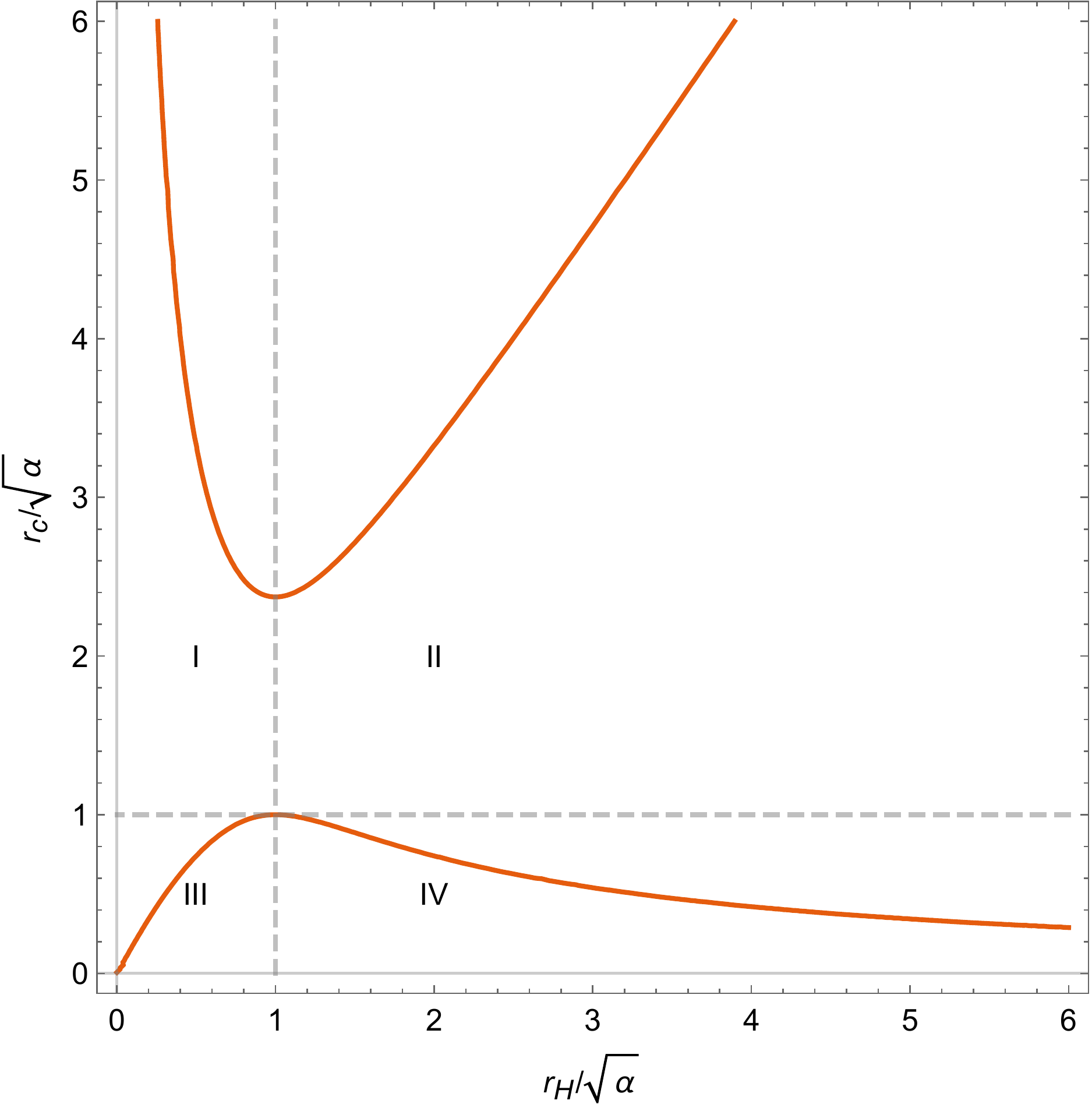}
    \caption{The BH-PS cone of the 4D EGB BH.
    The orange curves denote the BH-PS cones.
    The dashed gray lines correspond to the extremal value of the photon sphere radius and the horizon radius.}
    \label{fig:BH-PS-cone-GB}
\end{figure}

According to the equation of photon spheres \cite{cardoso2008},
the relationship between the dimensionless radius $x_c$ and parameter $a$ is
\begin{equation}
a^2 x_c^3-9 x_c+8 a=0.
\end{equation}
Replacing $a$ by the horizon $x_{\rm H}$ with the help of Eq.~(\ref{eq:a_xH}), we obtain the cone equation,
\begin{equation}
9 x_{\rm H}^4 x_c-16 x_{\rm H}^3-2x_{\rm H}^2(2  x_c^3-9  x_c)-16 x_{\rm H}+9 x_c=0,
\end{equation}
from which we solve the minimum of the upper cone, $x_c^{\rm min}=\left(\sqrt{33}-1\right)/2\approx 2.37$.
For the 4D EGB BH, the relation of the photon sphere radius versus the horizon radius is plotted in Fig.\ \ref{fig:BH-PS-cone-GB}.

\begin{figure}[h!]
    \centering
    \includegraphics[width=.4\textwidth]{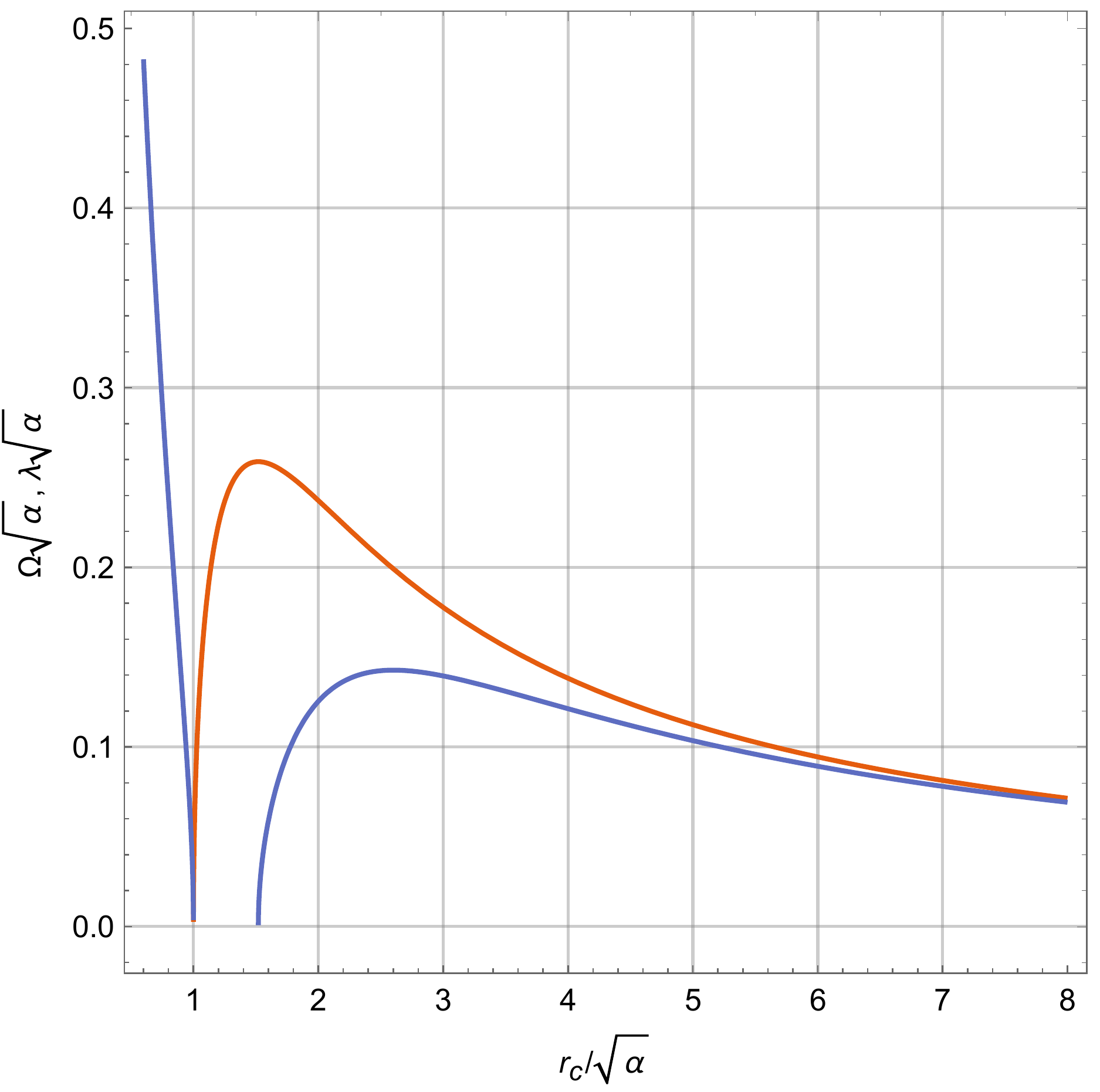}
    \caption{The QNMs
    with respect to the photon sphere radius in the unit of $\alpha$ for the 4D EGB BH.
    The orange curve denotes the real part of QNMs,
    while the blue curves denote the imaginary part defined in the range of $x_c>x_0=2/\sqrt[4]{3}$.}
    \label{fig:QNMs-GB-BH-alpha}
\end{figure}

Since there is a gap between the upper and lower cones,
we know that the imaginary part of QNMs must vanish in some range of the gap.
We compute the real and imaginary parts of QNMs, respectively, in the unit of $\alpha$,
\begin{equation}
\Omega^2 \alpha=
\left(\frac{\sqrt{9 x_c^4+16}+4}{9 \sqrt{2} x_c^5}\right)^2
\left[-9 x_c^4+8\left( \sqrt{9 x_c^4+16}-4\right)\right]
\left(-2-x_c^2+x_c^2\sqrt{1+\frac{8}{\sqrt{9 x_c^4+16}-4}} \right),
\end{equation}
and
\begin{eqnarray}
\lambda^2 \alpha&=&
\left[-9 x_c^4+8 \left(\sqrt{9 x_c^4+16}-4\right)\right]
\left(-2-x_c^2+x_c^2\sqrt{1+\frac{8}{\sqrt{9 x_c^4+16}-4}} \right)\nonumber \\
& &\times\left[-12+\sqrt{9 x_c^4+8 \left(\sqrt{9 x_c^4+16}+4\right)}\right]
\left(3 \sqrt{2} x_c^3 \sqrt[4]{9 x_c^4-8 \left(\sqrt{9 x_c^4+16}-4\right)}\right)^{-2}.
\end{eqnarray}
The corresponding graph of $\Omega$ and $\lambda$ versus $\alpha$ is shown in Fig.\ \ref{fig:QNMs-GB-BH-alpha}.
As mentioned in the above two sections, the imaginary part of QNMs vanishes in the range of photon sphere radii: $1\le x_c\le x_0=2/\sqrt[4]{3}\approx 1.52$, see the horizontal gap between the two blue curves in Fig.\ \ref{fig:QNMs-GB-BH-alpha}.
Because the minimal photon sphere radius $x_c^{\rm min}$ allowed by the physical region \rom{2} is larger than $x_0$, such a phenomenon of a vanishing imaginary part will never happen.

\begin{figure}[h!]
    \centering
    \includegraphics[width=0.4\textwidth]{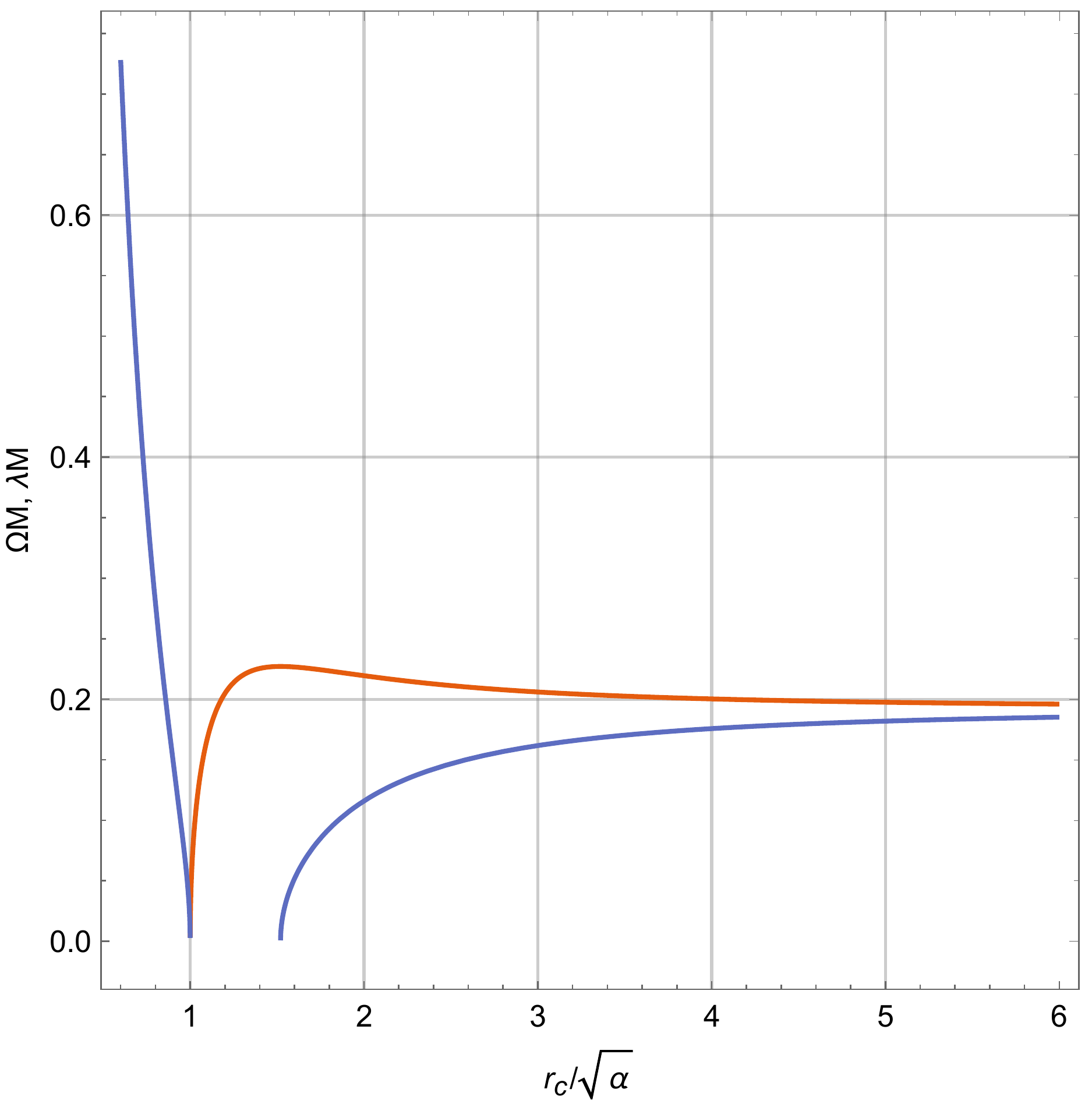}
    \caption{The QNMs
    with respect to the photon sphere radius in the unit of $M$ for the 4D EGB BH.
    The orange curve denotes the real part of QNMs,
    while the blue curves denote the imaginary part.
    There exists a gap for the imaginary part in the range of $1<x_c<2/\sqrt[4]{3}$, but this gap is outside the physical region.}
    \label{fig:QNMs-GB-BH-M}
\end{figure}

In addition, we compute the real and imaginary parts of QNMs, respectively, in the unit of $M$,
\begin{equation}
\Omega M
=
\frac{x_c^2}{\sqrt{6}}
\frac{ \sqrt{6+3 x_c^2-\sqrt{9 x_c^4+8\left( \sqrt{9 x_c^4+16}+4\right)}}}{\sqrt{9 x_c^4+16}-4},
\end{equation}
and
\begin{eqnarray}
\lambda^2 M^2 &=&
\left(6+3 x_c^2-\sqrt{9 x_c^4+8 \left(\sqrt{9 x_c^4+16}+4\right)}\right)\nonumber \\
& &\times\left(-12+\sqrt{9 x_c^4+8 \left(\sqrt{9 x_c^4+16}+4\right)}\right)
\left(54 \sqrt{9 x_c^4-8 \left(\sqrt{9 x_c^4+16}-4\right)}\right)^{-1}.
\end{eqnarray}
We plot the relation of the QNMs with respect to the photon sphere radius in the unit of $M$ for the 4D EGB BH in Fig.\ \ref{fig:QNMs-GB-BH-M}, where we can see the similar behaviors to those in the unit of $\alpha$ when we compare Fig.\ \ref{fig:QNMs-GB-BH-M} with Fig.\ \ref{fig:QNMs-GB-BH-alpha}.

\begin{figure}[h!]
    \centering
        \includegraphics[width=.7\textwidth]{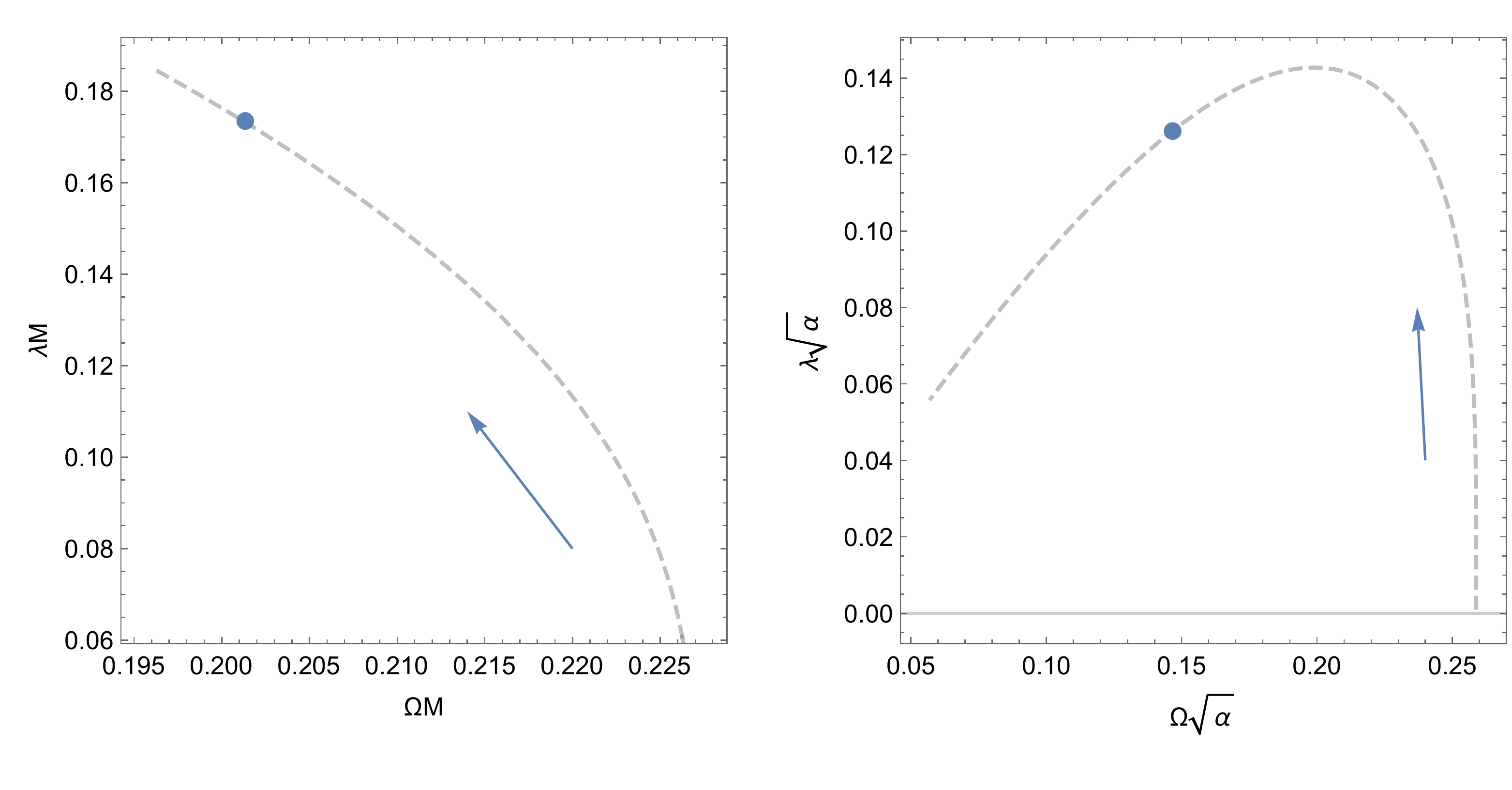}
    \caption{The relations of the real part versus the imaginary part of QNMs for the 4D EGB BH in the unit of $M$ (left) and $\alpha$ (right).
    The blue dots are Davies point. The arrows point to the direction of increasing parameter $x_c$.}
    \label{fig:Davies-GB-BH}
\end{figure}

\begin{figure}[h!]
    \centering
        \includegraphics[width=.7\textwidth]{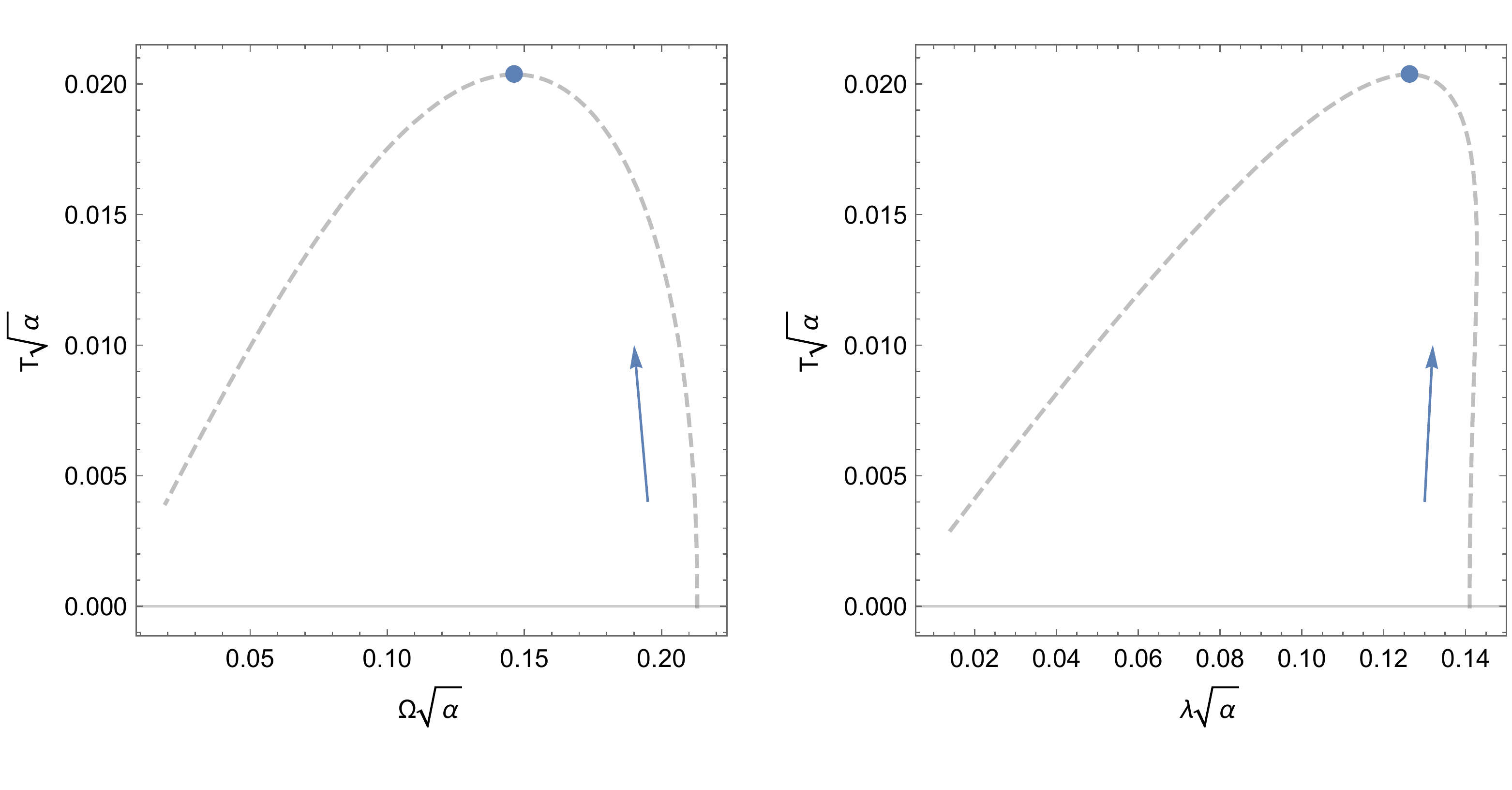}
    \caption{The dependence of the temperature on the real and imaginary parts
    of QNMs, respectively, in the unit of $\alpha$ for the 4D EGB BH.
    The arrows point to the direction of increasing parameter $x_c$.
    The Davies points (blue points) are located at the maxima of the curves.}
    \label{fig:OL-T-plane-GBBH}
\end{figure}

We also plot the relations of the real part versus the imaginary part of QNMs in the unit of $M$ and $\alpha$ for the 4D EGB BH in Fig.\ \ref{fig:Davies-GB-BH}. Note that we find the similar spiral-like shape to that in the two models we have considered in the above two sections. That is,  there exists a clear spiral-like behavior in the unit of $\alpha$, but no such a behavior in the unit of $M$.
Finally, we plot the dependence of the temperature on the real and imaginary parts of QNMs, respectively, in the unit of $\alpha$ for the 4D EGB BH in Fig.\ \ref{fig:OL-T-plane-GBBH}. We can see that the Davies points are located at the saddle points.

\section{Discussions and conclusions}
\label{eq:conclusion}

In this paper, starting with the 5D Myers-Perry BH as a sample,
we investigate the three models of regular BHs in terms of a dimensionless scheme.
They are the BV BH, the noncommutative Schwarzschild BH,
and the EGB BH in the 4-dimensional spacetime.
The regularity of the first two BHs is different from that of the last BH, that is, the former
has no singular points in background spacetimes, while the latter represents the fact that 
a particle can never reach the divergent region because the gravitational force becomes repulsive and tends to infinity.
What we focus on are the differences between regular BHs and singular (traditional) BHs,
in particular, in the aspects of QNMs and phase transitions,
and the relevant phenomena induced by QNMs and phase transitions.

\begin{itemize}
\item The entropy bounds are not universal in the models of regular BHs.

We have verified that the regularity of BHs is not compatible with the first law of BH mechanics.
By abandoning the linear correspondence between the mechanic and thermodynamic variables, we find that
the regularity of spacetime will lead to two results, where one is a broken first law of BH mechanics and the other  a deformation of entropy.

For the BV BH generated by nonlinear electrodynamics,
the entropy deviation equals
\begin{equation}
\delta S = \frac{A_{\rm Sch}}{4}  \left(\sqrt{\frac{A}{A_{\rm Sch}}}-\frac{A}{A_{\rm Sch}}\right),
\end{equation}
where $A_{\rm Sch}=16 \pi M^2$ is the area of the Schwarzschild BH.
Since $A< A_{\rm Sch}$, $\delta S$ is nonnegative.
When $A\to A_{\rm ext} = A_{\rm Sch}/e$,
\begin{equation}
\delta S \to \frac{A_{\rm Sch}}{4}
\left(\frac{1}{e}-\frac{1}{e^2}\right).
\end{equation}
For the noncommutative Schwarzschild BH,
the entropy deviation is also nonnegative,
\begin{equation}
 \delta S=8 \pi \theta
\int_{z_{\rm ext}}^{z} d \tilde z
\frac{\sqrt{\pi }-2 \gamma \left(\frac{3}{2},\tilde z\right)}{4 \gamma \left(\frac{3}{2},\tilde z\right)}.
\end{equation}
For the 4D EGB BH, the nonnegative correction to entropy takes the form,
\begin{equation}
\delta S =2 \pi \alpha \ln\left(\frac{A}{A_{\rm ext}}\right),
\end{equation}
where $A_{\rm ext}\le A\le A_{\rm Sch}$.
Therefore, the entropy bounds of the three regular BHs must be larger than
the Bekenstein bound of singular BHs, i.e. $S=A/4+\delta S > A/4$.
However, the origination of entropy corrections remains unclear at present.
$\delta S$ may be generated by some sort of entropy stored inside a BH, e.g. the Christodoulou-Rovelli (CR) entropy $S_{\rm CR}$ \cite{zhang2015}, or by the so-called quantum entropy,
\begin{equation}
\int \frac{P}{T}d V_{\rm CR},
\end{equation}
where $P$ is \emph{quantum} pressure caused by the vacuum polarization, and $V_{\rm CR}$ is the CR volume \cite{christodoulou2014}.
Here we just mention that $S_{\rm CR}-\int \frac{P}{T}d V_{\rm CR}$ should vanish for singular BHs, but may be nonzero for regular BHs. The detailed discussions will be reported elsewhere.

\item Davies point is a saddle point of temperature diagram.

Although the thermodynamics of regular BHs is quite different from that of singular BHs, the Davies points are always the maximum points on $\Omega$-$T$ and $\lambda$-$T$ planes if a proper parameter (except mass) is chosen to rescale quantities. For a BH with a single Davies point, the diagram of heat capacity is of two types, see Fig.\ \ref{fig:capacity-BH}.
\begin{figure}[h!]
    \centering
    \includegraphics[width=.7\textwidth]{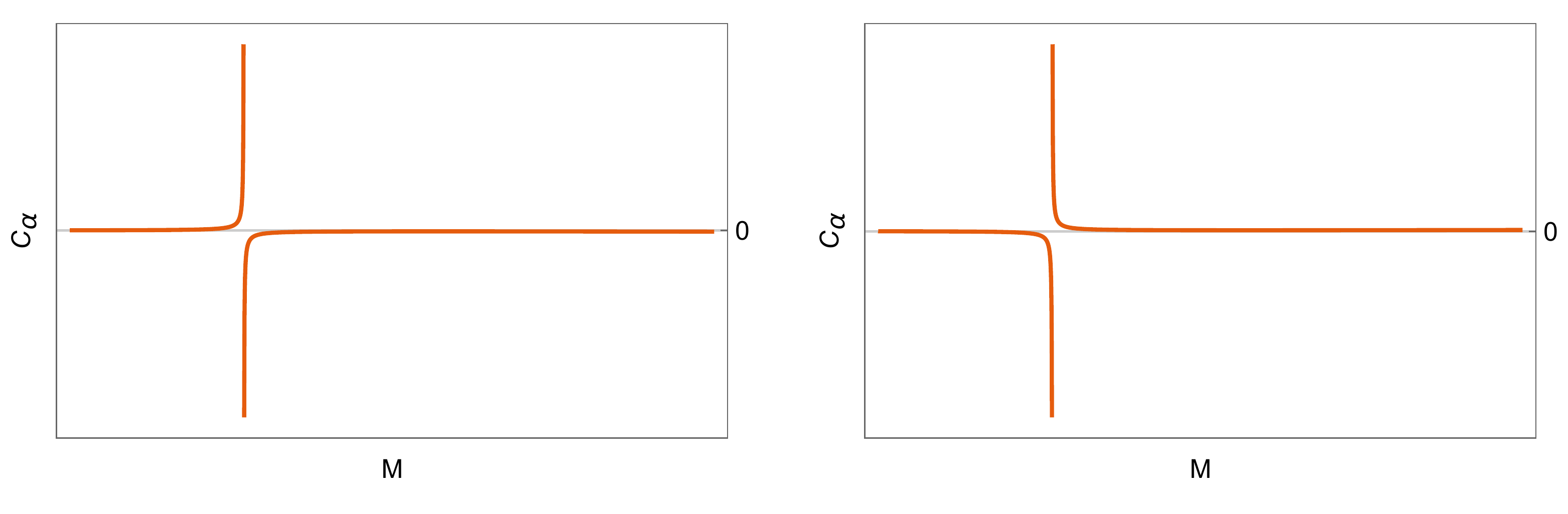}
    \caption{Two types of BH heat capacity.}
    \label{fig:capacity-BH}
\end{figure}

The left graph of Fig.\ \ref{fig:capacity-BH} depicts the case with $\partial C_\alpha/\partial M >0$,
which implies that the Davies point is the maximum of temperature.
Moreover, the Davies point separates the state of heat capacity into two phases. In the left phase,
the heat capacity is positive before the mass reaches its critical value at which the second order phase transition occurs, namely,
the temperature of BHs increases for a given ``heat''.
After the mass increases and exceeds its critical value, i.e. in the right phase,
the heat capacity becomes negative, namely,
for a given ``heat'' the temperature of BHs decreases and such a process is accompanied with the Hawking radiation.
The right graph of Fig.\ \ref{fig:capacity-BH} describes the inverse procedure, where
the Davies point is the minimum of temperature due to $\partial C_\alpha/\partial M<0$.
In the left phase, the heat capacity is negative and such a process is accompanied with the Hawking radiation;
then after the mass exceeds the critical point, i.e., in the right phase, the heat capacity becomes positive
and the Hawking radiation terminates. We note that
for the models of regular BHs we considered, $\partial C_\alpha/\partial M>0$,
i.e., the Davies points are the maximum of temperature.

\item QNMs of regular BHs in the eikonal limit have a spiral-like structure in a proper unit.

As we have observed, the spiral-like shapes discussed in Refs.\ \cite{jing2008,wei2019}
do not exist in the regular BHs if the QNMs are represented in the unit of $M$,
but they appear if the other parameters are adopted as units. The reason is that
the imaginary part of QNMs in the unit of $M$ is a monotonic function of $M$, but a non-monotonic function of the other parameters.

We apply the light ring/QNM correspondence to calculate QNMs.
According to the relationship between QNMs and photon sphere radii,
we can see that the imaginary parts of QNMs disappear in the range of $x_{\rm ext}<x_c<x_0$.
It seems that the regular BHs go into an oscillating stage without damping.
We further study this phenomenon from cone equations and discover
that such a phenomenon is in fact forbidden by the physical region of $x_{\rm H}-x_c$ graphs
in which the minimal photon sphere radius $x_c^{\rm min}$ is greater than $x_0$.
In addition, when the regular BHs decay to the final state,
where the horizon radius equals $x_{\rm ext}$ and the photon sphere radius $x_c^{\rm min}$,
neither the real nor the imaginary parts of QNMs vanish.
This is a universal property of the regular BHs considered in our current work,
which does not appear in singular BHs,
such as the Reissner-Nordstr\"om and Kerr BHs, etc.
From the classical point of view,
it may be difficult to understand why the BHs being in their final state still have damping contributions.
This puzzle gives the motivation for us to future investigate from the quantum point of view, e.g.
to study the canonical quantization of regular BHs in a minisuperspace \cite{christodoulakis2013}.

\end{itemize}


\section*{Acknowledgement}

C. L. would like to thank Huifang Geng (Nankai university) for the useful discussions.
This work was supported in part by the National Natural Science Foundation of China under grant No. 11675081.
C. L. is also supported by the Fundamental Research Funds for the Central Universities, Nankai university under the grant No. 63201006.
The authors would like to thank the anonymous referee for the helpful comments that improve this work greatly.

\appendix

\section{The Lagrangian for the anisotropic fluid}
\label{app:lagrangian-anisotropic}

The Lagrangian for the perfect fluid in general relativity has a long history \cite{Taub:1954zz,Schutz:1970my,ray1972lagrangian,jackiw2004perfect,Minazzoli:2012md,jackiw2013lectures}.
For the noncommutative matter introduced in Ref.~\cite{nicolini2005b},
the Lagrangian formalism considered in Ref.~\cite{Mann:2011mm}  explicitly depends on the spacetime coordinate.
In this appendix, we give an alternative one.

The stress-energy tensor is of the following form,
\begin{equation}
(T_\theta)^{\nu}_{\mu}={\rm diag}\left(-\rho_\theta, p_r, p_\perp, p_\perp\right),
\end{equation}
where $p_r=-\rho_\theta$ and $p_\perp=-\rho_\theta-\frac{r}{2}\frac{\partial\rho_\theta}{\partial r}$ for the anisotropic fluid considered in Ref.~\cite{nicolini2005b}, and
$\rho_\theta$ is mass density of gravitational source,
\begin{equation}
\rho_\theta = \frac{M}{(4\pi \theta)^{3/2}}\exp\left(-\frac{r^2}{4\theta}\right),
\end{equation}
which induces the noncommutativity of spacetime.
To derive the Lagrangian, we rewrite the stress-energy tensor by following Ref.~\cite{Mak:2001eb},
\begin{equation}\label{eq:TE-noncom}
(T_\theta)^{\mu\nu}=-(\rho_\theta+p_\perp) u^\mu u^\nu + p_\perp g^{\mu\nu}-\left(p_r-p_\perp\right)\chi^\mu \chi^\nu,
\end{equation}
where $u^\mu=f^{-1/2}(r)\delta^\mu_0$ is timelike four-velocity with normaliztion $g_{\mu\nu}u^\mu u^\nu=-1$;
$\chi^\mu=f^{1/2}(r)\delta^\mu_1$ is unit spacelike vector in the radial direction,
arising from the anisotropic character of the source.
Now following Jackiw's approach \cite{jackiw2013lectures},
we start with a Lagrange density,
\begin{equation}
\mathcal{L}=-j^\mu a_{\mu}+f\left(\sqrt{-j^{\mu} j_{\mu}}\right),
\end{equation}
where the auxiliary filed $a_\mu$ is irrotational,
i.e. $a_\mu=\partial_\mu\xi$,
 $\xi$ is a Clebsch parameter.
Meanwhile, we set the Lorentz current, $j_\mu=n u_\mu + c_1 \chi_\mu $, where 
$n$ is density and $c_1$ is a constant determined below.
Thus, the stress-energy can be derived,
\begin{equation}
(T_\theta)_{\mu\nu}
= g_{\mu\nu}\left[
f\left(j\right)+
j f'\left(j\right)
\right]+\frac{n^2 f'\left(j\right)}{j}
u_{\mu}u_\nu
+\frac{c_1^2f'\left(j\right)}{j}\chi_{\mu}\chi_\nu,\label{tel}
\end{equation}
where $j\coloneqq\sqrt{-j^{\mu} j_{\mu}}$,
and we have used $a_\mu =-f'(j)j_\mu/j $.
Then comparing Eq.~(\ref{tel}) with Eq. \eqref{eq:TE-noncom},
we find
\begin{equation}
c_1^2=n^2 \frac{p_r-p_\perp}{\rho+p_\perp},
\qquad
j^2=n^2 \frac{\rho_\theta-p_r+2p_\perp}{\rho+p_\perp},\qquad f\left(j\right)=-\rho_\theta+p_r-p_\perp.
\end{equation}
We note that $f\left(j\right)$ recovers the well-known result of a perfect fluid for the isotropic case ($p_r=p_\perp$).

\vskip 6mm
\noindent
{\bf\large References}


\bibliographystyle{model1a-num-names}

\bibliography{references}

\end{document}